\documentclass{aa}

\usepackage{graphicx}
\usepackage{graphbox}
\usepackage{xcolor}
\usepackage[position=top]{subfig}
\usepackage{amsmath}
\usepackage[utf8]{inputenc}
\usepackage{epsfig,amsmath,amssymb}
\usepackage[varg]{txfonts}
\usepackage{multirow}
\usepackage{enumitem}

\usepackage[rightcaption]{sidecap}
\usepackage{float}

\def\ga{\,\hbox{\hbox{$ > $}\kern -0.8em \lower 1.0ex\hbox{$\sim$}}\,}
\def\la{\,\hbox{\hbox{$ < $}\kern -0.8em \lower 1.0ex\hbox{$\sim$}}\,}
\def\beq{\begin{equation}}
\def\eeq{\end{equation}}

\definecolor{myblue}{rgb}{0.00, 0.0, 0.9}
\definecolor{myred}{rgb}{0.90, 0.0, 0.0}
\definecolor{mygreen}{rgb}{0.0, 0.7, 0.0}
\definecolor{Plum}{rgb}{0.5, 0.0, 0.6}

\usepackage[breaklinks,  citecolor=myblue, linkcolor=myred, urlcolor=purple, colorlinks=true, debug, baseurl=' ']{hyperref}

\titlerunning{Dust polarization from global MHD galaxies}
\authorrunning{Pelgrims et al.}

\begin{document}

%
\title{The effect of cosmic variance on the characteristics of dust polarization power spectra}

\author{V. Pelgrims \inst{1} \fnmsep \inst{2} \fnmsep \thanks{pelgrims@physics.uoc.gr},
E. Ntormousi \inst{3} \fnmsep \inst{1}
\and
K. Tassis \inst{1} \fnmsep \inst{2}}
\date{Received 27 July 2021 / Accepted 4 January 2022}

\institute{
Institute of Astrophysics, Foundation for Research and Technology-Hellas, Vasilika Vouton, GR-70013 Heraklion, Greece
\and
Department of Physics \& ITCP, University of Crete, GR-70013, Heraklion, Greece
\and
Scuola Normale Superiore,
Piazza dei Cavalieri, 7
56126 Pisa, Italy}

\abstract{
In the context of cosmic microwave background polarization studies and the characterization of the Galactic foregrounds, the power spectrum analysis of the thermal dust polarization sky has led to intriguing evidence of an $E/B$ asymmetry and a positive $TE$ correlation.
In this work, we produce synthesized dust polarization maps from a set of global magneto-hydrodynamic (MHD) simulations of Milky-Way-sized galaxies, and analyze their power spectra at intermediate angular scales (intermediate angular multipoles $\ell \in \left[60 ,\, 140\right]$). We study the role of the initial configuration of the large-scale magnetic field, its strength, and the feedback on the power spectrum characteristics. Using full-galaxy MHD simulations, we were able to estimate the variance induced by the peculiar location of the observer in the galaxy.
We find that the polarization power spectra sensitively depend on the observer's location, impeding a distinction between different simulation setups.
In particular, there is a clear statistical difference between the power spectra measured from within the spiral arms and those measured from the inter-arm regions. Also, power spectra from within supernova-driven bubbles share common characteristics, regardless of the underlying model. However, no correlation was found between the statistical properties of the polarization power spectra and the local (with respect to the observer) mean values of physical quantities such as the density and the strength of the magnetic field.
Finally, we find some indications that the global strength of the magnetic field may play a role in shaping the power spectrum characteristics; as the global magnetic field strength increases, the $E/B$ asymmetry and the $TE$ correlation increase, whereas the viewpoint-induced variance decreases.
However, we find no direct correlation with the strength of the local magnetic field that permeates the mapped region of the interstellar medium.
}

\keywords{ISM: dust, magnetic fields --
   submillimeter: ISM --
   polarization --
   MHD simulation --
   (cosmology) cosmic background radiation}

\maketitle

\section{Introduction}
In order to derive stringent constraints on the cosmological parameters from the study of the cosmic microwave background (CMB) radiation, the {\it Planck} satellite mapped the whole sky at 353~GHz in both intensity and polarization (e.g., \citealt{PlaI2020}). At this frequency, the polarized sky is dominated by the polarized thermal emission from dust grains of the magnetized interstellar medium (ISM) of the Galaxy.
Because most of the cosmological information can be obtained from the CMB radiation through the analysis of its $T$, $E$, and $B$ angular power spectra (intensity and linear polarization) and their correlation (e.g., \citealt{Kam1997a,Kam1997b,Zal1997}), the dust polarization sky has been characterized using that metric.
The characterization of the dust polarization auto- and cross-angular power spectra has led to several properties that intrigued the community (\citealt{PlaXXX2016}; \citealt{PlaXI2020}). Among those properties, the most significant ones are the observed values for the $E/B$ asymmetry and the $TE$ correlation; that is to say, that the power in the $E$-mode polarization is about twice than that of the $B$ modes and that the $E$ modes show a positive cross-correlation with the total intensity ($T$).

It was quickly hypothesized that these features in the polarization power spectra should be related to the physics of the ISM, at least to some extent and on certain angular scales. In particular, it has been argued that they might reflect the correlation between the orientations of anisotropic density structures and magnetic field lines, which is expected from magneto-hydrodynamic (MHD) physics in some regimes. This is an interpretation that is supported by observational evidence of such alignments (e.g., \citealt{Cla2014,PlaXXXVIII2016}).

\smallskip

Several authors have invoked the turbulent properties of the ISM as the physical origin of these statistical features. 
\cite{Cal2017} and \cite{Kan2017} developed theoretical methods to derive polarization power spectra of thermal dust emission from MHD turbulence theory.
Both works suggest that the observed $E/B$ power asymmetry and $TE$ correlation could be predictive of the MHD turbulence parameters in the ISM.
In fact, \cite{Kri2018} demonstrate that dust polarization maps synthesized from multiphase MHD turbulent simulations of the local ISM 
\citep{Kri2017} can naturally lead to the observed $E/B$ asymmetry.
Along the same lines, \cite{Kim2019} analyzed the power spectra of synthetic dust polarization maps from simulations of multiphase, supernova-driven MHD turbulence in a kiloparsec-sized stratified shearing box \citep{Kim2017}.
They report an $E/B$ asymmetry and a $TE$ correlation roughly consistent with observations, although at somewhat lower values. Interestingly, they also report fluctuations of the power spectrum characteristics with the observer's galactic environment (simulated by varying the differential rotation rate of the box) and with temporal variations of the ISM properties due to bursts of star formation.
\cite{Bran2019} show that helical turbulence induces strongly asymmetric distributions of $E$-mode contributions along the line of sight whereas $B$-mode contributions are distributed more symmetrically. In this view, the $E/B$ asymmetry naturally stems from the fact that, on average, for a given line-of-sight, there is less cancellation of $E$- compared to $B$-modes.
This observation remains true on large scales if the Sun is embedded into a large-scale helical field \citep{Bra2019a}.
According to the evidence gathered so far, the dust polarization power spectra contain information about the properties of the magnetized ISM.
However, the variance induced by the observer's viewpoint (also referred to as cosmic variance) might be strongly diluting this information.

In this paper, we quantify the effect of cosmic variance on the $E/B$ power asymmetry and the $TE$ correlation. For this purpose, we use a set of global MHD simulations of Milky-Way-sized galaxies to synthesize full-sky observations of polarized thermal dust emission at 353~GHz from different observer positions and study the corresponding power spectra.
With these calculations we are not aiming to replicate (or model) the \textit{Planck} observations. Instead, we are looking for observable trends based on physical processes, which we can then use to interpret polarization power spectra of Galactic observations.

Our input galaxy models are the six simulations presented in \cite{EN2018} (hereafter N18). They include different initial magnetic fields (both in topology and strength) and are realized with or without feedback.
Although the selected models are not exact replicas of the Milky Way, they cover a range of parameters wide enough to address our objective: different initial conditions for the magnetic field, which represent different realizations of the ordered component of the magnetic field; the presence or not of feedback, which allows us to explore the impact of the turbulent component of the magnetic field on the observables; and a complex three-dimensional environment in which to move the hypothetical observer.
Among the (few) galactic models with magnetic fields available to the community, these models have the additional benefit of having a divergence-free magnetic field.

\smallskip

In Sect.~\ref{sec:2} we present the galaxy simulations, the polarized thermal dust emission model, and our method to synthesize dust polarization maps from the simulation grid.
In Sect.~\ref{sec:3} we present the computation of the polarization power spectra and define the quantities of interest.
Section~\ref{sec:4} contains our main results: the measurements of the $E/B$ asymmetry and $TE$ correlation in the whole sample and per galaxy setup. Their we also explore the viewpoint-induced variance in more details.  
Finally, Sect.~\ref{sec:ccl} contains summarizing comments and  conclusions.

This document contains appendices: Appendix~\ref{app:conventions} illustrates the convention used in this work to synthesize maps from a simulation performed on an Adaptive Cartesian grid. Appendix~\ref{app:raytracing} explains details of the ray-tracing algorithm we developed. Finally, in Appendix~\ref{app:interpolation} we show that creating synthetic maps using interpolation schemes can lead to artifacts in the resulting power spectra.
%
\section{Synthetic dust polarization maps}
\label{sec:2}

\subsection{MHD galaxies}
\subsubsection{Numerical code}
The galaxy simulations, presented in N18, are performed with the publicly available MHD code RAMSES \citep{Teyssier_02}, which solves the ideal MHD equations on a Cartesian grid and has Adaptive Mesh Refinement (AMR) capabilities. RAMSES uses a constrained transport scheme to evolve the magnetic field $\mathbf{B}$, which guarantees $\nabla\cdot{\bf{B}}=0$ always \citep{Fromang_2006}.  
This provides a significant advantage as compared to codes that rely on divergence-cleaning algorithms which are known to create spurious effects in studies of turbulent environments \citep{Balsara_2004}.  

\begin{figure}
    \centering
    \rotatebox{90}{ \hspace{1.4cm} {\large{Distribution Function}}} \hspace{.2cm}
    \includegraphics[trim={1.25cm 1.25cm .6cm 0.0cm},clip,width=.90\columnwidth]{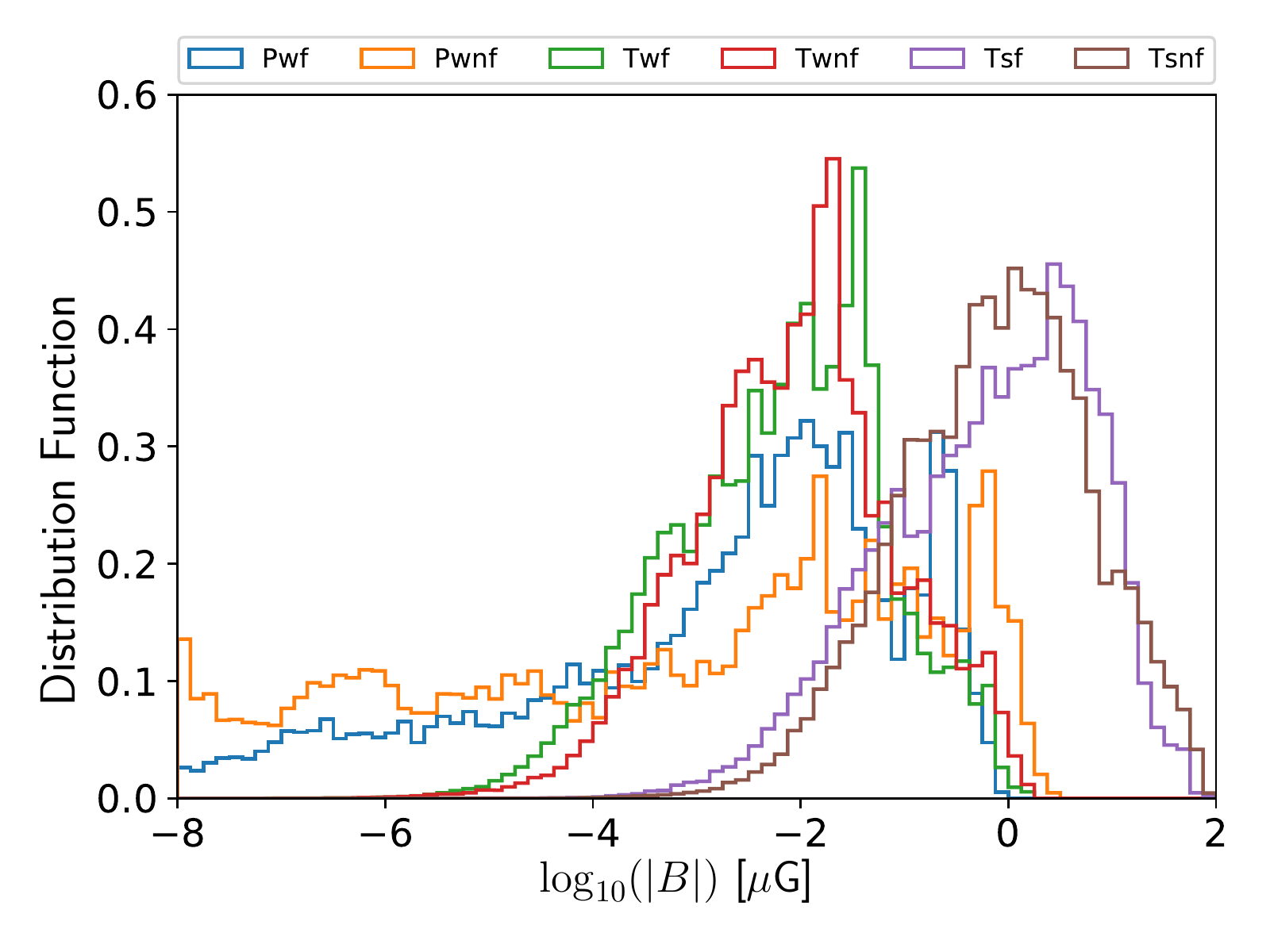} \\
    \vspace{.1cm}
    \rotatebox{90}{ \hspace{2.cm} {\large{Cumulative}}} \hspace{.2cm}
    \includegraphics[trim={1.25cm 1.25cm .6cm 1.25cm},clip,width=.90\columnwidth]{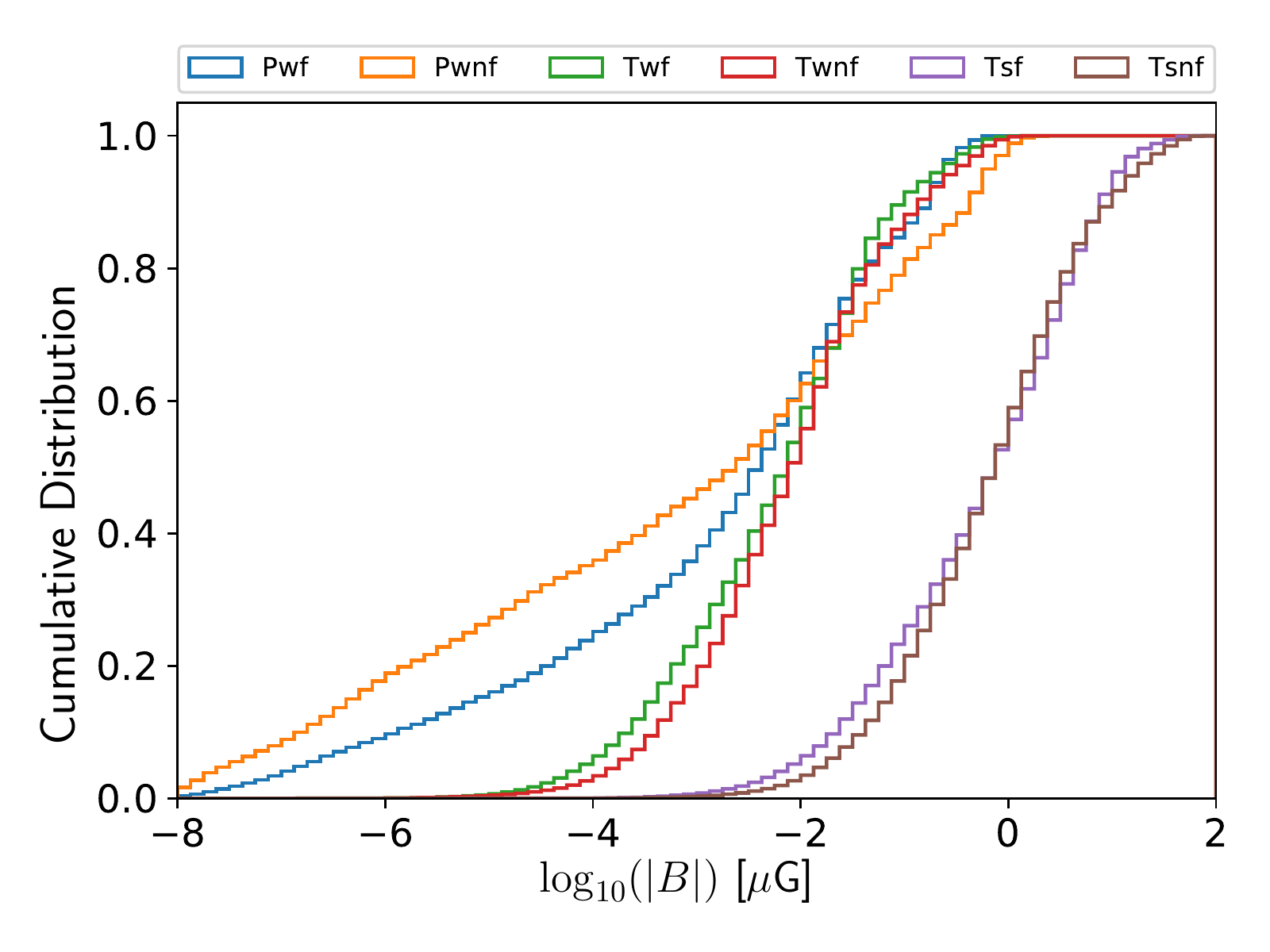} \\
    {\hspace{1.cm} \large $\log_{10}(|\mathbf{B}|)$ [$\mu$G]}
    \caption{Normalized distribution (top) and cumulative distribution (bottom) of the mass-weighed magnetic field strength ($\log_{10}(|\mathbf{B}|)$) in the disk region
    for each simulation as indicated by the legend. We define the disk region as a vertical cylinder centered on the center of the simulation box, with radius of 15 kpc and extending up to 500 pc above and below the plan $Z=0$.}
    \label{fig:Bn_histo}
\end{figure}
\begin{table*}
    \caption{Summary of the galaxy models.}
    \label{tab:sim_table}
    \centering
    \begin{tabular}{lcccl | c c }
      \hline
    \hline \\[-.5ex]
    Label  &  $\mathbf{B}_{t_0}(0)$ [$\mu$G]  & $\mathbf{B}$ morphology &  feedback  & N18 model & age [Myr] & $\tilde{\mathbf{B}}_t$ [$\mu$G]\\ \\[-1.5ex]
    \hline \\[-.5ex]
     Pwf    & 0.1   & poloidal  & yes   & fb\_b100\_P   & 21.2  & $0.003$  \\
     Pwnf   & 0.1   & poloidal  & no    & nofb\_b100\_P & 20.8  & $0.002$  \\
     Twf    & 0.1   & toroidal  & yes   & fb\_b100\_T   & 21.3  & $0.006$  \\
     Twnf   & 0.1   & toroidal  & no    & nofb\_b100\_T & 20.7  & $0.007$  \\
     Tsf    & 1     & toroidal  & yes   & fb\_b1\_T     & 25.7  & $0.562$  \\
     Tsnf   & 1     & toroidal  & no    & nofb\_b1\_T   & 24.1  & $0.562$  \\ \\[-.5ex]
    \hline
  \end{tabular}
  \tablefoot{ $\mathbf{B}_{t_0}(0)$ and morphology refer to the initial conditions. N18 model refers to the model names given in \cite{EN2018}. $\tilde{\mathbf{B}}_t$ is the value at which the mass-weighted cumulative histogram of magnetic field values in the disk region of the simulation snapshot equals 0.5. (see caption of Fig.~\ref{fig:Bn_histo}).}
\end{table*}
\subsubsection{Setup}
The initial conditions for the MHD fluid, the stellar and dark matter particles were created using DICE \citep{Perret_2014,Perret_2016}. They represent a Milky-Way-like galaxy (total mass $M_{tot}=2\cdot10^{12}~M_{\odot}$) at redshift zero with different initial morphologies and strengths of the magnetic field. The virial velocity of the galaxy is 200 km/s, the mass fraction in stars is about $4.5\%$ (including a stellar bulge with a mass fraction of 0.5$\%$) and the mass fraction in gas is about $1\%$.  
The dark matter halo follows a NFW profile \citep{NFW96}, while the gas and stars are initially placed in exponential disks with a scale length of 9~kpc.  The galactic disk contains two spiral arms, starting at 2~kpc and ending at 12~kpc. The mean gas temperature is set at $10^4$ K and subsonic turbulence, with an $rms$ value of 8~km/s, is introduced throughout the disk.

The base (coarsest) simulation grid has a resolution of $256^3$ cells filling the box of 60~kpc$^3$. Four AMR levels are used to capture the disk dynamics, and an additional AMR level in regions of star formation. In physical units, the highest resolution corresponds to 29~pc and the lowest to 234~pc.

Table~\ref{tab:sim_table} contains the initial parameters of the different galaxy models. The parameters varied are the strength and the initial topology of the magnetic field.
The initial field is either toroidal (labels start with T), with a scale height and scale length of 1~kpc, or a poloidal  field (labels start with P), with a scale height of 1~kpc and a scale length of 2~kpc.  The poloidal magnetic field is model C from \citet{Ferriere_Terral_2014}.
Star formation is simulated in all models by forming sink particles when the density exceeds 1000 cm$^{-3}$. Models whose label contain ("nf") "f" (do not) include stellar feedback from supernovae, resulting from previously formed sink particles with a time delay of 3~Myrs. Supernovae are implemented by injecting thermal energy into the cells around the sink particle according to the number of supernovae estimated for the predicted size of the formed stellar cluster.
Table~\ref{tab:sim_table} also contains the physical time of the outputs we use in this work and the mass-weighed median of the magnetic field strength at that time. Unfortunately, evolving these galaxies at this resolution is very computationally demanding, so we only have one stellar generation (30~Myrs). 

The power spectra of magnetic and kinetic energy of these galaxies are given in N18. Histograms and cumulative histograms of the mass-weighed magnetic field strength are shown in Fig.~\ref{fig:Bn_histo} for the six galaxies.

\subsection{Synthetic dust polarization maps}
\subsubsection{Synthesis}
Mapping 3D data to 2D through line-of-sight integration generally leads to artifacts and systematics that depend on the integration method.
In the specific case of mapping a (nonuniform) Cartesian onto a spherical grid, certain regions are over- or under-sampled and spurious effects appear.

One approach for removing these artifacts is to interpolate the values of neighboring cubic cells at the positions of the spherical grid. However, this process can lead to a mixing of angular scales, which in turn has a measurable effect on the power spectra (see Appendix~\ref{app:interpolation} for a detailed explanation). Therefore, instead of interpolating, here we sample the Cartesian grid with a spherical grid sufficiently dense to include all the mass in the simulation.
Going one step further in this direction, we develop a ray-tracing algorithm that computes, for each line of sight, the path length through each intervening cell of the Cartesian grid. Our algorithm, presented in more detail in Appendix~\ref{app:raytracing}, maps 3D Cartesian grids on to 2D maps that follow an HEALPix tessellation with a resolution parameter $N_{\rm{side}}$ (\citealt{Gor2005}). $N_{\rm{side}}$ is the only free parameter of our map-making process. 
\smallskip

To produce the synthetic maps used in this work, we place an observer in the $xy$ plane of the simulated galaxies at a radial distance of 8~kpc from the center, that is roughly the distance from the Sun to the Galactic center. From each observer location (see Sect.~\ref{subsec:map_making}) we integrate the simulations to produce 2D maps of the required observables using the algorithm described above.

We fix the targeted map resolution setting the HEALPix parameter $N_{\rm{side}} = 128$. Technically, because the minimum side length of the cells populating the galactic disk is 29 pc and the maximum distance along the line-of-sight is about 50 kpc, ensuring that every single cell is crossed by at least one line of sight would require an $N_{\rm{side}}$ of 2048.
However, by construction, the smallest cells are found toward high density regions and, therefore, mostly in the galactic disk. Since we are about to disregard the disk by masking out the bright (mostly equatorial) regions, an $N_{\rm{side}}$ of $128$ is sufficient for the purposes of our analysis. Indeed, an $N_{\rm{side}}$ of 128 allows us to ray-trace the smallest cells (29pc) up to a distance of about 3.5~kpc (i.e., at least one sightline goes through each cell).
Toward the poles, such a distance is about an order of magnitude larger than the scale height of the disk inferred in our simulations. Of course, going at lower latitudes, material at a lower distance from the disk (lower $|Z|$), could be missed. We checked that synthetic maps computed with an $N_{\rm{side}}$ of 512 (effectively ``missing'' no material up to 15~kpc distance, but also tracing each voxel with 16 times more sightlines) do not show difference in the sky regions of interest.
Subtle differences were spotted only in the disk and toward the galactic center, both of which we disregard in this paper.

\subsubsection{Polarized dust emission model}
Each cell of the grid contains information on the matter density and the magnetic field. We use these physical quantities to produce synthetic maps of the thermal dust polarized emission as seen at $353$~GHz.

We start from the integral equations for the Stokes $I$, $Q$, and $U$ similar to those given by \cite{Lee85}, \cite{PlaXX2015} (Appendix~B) and reviewed in \cite{Pel2020}.
For optically thin emission at frequency $\nu$ and following the HEALPix convention for the polarization position angle\footnote{\url{https://healpix.jpl.nasa.gov/html/intronode12.htm}}, we write:
\begin{equation}
\label{eq:simulated_I}
I=\int S_\nu\,\left[1- p_0\left(\cos^2\gamma-\frac{2}{3}\right)\right]\,
n_\mathrm{H} \, \sigma_\mathrm{H} \, \mathrm{d}s \, ,
\end{equation}

\begin{equation}
\label{eq:simulated_Q}
Q=\int p_0\,S_\nu\,\cos\left(2\phi\right)\cos^2\gamma\,
n_\mathrm{H} \, \sigma_\mathrm{H} \, \mathrm{d}s \, ,
\end{equation}
\begin{equation}
\label{eq:simulated_U}
U=\int p_0\,S_\nu\,\sin\left(2\phi\right)\cos^2\gamma\,
n_\mathrm{H} \, \sigma_\mathrm{H} \, \mathrm{d}s \, ,
\end{equation}
where the integrals are computed along the line of sight over the emitting region (here, the full simulation box); $S_\nu$ is the source function, $p_0$ a parameter related to dust polarization properties combining grain asymmetric cross sections and the degree of alignment with the magnetic field, $n_\mathrm{H}$ the gas density, $\sigma_\mathrm{H}$ the dust cross-section per hydrogen atom averaged over angles, $\gamma$ the angle of the local magnetic field to the plane of the sky, and $\phi$ the local polarization angle (see Fig.~14 in \citealt{PlaXX2015}).
Following \cite{PlaXLIV2016} and \cite{Kim2019}, we adopt a universal value for $p_0 = 0.2$, the black-body source function $B_\nu$ with a constant dust temperature of $18$ K and a dust opacity at $353$ GHz of $\sigma_\mathrm{d,353} = 1.2 \times 10^{-26} \, \rm{cm}^{-2}$. A constant dust-to-gas ratio is assumed.

In this work, we want to understand the influence of the ISM conditions on the observables. Therefore, we neglect any variations in the emissivity of the dust grains throughout the Galaxy (despite known observational evidence for slight variations in the Milky Way; \cite{Fin1999}; \citealt{PlaXI2014}; \citealt{Pel2021a}). We also assume that the dust grain alignment physics is the same throughout the Galaxy, an assumption that is valid on Galactic scales (\citealt{Rei2020}; \citealt{Sei2019}; \citealt{Van2021}).

It is important to notice that, unlike in works using parametric modeling, such as \cite{PlaXLIV2016} and \cite{Pel2020}, the parameter $p_0$ is not intended to account for variations in the magnetic field orientation along the line of sight (in the integral step). 
The reason is that, unlike parametric modeling, direct numerical simulations follow the evolution of fluctuations self-consistently.
Additional fluctuations at the subgrid level would interfere with the physics encoded in the simulations, and they would alter the power spectra and presumably their synthesis characteristics (\citealt{Rei2019}) in a way that depends on the type of fluctuations and on the implementation (see, e.g., Sect.~4.3 of \citealt{Wan2020}).

\begin{figure}
    \centering
    \vspace{.5cm}
    $\log_{10}(I_{\rm{353}})$ \\ [1.ex]
    \includegraphics[trim={0.0cm .0cm 0.0cm .8cm},clip,width=.98\columnwidth]{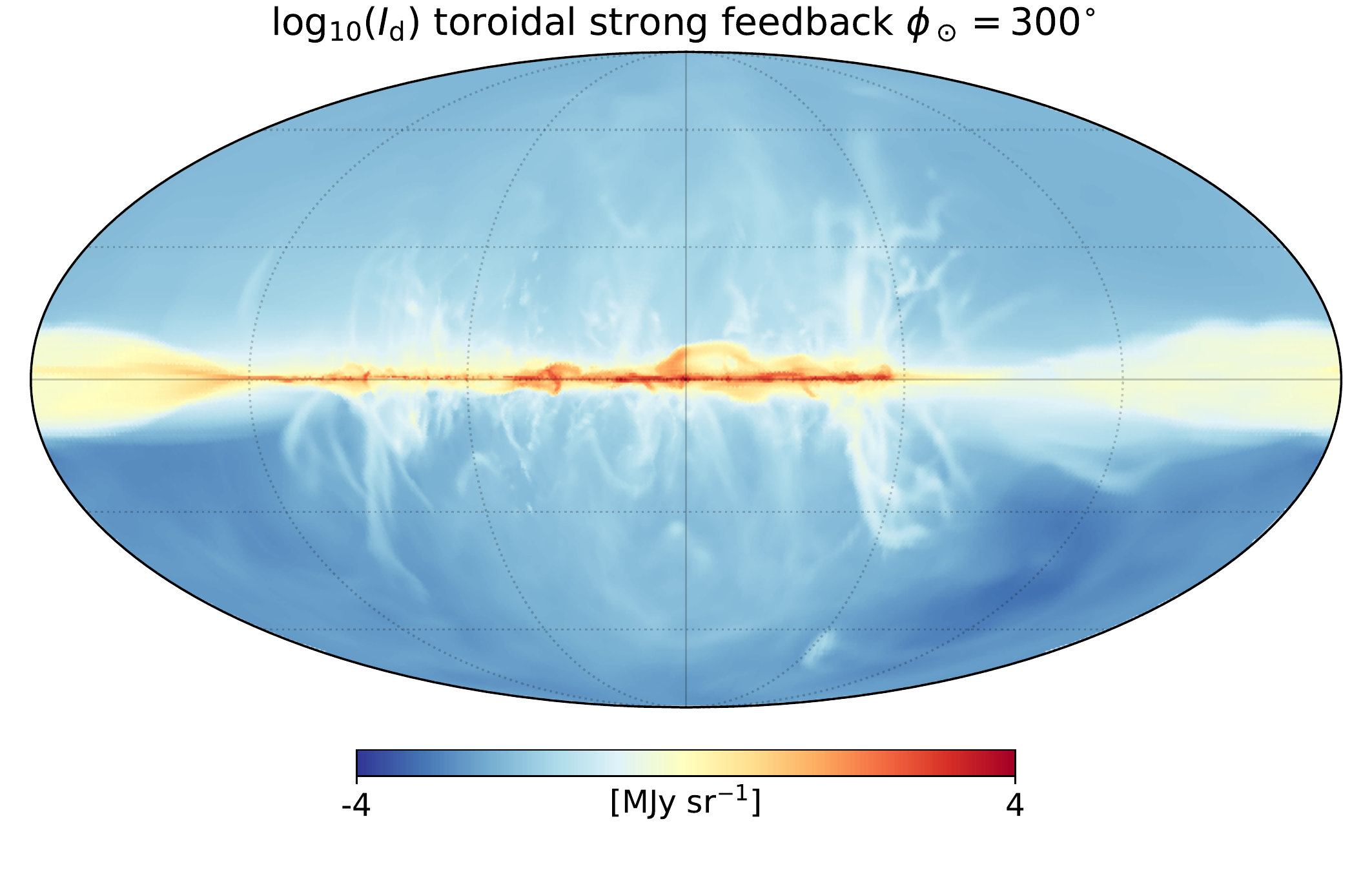}\\ [1.5ex]
    $Q_{\rm{353}}$ \\ [1.ex]
    \includegraphics[trim={0.0cm .0cm 0.0cm .8cm},clip,width=.98\columnwidth]{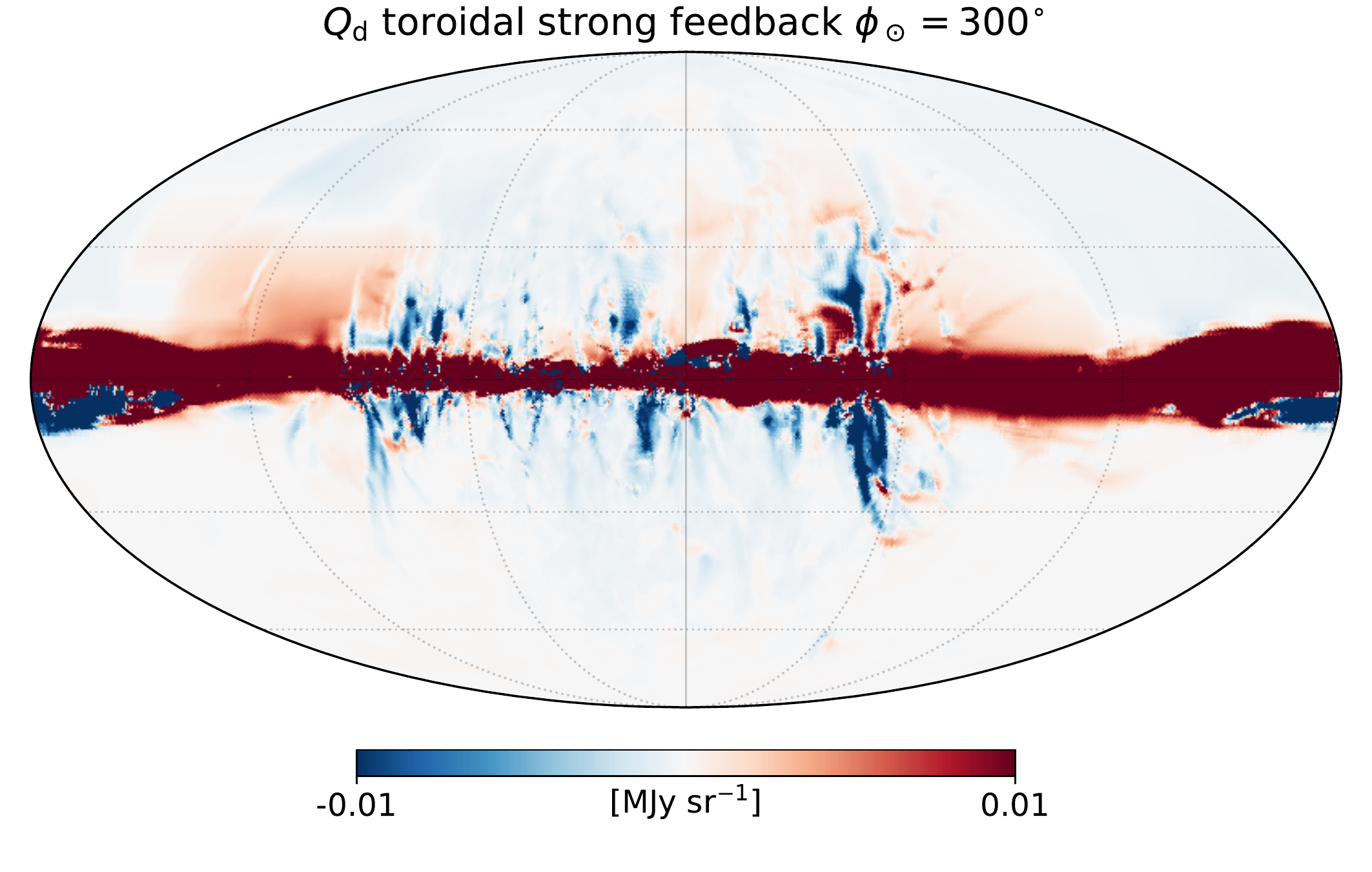}\\ [1.5ex]
    $U_{\rm{353}}$ \\ [1.ex]
    \includegraphics[trim={0.0cm .0cm 0.0cm .8cm},clip,width=.98\columnwidth]{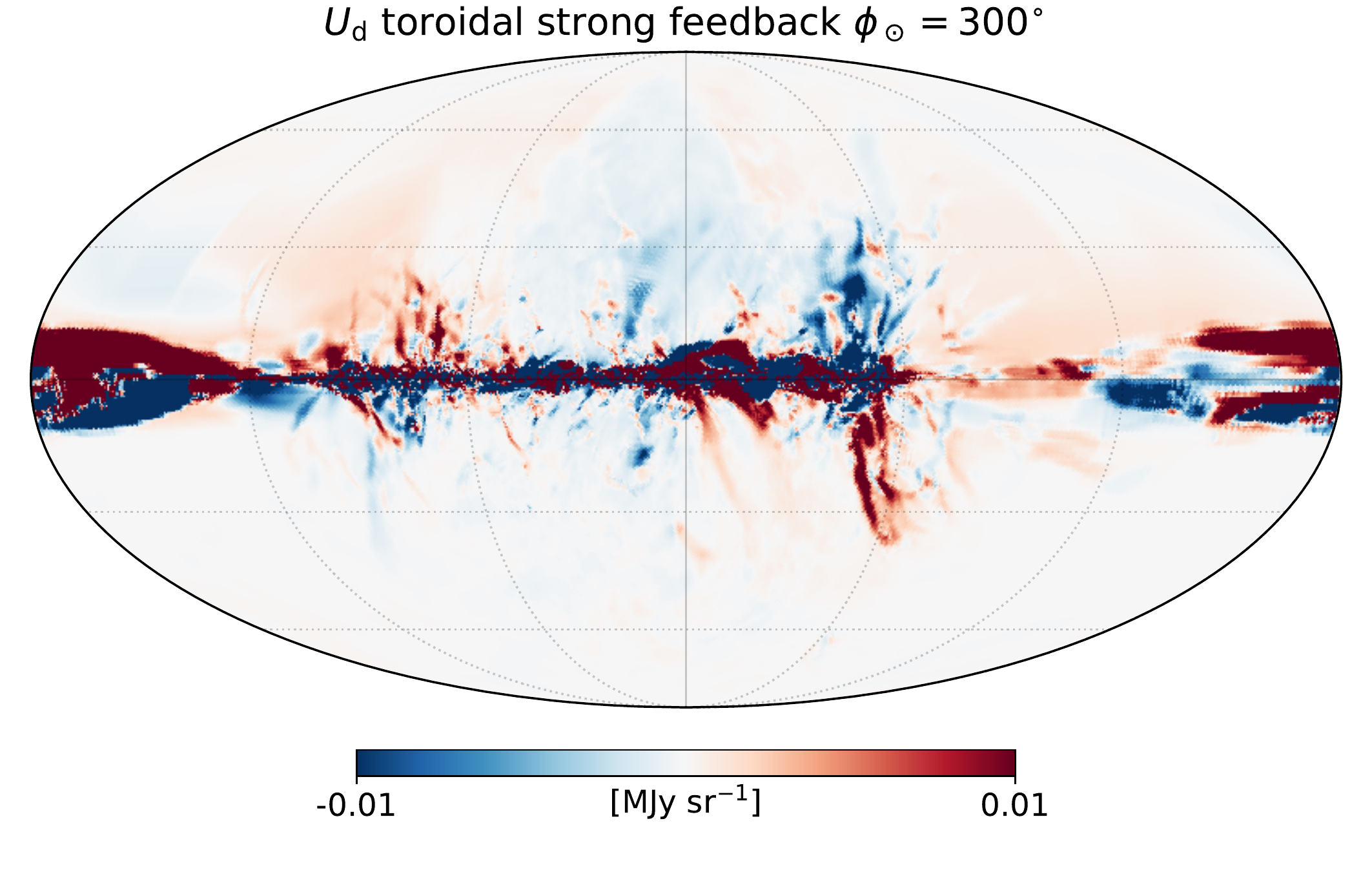}\\
    \caption{Example of synthetic polarization maps of thermal dust emission. The observer is immersed in the Tsf galaxy at 8 kpc from the the galactic center and at angular coordinate $\phi_\odot = 300^\circ$.}
    \label{fig:IQUmap_example}
\end{figure}

\subsubsection{Map making}
\label{subsec:map_making}
For each of the six simulation setups discussed above, we synthesize 72 sets of polarization maps ($I$, $Q$, and $U$) by allowing the observer to move in the $xy$ plane and make a complete circular excursion around the galactic center, at 8 kpc distance and with an angular step of $5^\circ$. In total we generate 432 sets of polarization maps. The used conventions for the rotation of the observer are explained and illustrated in Appendix~\ref{app:conventions}.

We use our ray-tracing algorithm, discussed in Appendix~\ref{app:raytracing}, to generate the synthetic dust polarization maps at the resolution of $N_{\rm{side}} = 128$. This choice of resolution fixes a set of 196608 lines of sight, corresponding to pixels with angular size of about 0.46 degrees.
To compute the integrated observables (Eqs.~\ref{eq:simulated_I}$-$\ref{eq:simulated_U}) we split the integrals into the contribution from each cell.
For each cell the algorithm computes the lengths of all the line-of-sight segments (corresponding to a map at fixed $N_{\rm{side}}$) that cross it. The lengths are used to weigh the observables using the midpoint rule.
After rotation and projection of the AMR grid on the observer spherical coordinate system (see Appendix~\ref{app:conventions}), the contribution from a given cell, labeled $i$, to the integrated Stokes parameters for the line of sight subtended by $\mathbf{e}_r$ is given by
\begin{eqnarray}
I^i_{\mathbf{e}_r} &=& \lambda^i_{\mathbf{e}_r} \, n^i_{\rm{d}} \left(1 - p_0 \left(\frac{{B^i_\theta}^2 + {B^i_\phi}^2}{{|\mathbf{B}^i |}^2} - 2/3\right) \right) \label{eq:I_model} \\
Q^i_{\mathbf{e}_r} &=& \lambda^i_{\mathbf{e}_r} \,  n^i_{\rm{d}} \, p_0 \, \frac{\left({B^i_\theta}^2 - {B^i_\phi}^2  \right)} {{|\mathbf{B}^i |}^2} \label{eq:Q_model} \\
U^i_{\mathbf{e}_r} &=& - 2\,  \lambda^i_{\mathbf{e}_r} \, n^i_{\rm{d}} \, p_0 \, \frac{\left(B^i_\theta \, B^i_\phi \right)}{{|\mathbf{B}^i |}^2}
\label{eq:U_model}
\end{eqnarray}
where $n^i_{\rm{d}}$ is the dust density in cell $i$ and $\lambda^i_{\mathbf{e}_r}$ is the path length of the given line of sight through that cell $i$.
The path lengths $\lambda^i_{\mathbf{e}_r}$ are computed through our ray-tracing algorithm as explained in Appendix~\ref{app:raytracing}. No further post-processing is applied to the synthetic maps for the power spectrum analysis.
As a result, projected cubic patterns can be spotted on some sets of maps. Interpolation would produce visually better maps but not without affecting the power spectra (Appendix~\ref{app:interpolation}). An example set of synthetic polarization maps is shown in Fig.~\ref{fig:IQUmap_example} for the Tsf galaxy.
        
\section{Power spectrum analysis}
\label{sec:3}
In this section we follow the same statistical characterization as that of the Galactic thermal dust emission performed in the context of CMB foregrounds by producing polarization power spectra of the synthetic dust maps.
We also rename the intensity map $T \equiv I$ to conform with the relevant literature.

\subsection{Formalism}
To the Stokes $Q$ and $U$ maps correspond rotation-invariant quantities, the $E$ (gradient) modes and $B$ (curl)
modes which are the even and odd parts of the polarization vector field under parity transformation (\citealt{Kam1997a}; \citealt{Zal1997}; \citealt{Hu1997}).
The statistical description of these three scalar and pseudo-scalar quantities ($T$, $E$, and $B$) is commonly based on their auto- and cross-angular power spectra as a function of multipole, $C_\ell^{XY}$, where $X$ and $Y$ refer either to $T$, $E$, and or $B$ and $\ell$ is the multipole number. A given $\ell$ roughly corresponds to an angular scale $\alpha \approx 180^\circ / \ell$.
Following the commonly adopted formalism, we carry out our analysis using the pseudo power spectra:
$\mathcal{D}_\ell^{XY} = \ell (\ell +1) \, C_\ell^{XY} / (2\pi)$. (See e.g., \cite{Bra2019} for a recent review of the formalism.)

\medskip

The $E/B$ power asymmetry is measured through the $\mathcal{R}_{EB}$ ratio, which is obtained by averaging the ratio of the auto-power spectra $\mathcal{D}_\ell^{EE}$ and $\mathcal{D}_\ell^{BB}$ over a specified multipole range
\begin{eqnarray}
    \mathcal{R}_{EB} \equiv \left< \frac{\mathcal{D}_\ell^{EE}}{\mathcal{D}_\ell^{BB}} \right>
\end{eqnarray}
where $\left\langle \cdot \right\rangle$ stands for the mean over multipole bins.

\medskip

To quantify the correlation between power spectra we use the normalized parameter $r^{XY}_\ell$ introduced by \cite{Cal2017}. $r^{XY}_\ell$ takes values $1$, $-1$ and $0$ in case of perfect correlation, perfect anticorrelation and absence of correlation, respectively, and is defined as
\begin{equation}
\label{eq:r_XY}
r^{XY}_\ell = \frac{C_\ell^{XY}}{\sqrt{C_\ell^{XX} \, C_\ell^{YY}}} \; .
\end{equation}

Even though we used all three $r^{XY} = \left\langle r^{XY}_\ell \right\rangle$ (with $XY = \{ TE,\, TB,\,EB \} $) to generate the idealized power spectra for the mask validation (see Sect.~\ref{subsec:maskvalid}), we only consider the $r^{TE}$ parameter to characterize our power spectra.

\subsection{Computation from synthetic maps}
\label{subsec:PS_computation}
We use the \texttt{Xpol} code \citep{Tri2005}, utilized in {\it Planck} analyses (e.g., \citealt{PlaXXX2016}; \citealt{PlaXI2020}), to compute the pseudo power spectra of the dust polarized sky and account for incomplete sky coverage.
Since our synthetic maps do not contain noise, we consider only the sampling variance as a source of uncertainty in the power spectrum estimates. The uncertainty is thus linked to the number of unmasked pixels and their spatial arrangement on the sky. Under Gaussian approximation these uncertainties estimated for a defined multipole bin ($\ell_{\rm{bin}}$) and retained sky fraction ($f_{\rm{sky}}$) can be computed analytically using
\begin{equation}
\label{eq:Cl_variance}
{\left(\sigma_{\mathcal{D}_\ell^{XX}} \right)}^2 = \frac{2}{(2 \ell_{\rm{bin}}) f_{\rm{sky}} \Delta\ell_{\rm{bin}}} \, \left(\mathcal{D}_{\ell_{\rm{bin}}}^{XX}\right)^2
\end{equation}
where $X = \{T; E; B\}$ and $\Delta \ell_{\rm{bin}}$ is the width of the multipole bins (\citealt{PlaXXX2016}).

\smallskip

\begin{figure}[!ht]
    \centering
    \vspace{.1cm}
    \includegraphics[trim={0.0cm 1.72cm 0.0cm 1.2cm},clip,width=.98\columnwidth]{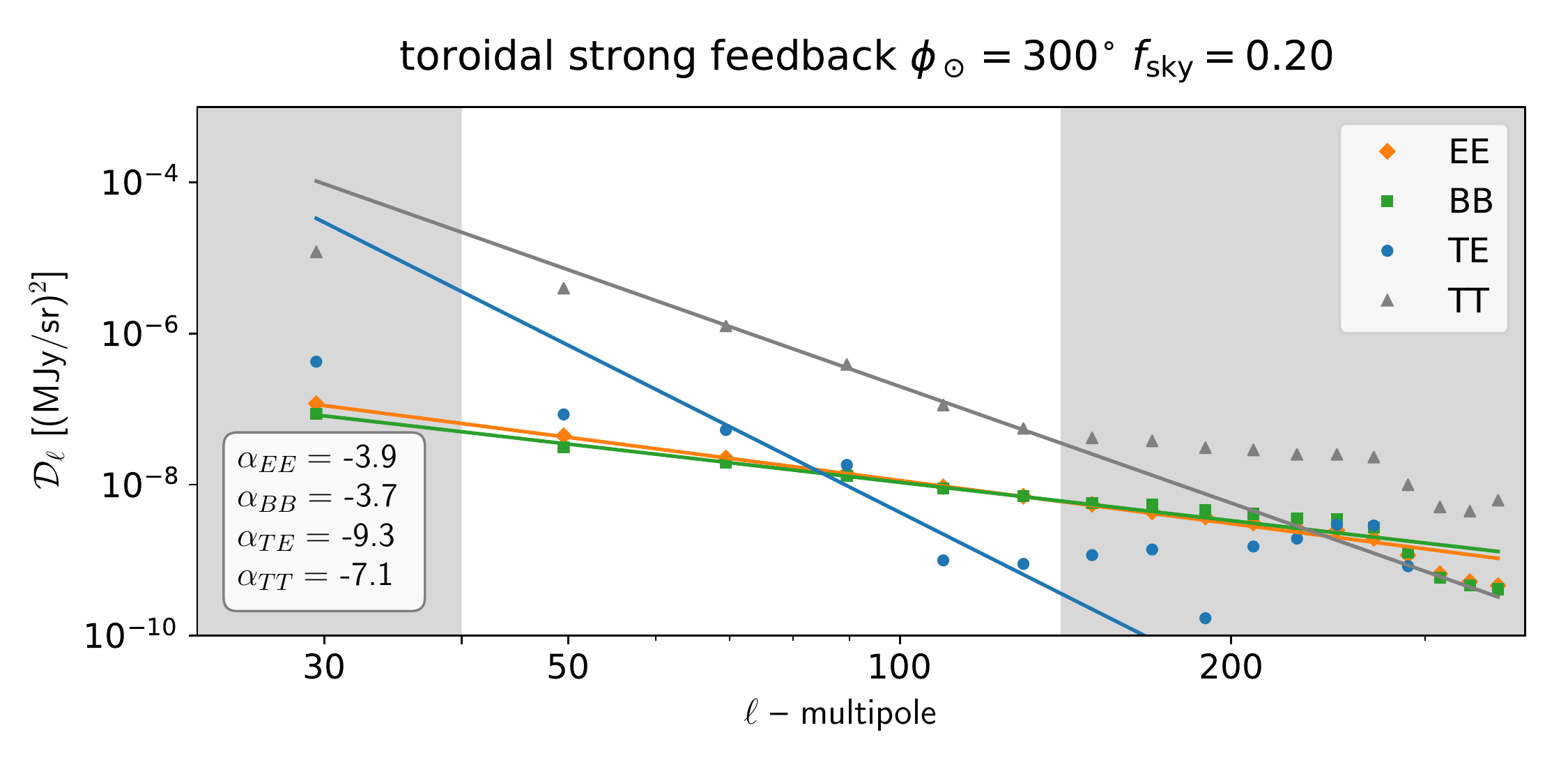}\\
    \includegraphics[trim={0.0cm 1.72cm 0.0cm 1.4cm},clip,width=.98\columnwidth]{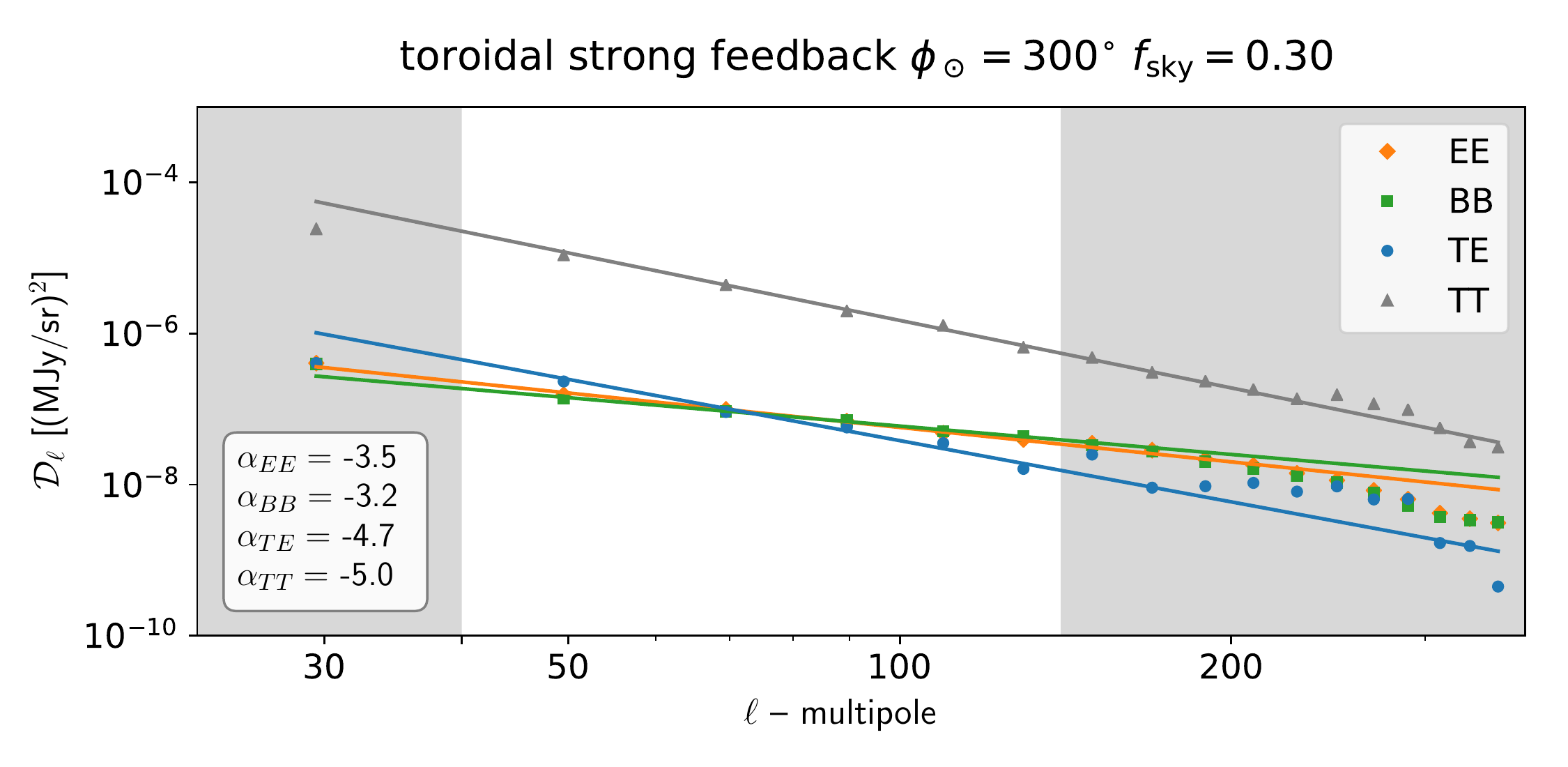}\\
    \includegraphics[trim={0.0cm 1.72cm 0.0cm 1.4cm},clip,width=.98\columnwidth]{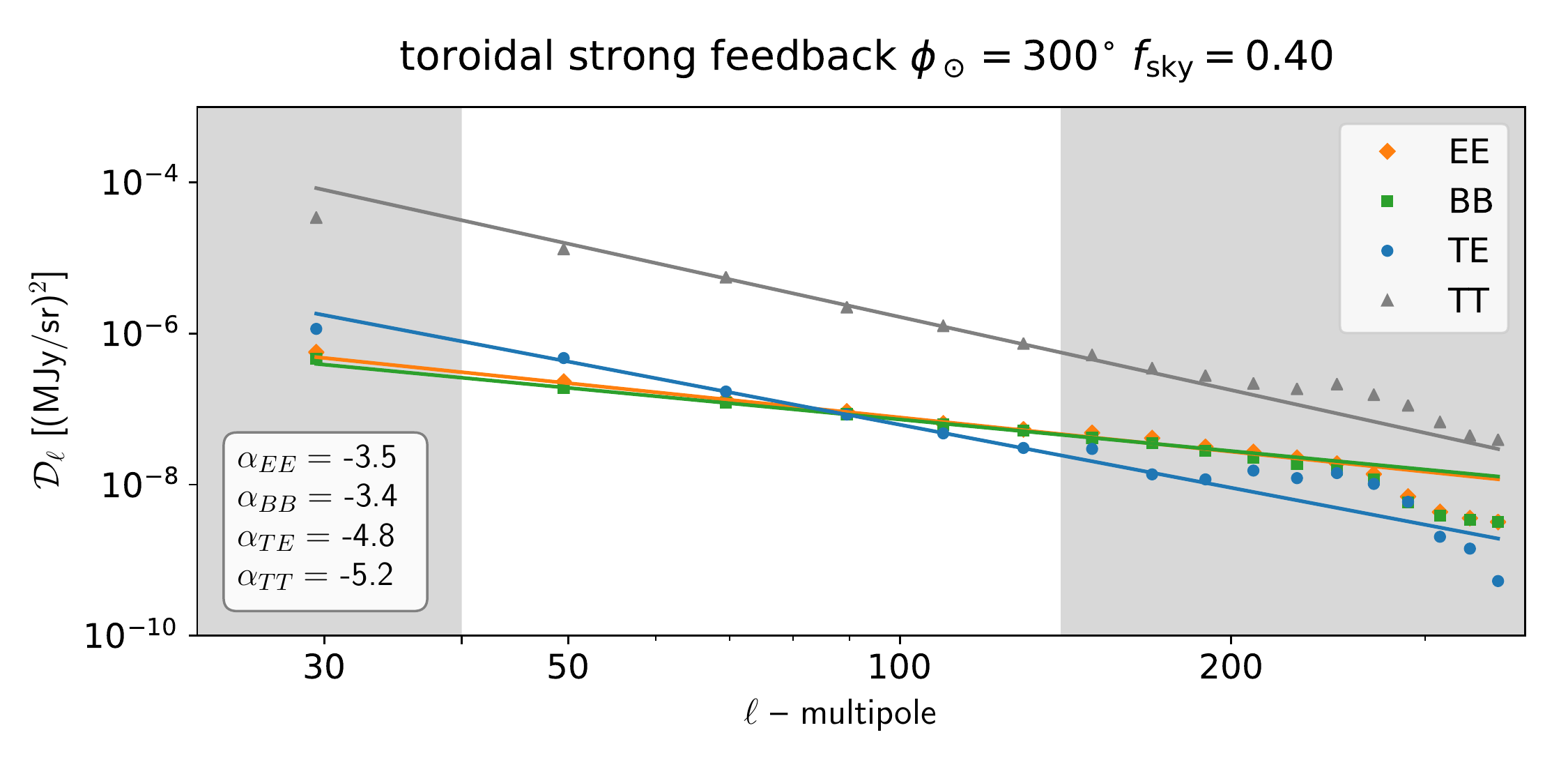}\\
    \includegraphics[trim={0.0cm 1.72cm 0.0cm 1.4cm},clip,width=.98\columnwidth]{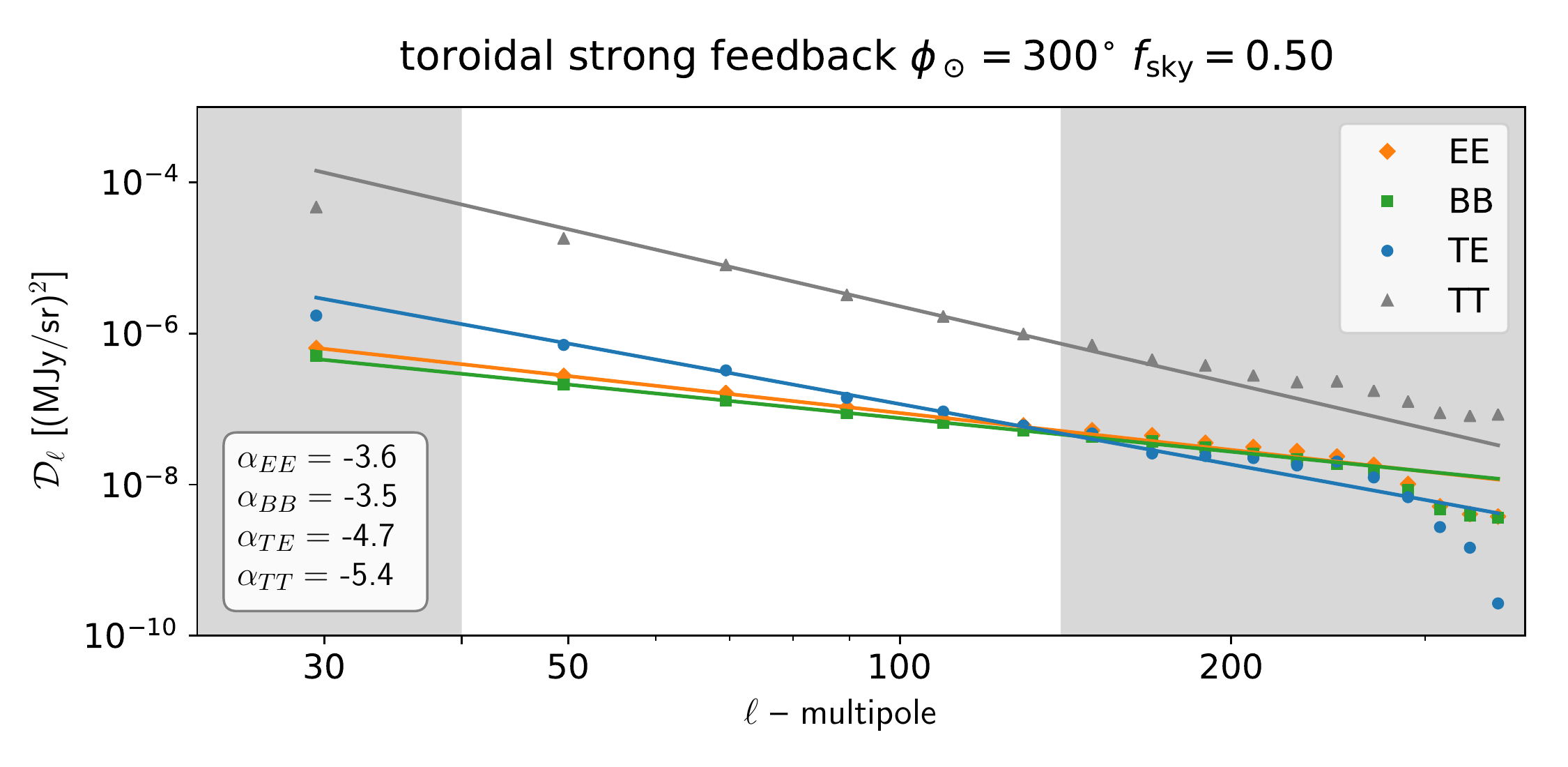}\\
    \includegraphics[trim={0.0cm 1.72cm 0.0cm 1.4cm},clip,width=.98\columnwidth]{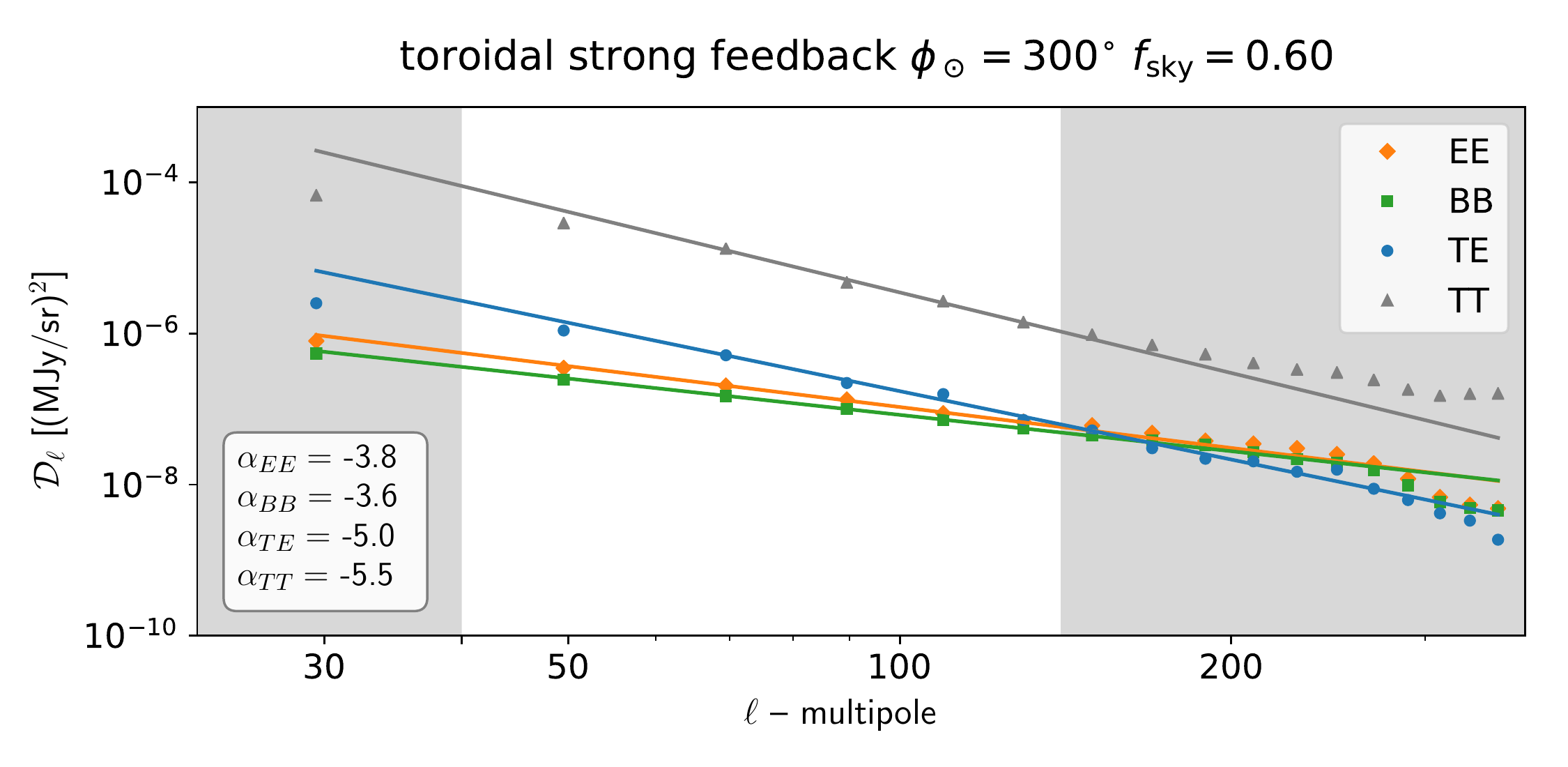}\\
    \includegraphics[trim={0.0cm 1.05cm 0.0cm 1.4cm},clip,width=.98\columnwidth]{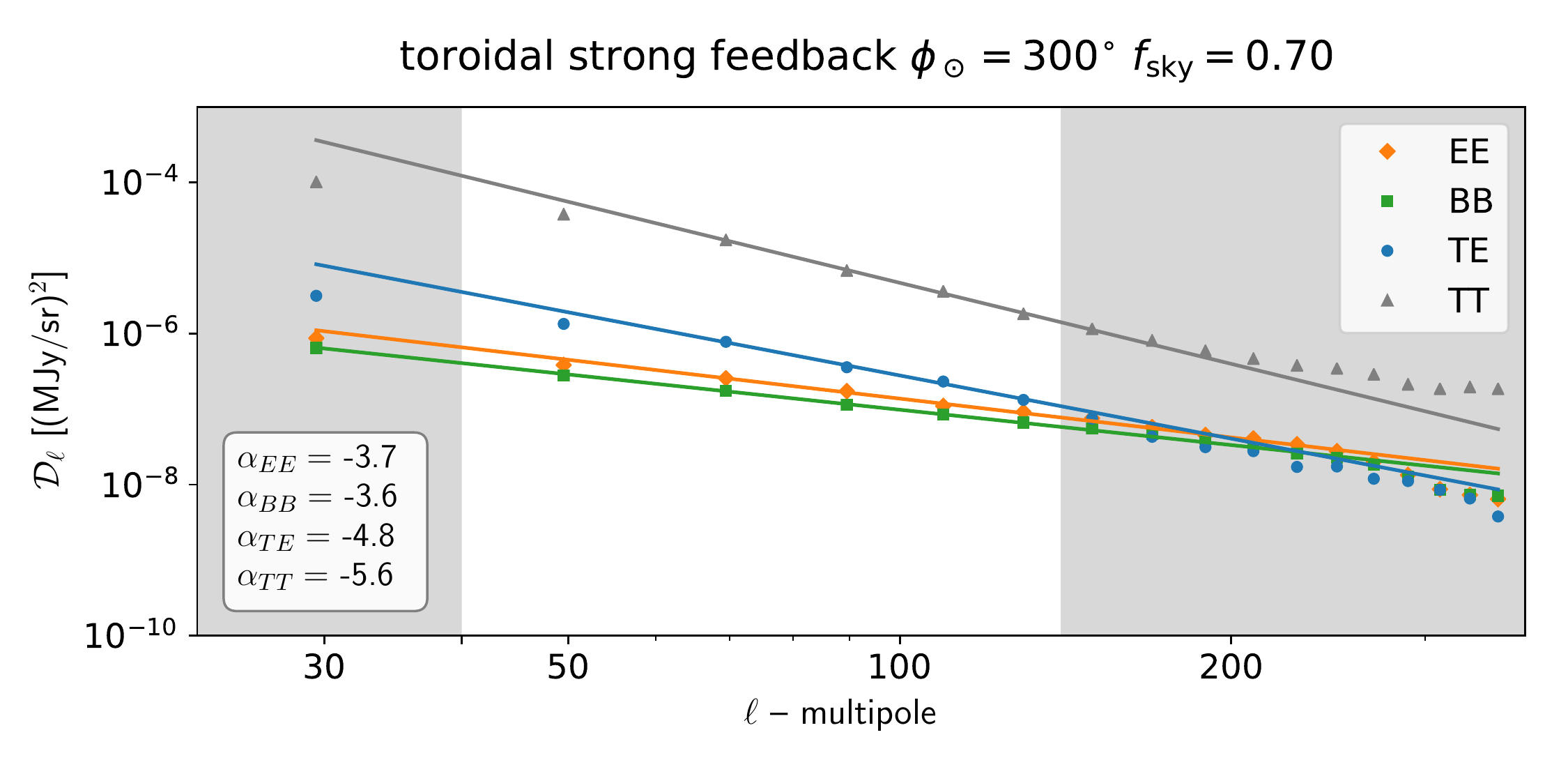}\\
    $\ell$ -- multipole \\
    \caption{Examples of polarization (pseudo) power spectra ($\mathcal{D}_\ell$) obtained for the polarization maps shown in Fig.~\ref{fig:IQUmap_example} and for the six values of the retained sky fraction ($f_{\rm{sky}}$) in the analysis; from 0.2 to 0.7 (top to bottom). The gray-shaded areas are excluded from our analysis as explained in the text. The straight lines are power-law power spectra fitted in the white area, i.e. the range  $60 \leq \ell \leq 140$.}
    \label{fig:PS_example}
\end{figure}
For each set of polarization maps we consider six nested regions at high galactic latitudes with retained sky fractions ranging from 20\% to 70\% by step of ten per cent. Those regions are determined independently for each set of polarization maps according to the masking procedure described below. Then, for each of the 2592 sets of masked polarization maps we evaluate the polarization power spectra using \texttt{Xpol}.
The power spectra are estimated in the multipole range of $\ell \in \left[40,\, 370\right]$ which are split in bins with a width of 20.
However, as discussed below, a shorter range of $\ell$ is determined and used in our analysis so as to allow for a systematic and homogeneous study of our sample and to easily perform any comparison.

To determine the high-galactic latitude sky regions on which to estimate the power spectra, we proceed as follows. For each set of maps, we start with the intensity map smoothed to a five degree resolution (FWHM) using a symmetric Gaussian beam. We then successively mask all pixels brighter than threshold values defined such that the retained sky fractions range from 0.2 to 0.7 by step of 0.1. To avoid power leakage we apodize the masks using a $2.5^\circ$ FWHM beam. Each mask is then applied to the corresponding set of synthetic polarization maps and the polarization power spectra computed through the use of \texttt{Xpol}.

\subsection{Full sample of polarization power spectra}
\subsubsection{Power-law power spectra}
Upon visual inspection, most of the $EE$ and $BB$ auto-power spectra present a steep power-law dependence on $\ell$, at least within a certain multipole range (see Fig.~\ref{fig:PS_example} for an example).
Such a power-law dependence is also observed in power spectra of the real sky (e.g., \citealt{PlaXXX2016}) and of MHD simulations (e.g. \citealt{Kim2019}). We make use of least-square fits to the power spectra with a power-law function of the form\footnote{We note that Eq.~12 of \cite{Kim2019} reads $(80/\ell)$ instead of $(\ell/80)$, which is likely a misprint.}
\begin{equation}
\hat{\mathcal{D}}^{XY}_\ell = A_{80}^{XY} \left(\frac{\ell}{80} \right)^{\alpha_{XY} + 2}
\end{equation}
where $XY$ can be $TE$, $EE$, and $BB$. $A_{80}^{XY}$ and $\alpha_{XY}$ are the amplitudes of $\mathcal{D}^{XY}_\ell$ at $\ell = 80$, and the spectral index of the spectrum ($C_\ell$), respectively.

We quantify the quality of the fits by computing the reduced $\chi^2$ over the same multipole range as
\begin{equation}
\chi^2_{XX} \equiv \frac{1}{N_{\rm{bin}}} \sum_{\ell_b} \frac{\left(\mathcal{D}^{XX}_{\ell_b} -  \hat{\mathcal{D}}^{XX}_{\ell_b} \right)^2}{{\sigma_{\mathcal{D}^{XX}_{\ell_b}}}^2},
\end{equation}
in which the uncertainties are from Eq.~\ref{eq:Cl_variance} and $N_{\rm{bin}}$ is the number of multipole bins involved.

\subsubsection{Choice of $\ell$ range}
In order to compare polarization power spectra from different sets of maps and masks which may look very different, it is necessary to define a common range of multipoles within which to characterize the power spectra.

By visually inspecting the power spectra, we notice that most of them show power-law behavior up to large multipole bins ($\ell \sim 360$), especially when the maps contain a significant fraction of the sky. However, some power spectra show a sharp drop of power or a flattening in the high-$\ell$ range of the spectrum. These features either originate from the limited resolution of the 3D AMR grid (e.g., in situations where only a few low-resolution voxels contribute to the simulated unmasked sky) or from sharp edges in the maps created by the projection.
We find that cutting the $\ell > 140$ mitigates both effects homogeneously over our entire sample of power spectra.

At low-$\ell$ values, our reliability test discussed below indicates that it is safer to discard the bin centered at $\ell = 50$ because some masks may lead to biased estimates of the power spectra. Therefore, we decide to constrain the characterization of all the power spectra in the multipole range $60 \leq \ell \leq 140$.
This low-$\ell$ cut mitigates the spurious effects of the projected cubic voxels (visible in some maps) on the power spectrum estimate, because it avoids the range of angular scales corresponding to the projected voxels at high and intermediate latitudes.

For this multipole range, the $EE$ and $BB$ power spectra can be described by power laws. Out of the 2592 sets of power spectra, about 82\% show reduced $\chi_{XX}^2$ lower than 10, using Eq.~\ref{eq:Cl_variance} as uncertainties, for both $EE$ and $BB$ simultaneously.

\subsubsection{Reliability of the power spectra}
\label{subsec:maskvalid}
In order to check the estimates of the uncertainties, but more importantly, to make sure that the masks used do not introduce biases in our polarization power spectra, we rely on a simulation-based validation for a subset of observer locations.

For each galaxy model we produce 12 sets of polarization maps from positions $30^\circ$ apart.  We apply the six masks to each set and calculate the power spectra, like above. We fit the resulting $TT$, $EE$, and $BB$ auto-power spectra with power-laws in the multipole range $40 \leq \ell \leq 140$ (see below). Finally, we compute the three correlation coefficients between the spectra using Eq.~\ref{eq:r_XY}.
The fitted power-law auto-power spectra and the correlation coefficients determine a set of (idealized) polarization power spectra. 
Then, using the \texttt{synfast} functionality of the HEALPix Python package, we use the idealized power spectra to create 300 sets of full-sky polarization maps through Gaussian (Monte Carlo) realizations.
We then apply the sky-fraction masks to these simulated maps and compute their power spectra with \texttt{Xpol}.
To detect and quantify possible biases, we compare the value of the idealized spectra to the distribution (mean and standard deviation) of the Monte Carlo (MC) power spectra at each multipole bin.

For all the tested cases, we find that the MC and analytic uncertainties generally agree in the range $40 \leq \ell \leq 140$. 
Therefore, to avoid the additional computational cost of running MC simulations, in what follows we only consider the analytic uncertainties.
We then verify that our power spectra do not suffer from biases. 
We find that several polarization maps lead to a significant bias for the multipole bin centered at $\ell=50$. 
However, no significant bias is detected in the range $60 \leq \ell \leq 140$.
Therefore, we focus the comparison on the multipole range $60 \leq \ell \leq 140$.

\medskip

The  polarization power spectra corresponding to the maps shown in Fig.~\ref{fig:IQUmap_example} are presented in Fig.~\ref{fig:PS_example} for the six $f_{\rm{sky}}$ values. On those plots we are also presenting results from fitted power-law power spectra on $TT$, $EE$, $BB$ and $TE$ in the range of $60 \leq \ell \leq 140$.
\begin{figure}
    \centering
    \includegraphics[trim={8.cm 5.cm 9.8cm 4.4cm},clip,width=.95\columnwidth]{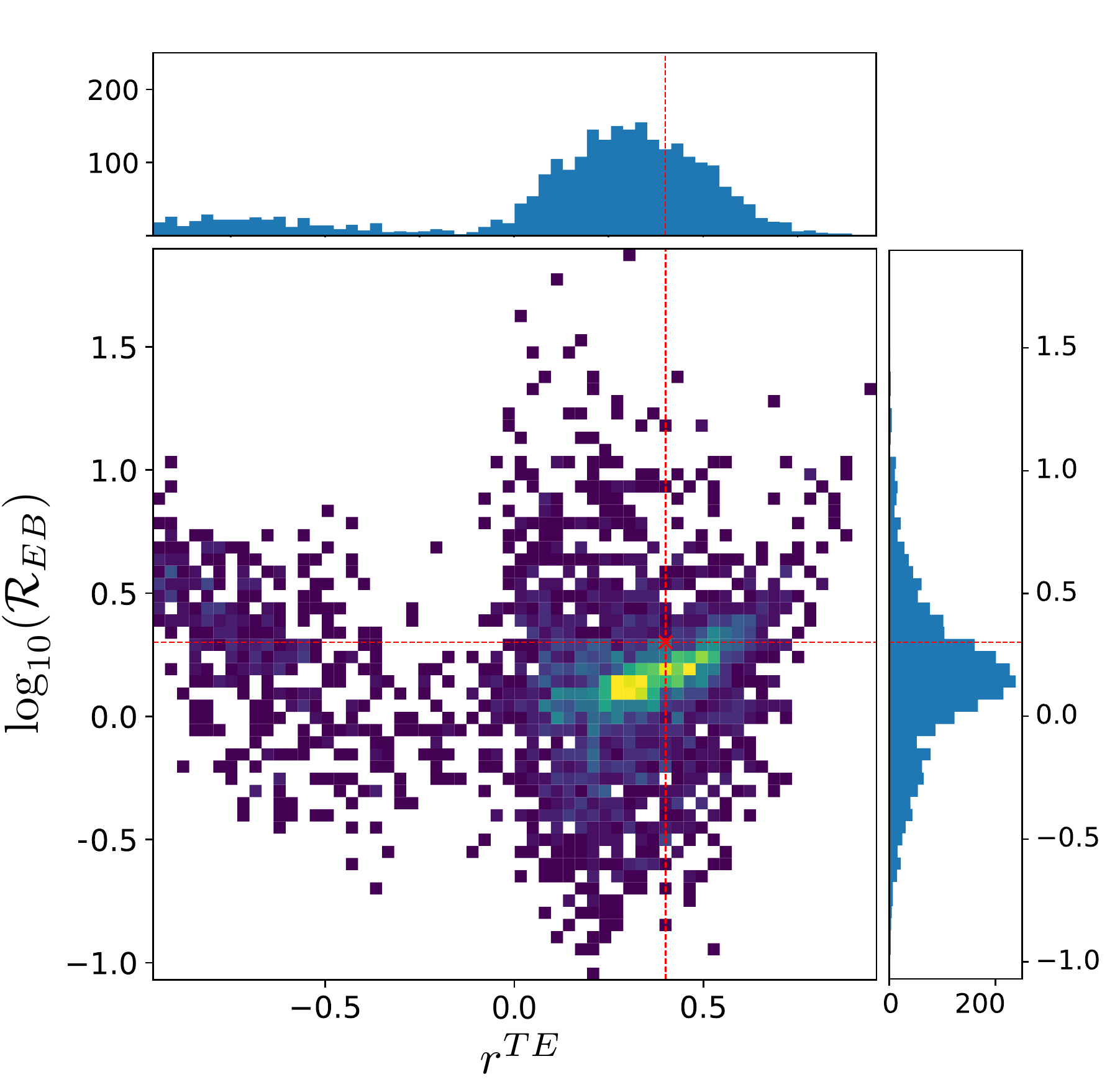}
    \caption{Histograms of $R_{EB}$ and $r^{TE}$ as measured from all galaxy setups, observer positions and sky fraction together. The red cross marks the location $(r^{TE},\,\mathcal{R}_{EB}) = (0.4,\,2.0)$, typical values deduced from {\it Planck} data at 353~GHz.
    }
    \label{fig:REBrTE_fullsample}
\end{figure}

\section{Results}
\label{sec:4}

In this section we first present the characteristics of the full sample of power spectra. Then we examine the differences arising from the underlying galaxy model.
We characterize the properties of the power spectra through the $\mathcal{R}_{EB}$ ratio, the correlation coefficient $r^{TE}$ and power-law fits of the $EE$ and $BB$ spectra.
We quantify the variance induced by the observer's location and since we find it to be significant, we then explore our sample and simulations in more detail in the attempt to identify the dominant factors that determine the shape of the power spectra and their correlations.

\subsection{Full sample inspection}
We use the parameters $A_{80}^{XX}$, $\alpha_{XX}$, and $\chi^2_{XX}$ in addition to $\mathcal{R}_{EB}$ and $r^{TE}$ to characterize the synthetic polarization power spectra.
Figure~\ref{fig:REBrTE_fullsample} presents histograms of $\mathcal{R}_{EB}$ and $r^{TE}$ as measured from all galaxy setups, observer positions and sky fractions.

\smallskip

The $\log_{10}(\mathcal{R}_{EB})$ distribution is slightly shifted toward positive values, which means that $E$ modes dominate most of the polarization signal in our synthetic maps. However, there is a large scatter about the mean and several sets of polarization maps are dominated by $B$ modes. The 16, 50, and 84 percentiles of $\log_{10}(\mathcal{R}_{EB})$ are $-0.17$, $0.15$, and $0.43$ respectively (the median of $\mathcal{R}_{EB}$ is 1.42).
The correlation coefficient $r^{TE}$ is mostly positive, with a second minor peak at negative values, revealing a general positive correlation between $T$ and $E$ modes.
Interestingly, the pair of typical values reported for the real sky by {\it Planck} (marked by the red cross in Fig.~\ref{fig:REBrTE_fullsample}) is covered by the 2D distribution obtained from the synthetic maps.

\medskip

\begin{figure*}
    \centering
    \begin{tabular}{cc}
    (a) & (b) \\
    \includegraphics[align=c,trim={2.5cm 1cm 2.8cm 1cm},clip,width=.57\linewidth]{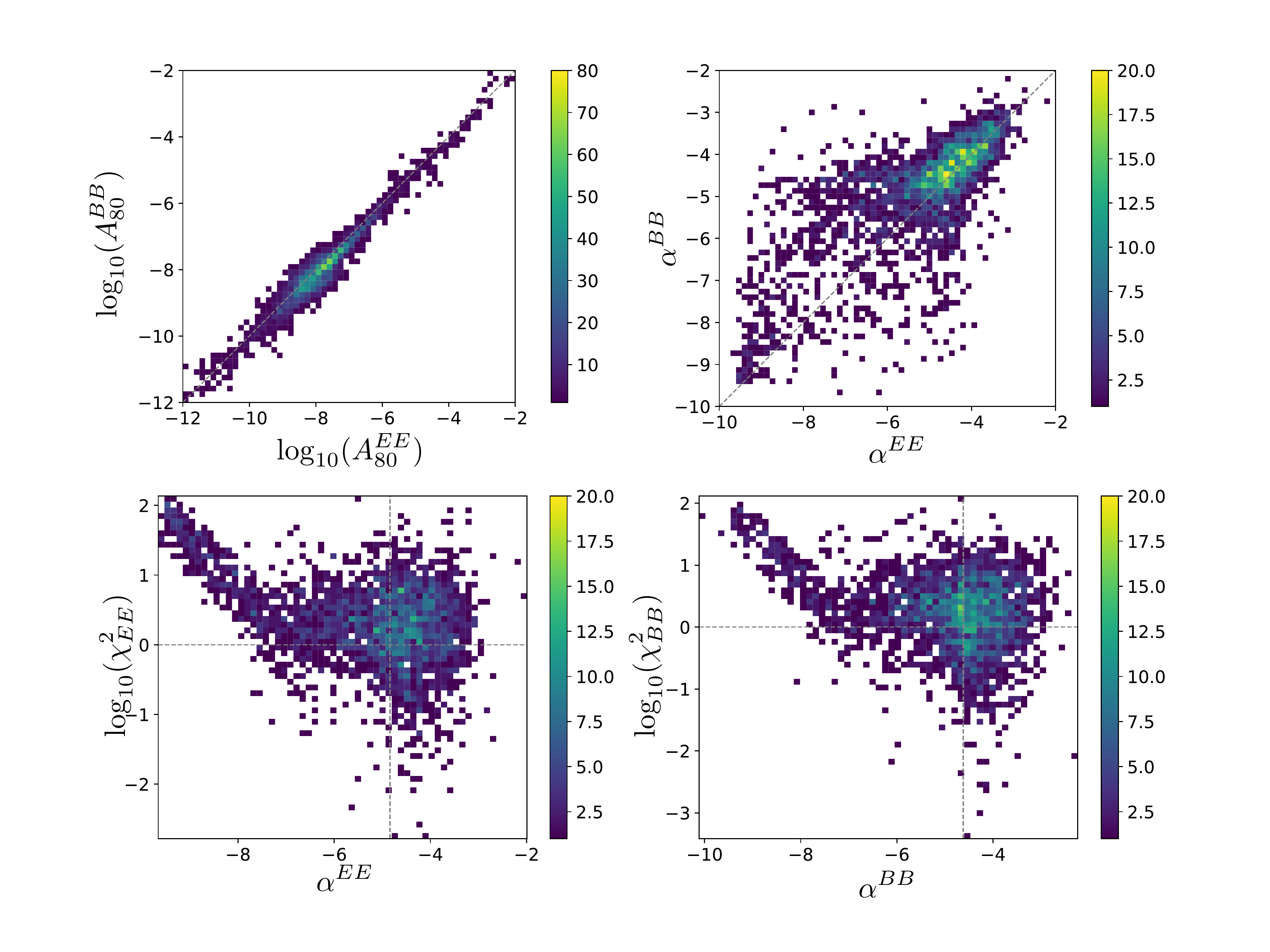} &
    \includegraphics[align=c,trim={.9cm .5cm 1.1cm .5cm},clip,width=.38\linewidth]{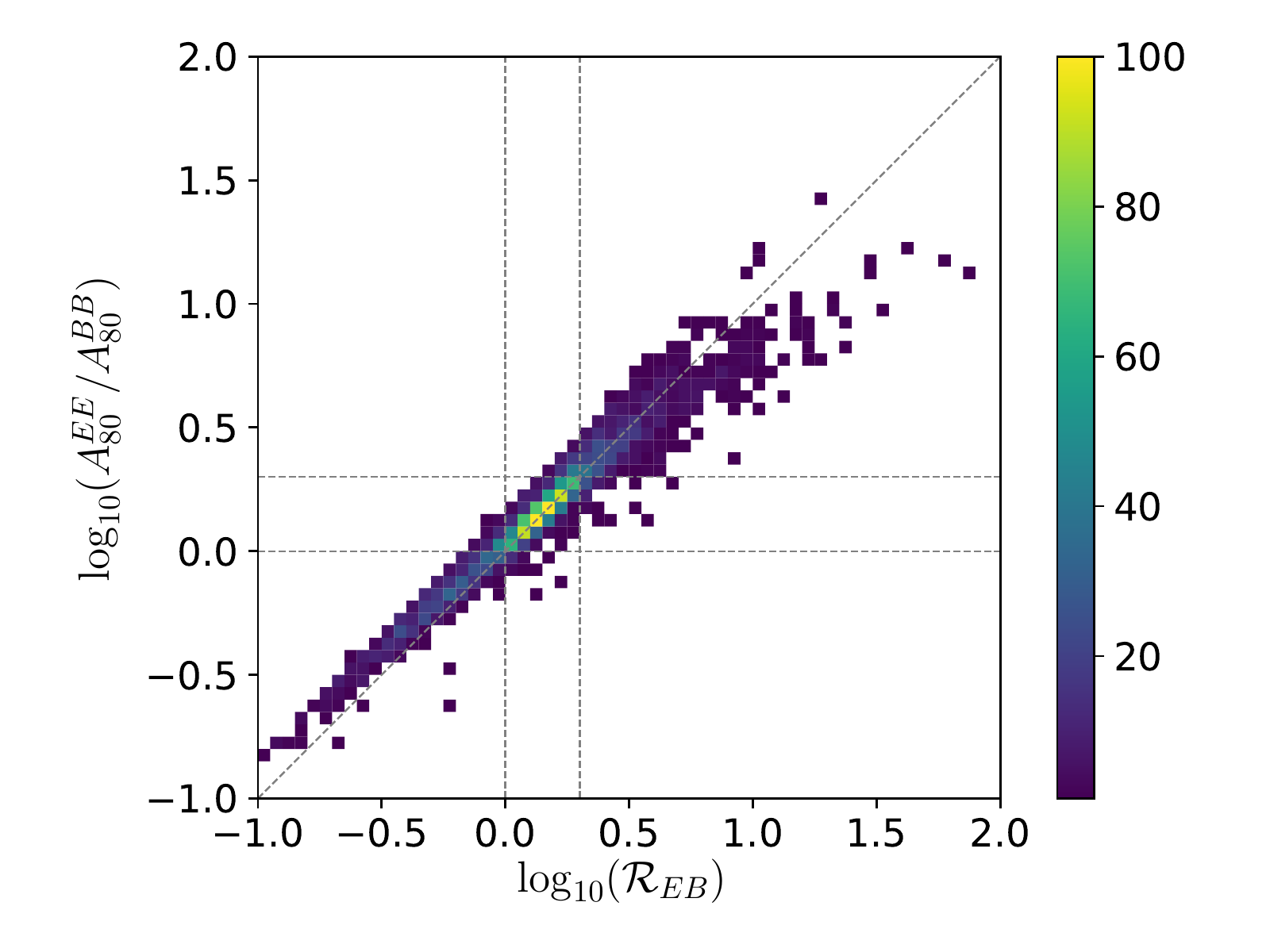}
    \end{tabular}
    \caption{2D histograms of combinations of parameters from power-law fits used to characterize the shape of the power spectra for the full sample; all galaxy setups, observer position and sky fractions merged together.
    Panel (a): From top left to bottom right we have $(\log_{10}(A_{80}^{EE}),\,\log_{10}(A_{80}^{BB}))$, $(\alpha^{EE},\,\alpha^{BB})$, $(\alpha^{EE},\,\chi_{EE}^2)$ and $(\alpha^{BB},\,\chi_{BB}^2)$.
    The main diagonals are shown in top panels for reference. In the bottom panels the horizontal lines mark values of $\chi_{XX}^2 = 1$ and the vertical lines show the median of the spectral indices.
    Panel (b): $\log_{10}(\mathcal{R}_{EB})$ versus $\log_{10}(A_{80}^{EE}/A_{80}^{BB})$.
    The horizontal and vertical lines mark ratio values of 1 and 2 and are given for visual reference. The main diagonal is also shown.}
    \label{fig:PLfit_fullsample}
\end{figure*}

Figure~\ref{fig:PLfit_fullsample} (panel a) presents 2D histograms for some combinations of parameters from the power-law fits as measured in the multipole range $60 \leq \ell \leq 140$ from all galaxy setups, observer positions and sky fractions.

The $EE$ and $BB$ auto-power spectra are very steep ($\alpha_{XX} \approx -4.5$); much steeper than those observed in the {\it Planck} data ($\alpha_{XX} \approx -2.5$) or in kpc-scaled MHD simulations (\citealt{Kim2019}) ($\alpha_{XX} \approx -3.6$).
The reason for this behavior is that most of the high-resolution areas of the AMR grid are excluded from the analysis by masking. This prevents our synthetic skies from being populated by small angular-scale variations. The limited resolution of the retained sky also reduces the variance in the maps, and is therefore responsible for the low values of the $A_{80}^{XX}$ parameters compared to the real sky.

As illustrated in Fig.~\ref{fig:PLfit_fullsample}, the amplitudes at $\ell = 80$ of the $EE$ and $BB$ auto-power spectra span several orders of magnitude but appear to be correlated. Generally, $A_{80}^{EE}$ is found to be larger than $A_{80}^{BB}$. The distribution of $\log_{10}(A_{80}^{EE}/A_{80}^{BB})$ is roughly symmetric with percentile 16, 50, and 86 of $-0.12$, $0.16$, and $0.42$ (the median of $A_{80}^{EE}/A_{80}^{BB}$ is $1.45$).
The spectral indices, $\alpha_{XX}$, of the $EE$ and $BB$ power spectra are also positively correlated and have similar values for the entire set of maps.
The two bottom panels of the figure show that very steep power-law fits to the spectra have large $\chi_{XX}^2$ values. These poorly-fitted spectra with a lack of power at small scales likely correspond to quiescent sky realizations.

\medskip

Figure~\ref{fig:PLfit_fullsample} (panel b) shows that $\log_{10}(\mathcal{R}_{EB})$ and $\log_{10}(A_{80}^{EE}/A_{80}^{BB})$
are very well correlated. This correlation implies that, despite the small differences in the spectral indices of the $EE$ and $BB$ auto-spectra, the $\mathcal{D}_\ell^{EE}/\mathcal{D}_\ell^{BB}$ ratios are not strongly scale-dependent. Therefore, the $\mathcal{R}_{EB}$ parameter can be used to derive meaningful information.

\subsection{Dependence of $\mathcal{R}_{EB}$ and $r^{TE}$ on the galaxy model}
\begin{table}
    \caption{General statistics in $\mathcal{R}_{EB}$ and $r^{TE}$.}%
    \label{tab:REB-rTE_dist}%
    \centering
    \begin{tabular}{r ccc | ccc}
    \hline
    \hline
    \\[-1.5ex]
    \multirow{2}{*}{Setup} & \multicolumn{3}{c}{$\log_{10}(\mathcal{R}_{EB})$} &
        \multicolumn{3}{c}{$r^{TE}$} \\
        &   \multicolumn{3}{c}{percentile}  & \multicolumn{3}{c}{percentile}  \\
         &  16  & 50 & 84 & 16 & 50 & 84  \\
    \hline
    \\[-.5ex]
    Pwf     &   0.78	& 1.31	& 1.96 	&  0.11	&	0.25    &	0.43	\\
    Pwnf    &   0.78	& 1.77	& 3.51  & -0.84	&	-0.64	&	-0.32	\\
    Twf     &   0.97	& 1.40	& 2.09  &  0.16	&	0.32	&	0.51	\\
    Twnf    &   0.47	& 1.25	& 5.27  &  0.08	&	0.29	&	0.54	\\
    Tsf     &   1.17	& 1.49	& 1.93  &  0.24	&	0.39	&	0.51	\\
    Tsnf    &   0.42	& 1.17	& 3.06  &  0.17	&	0.33	&	0.55	\\ \\[-.5ex]
    \hline
    \end{tabular}
    \tablefoot{16, 50, and 84 percentile of the $\log_{10}(\mathcal{R}_{EB})$ and $r^{TE}$ distributions per galaxy, merging data from all $f_{\rm{sky}}$ values.}
\end{table}

In this section we infer a possible dependence of the $\mathcal{R}_{EB}$ and $r^{TE}$ parameters on the simulation setup and on the retained sky fraction. We find that different galaxy models lead to different distributions of $\mathcal{R}_{EB}$ and $r^{TE}$ and a small but systematic increase of $\mathcal{R}_{EB}$ and $r^{TE}$ values with increasing sky fraction.

\begin{figure*}
    \centering
    \begin{tabular}{cc}
    \hspace{1cm} $\mathcal{R}_{EB}$  & \hspace{1cm} $r^{TE}$\\
    \includegraphics[trim={1.2cm 1.cm 0.6cm 1.5cm},clip,width=.95\columnwidth]{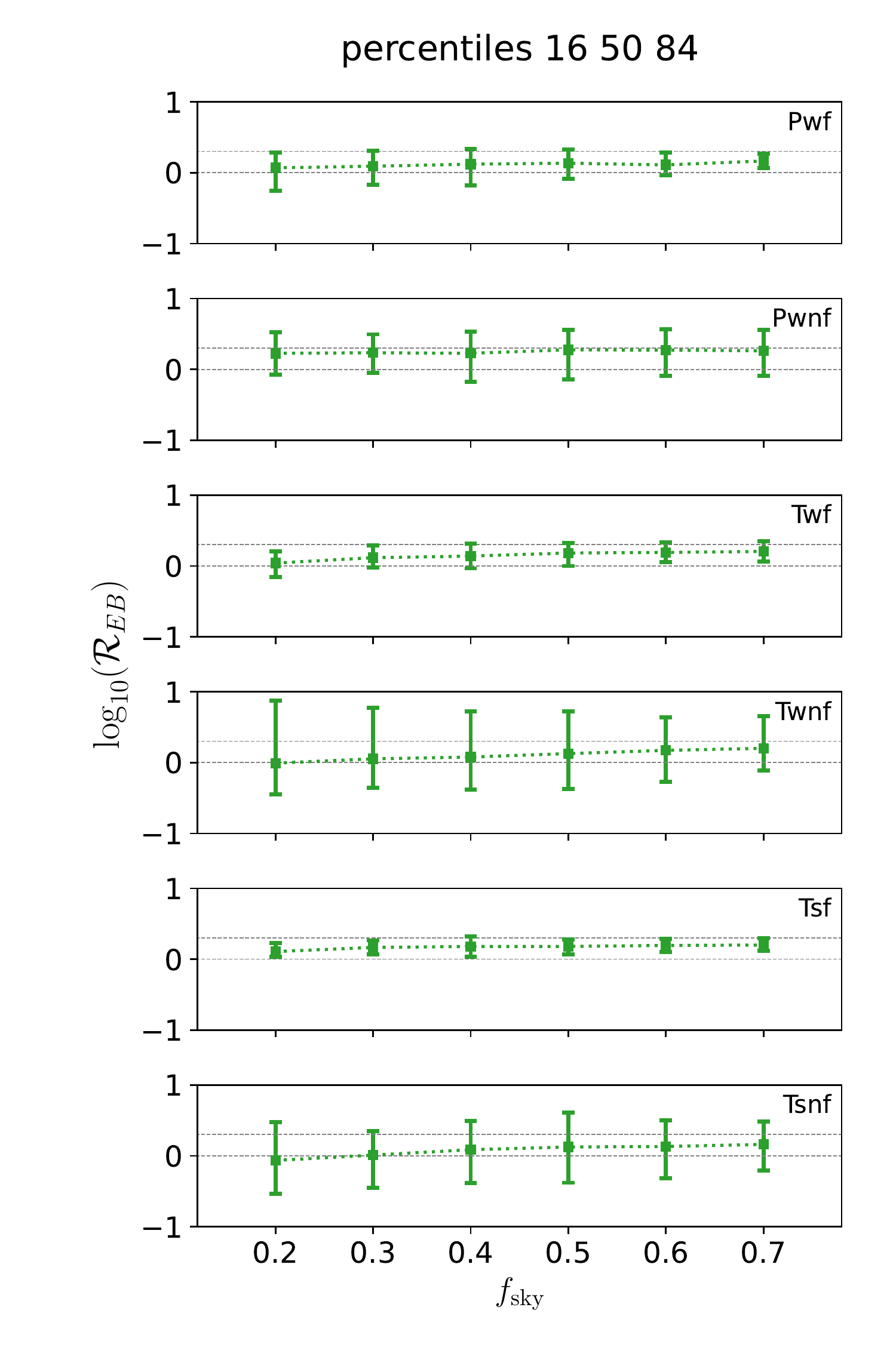}
    &
    \includegraphics[trim={1.2cm 1.cm 0.6cm 1.5cm},clip,width=.95\columnwidth]{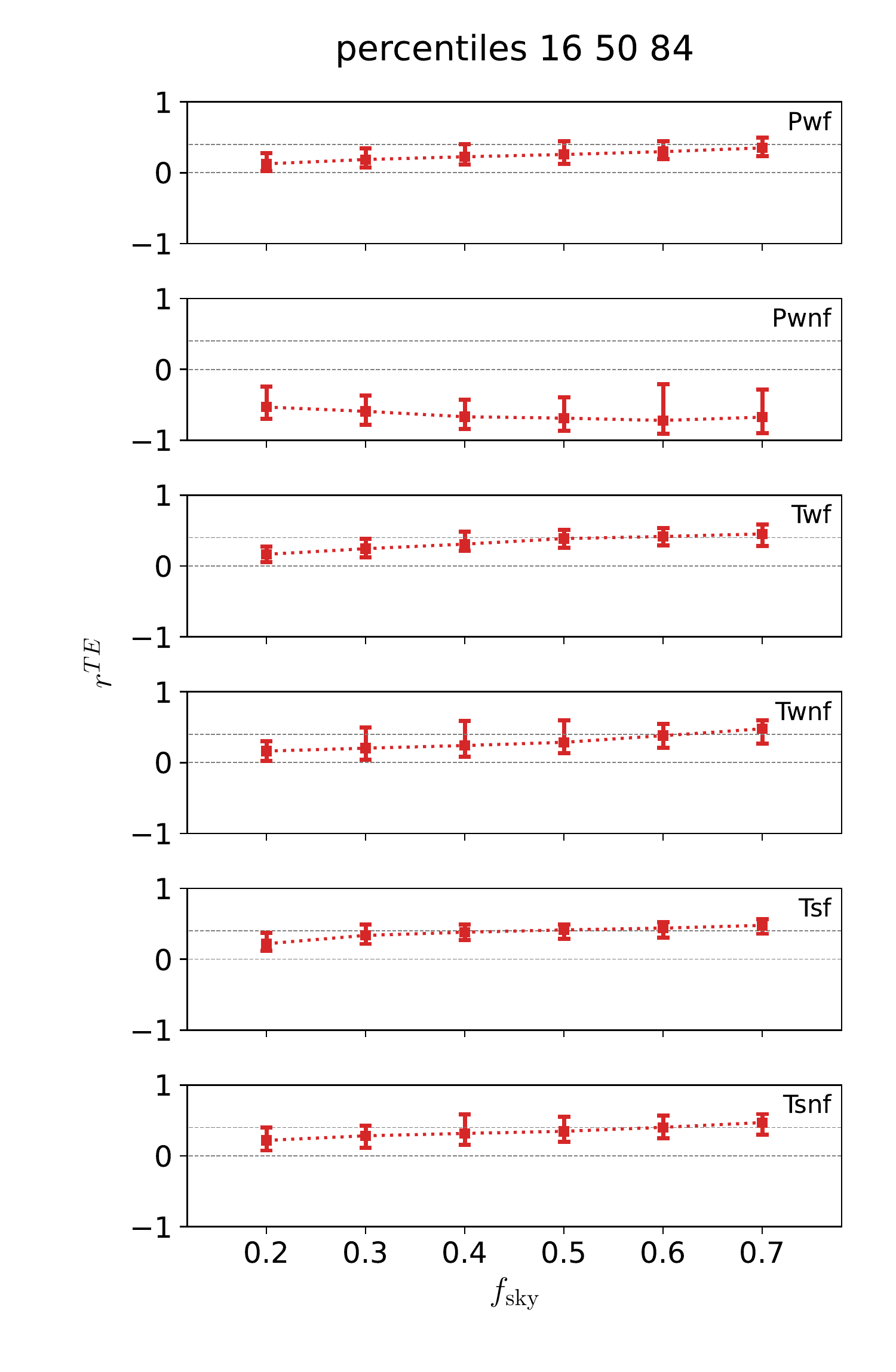}
    \end{tabular}
    \caption{Summarizing statistics of $\mathcal{R}_{EB}$ (left) and $r^{TE}$ (right) per galaxy and per sky fraction $f_{\rm{sky}}$. Asymmetric error bars around the median encompass 68\% of the data points.
    For visual reference, the horizontal dashed lines mark the values of $\mathcal{R}_{EB} = 1$ and $2$ to the left and $r^{TE} = 0$ and $0.4$ to the right.
    }
    \label{fig:REB-rTE_fsky}
\end{figure*}
Figure~\ref{fig:REB-rTE_fsky} summarizes the distributions of $\log_{10}(\mathcal{R}_{EB})$ (left) and $r^{TE}$ (right) per galaxy (Pwf, Pwnf, Twf, Twnf, Tsf and Tsnf, from top to bottom panel respectively) as a function of retained sky fraction ($f_{\rm{sky}}$). The central point is the median and the asymmetric error bars show the 1$\sigma$ spread of the 72 data points we have per galaxy and $f_{\rm{sky}}$.

\subsubsection{The $\mathcal{R}_{EB}$ parameter}
For all six galaxy setups, $\log_{10}(\mathcal{R}_{EB})$ is found to be mainly positive with medians in the range from $\mathcal{R}_{EB} \sim 1$ to $\mathcal{R}_{EB} \sim 2$. Although the spread of the distributions are larger for low $f_{\rm{sky}}$ values, a striking difference between galaxies with and without feedback appears: Galaxies for which the feedback is disabled show a much larger dispersion of the $\mathcal{R}_{EB}$ ratio.

\begin{figure}
    \centering
    \includegraphics[trim={1.2cm .0cm 0.6cm .0cm},clip,width=.98\columnwidth]{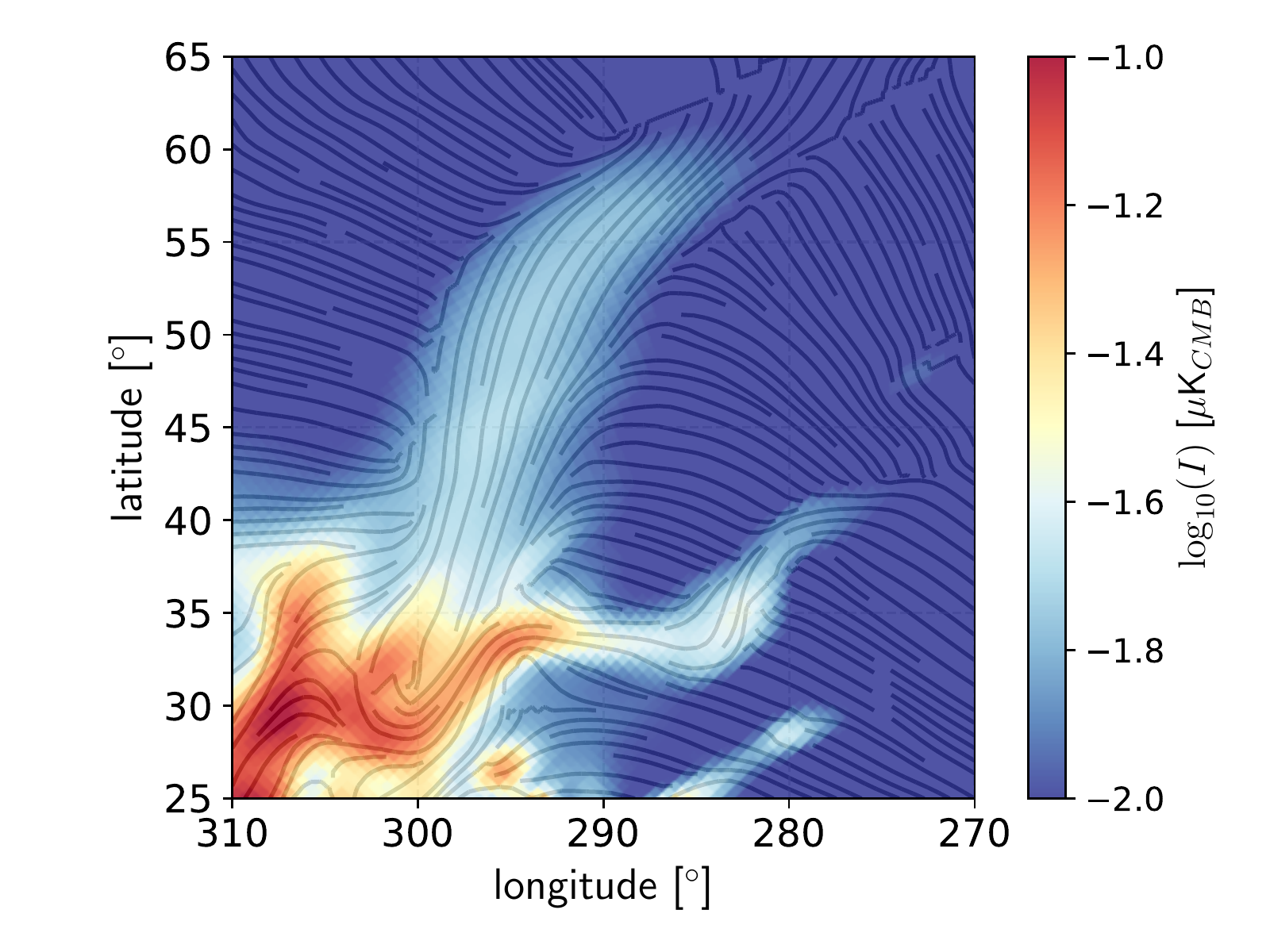}
    \caption{Zoom in the set of maps presented in Fig.~\ref{fig:IQUmap_example} to illustrate the alignment between the projected plane-of-the-sky magnetic field orientation and projected density structure orientation. The background color shows the intensity map while the streamlines show the magnetic field orientation as inferred from the $Q$ and $U$ Stokes parameters.}
    \label{fig:zoomIN}
\end{figure}

Figure~\ref{fig:REB-rTE_fsky} shows a small but systematic increase of $\log_{10}(\mathcal{R}_{EB})$ with $f_{\rm{sky}}$.
This trend reflects the fact that, as $f_{\rm{sky}}$ increases, more of the bright filamentary patterns clearly seen above and below the disk in Fig.~\ref{fig:IQUmap_example} are included in the spectra.
These features are generally well-aligned with the magnetic field (see Fig.~\ref{fig:zoomIN} where the integrated and projected magnetic field orientation is superimposed over the intensity map close to one of these features), and therefore produce more E-modes than B-modes (e.g. \citealt{Zal2001,Cla2021,Kon2021}).

\begin{figure*}
    \centering
    \begin{tabular}{cc}
    \hspace{1cm} $\mathcal{R}_{EB}$  & \hspace{1cm} $r^{TE}$\\
    \includegraphics[trim={0.5cm 0cm 0.5cm 0.2cm},clip,width=.95\columnwidth]{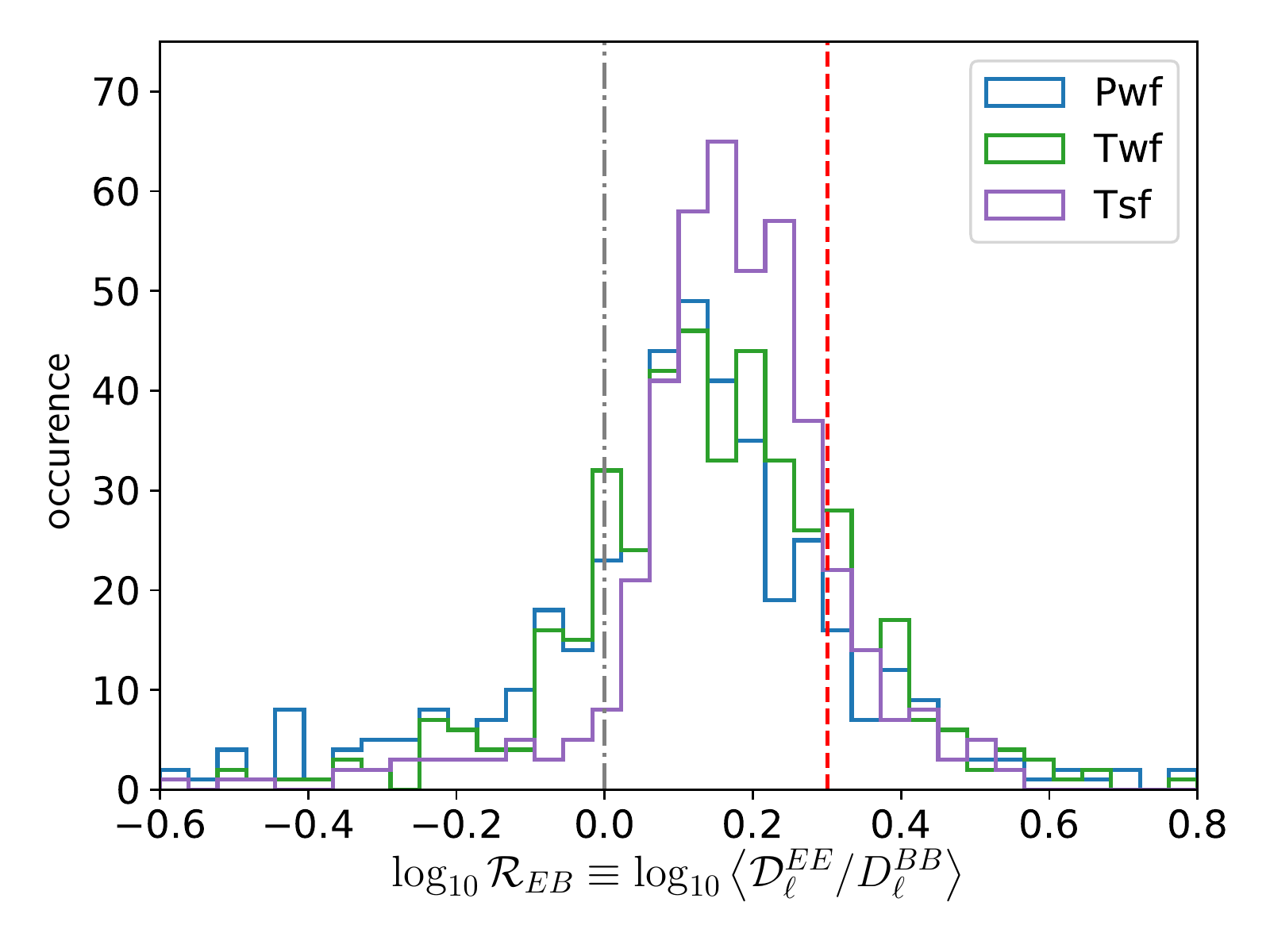}
    & \includegraphics[trim={0.5cm 0cm 0.5cm 0.2cm},clip,width=.95\columnwidth]{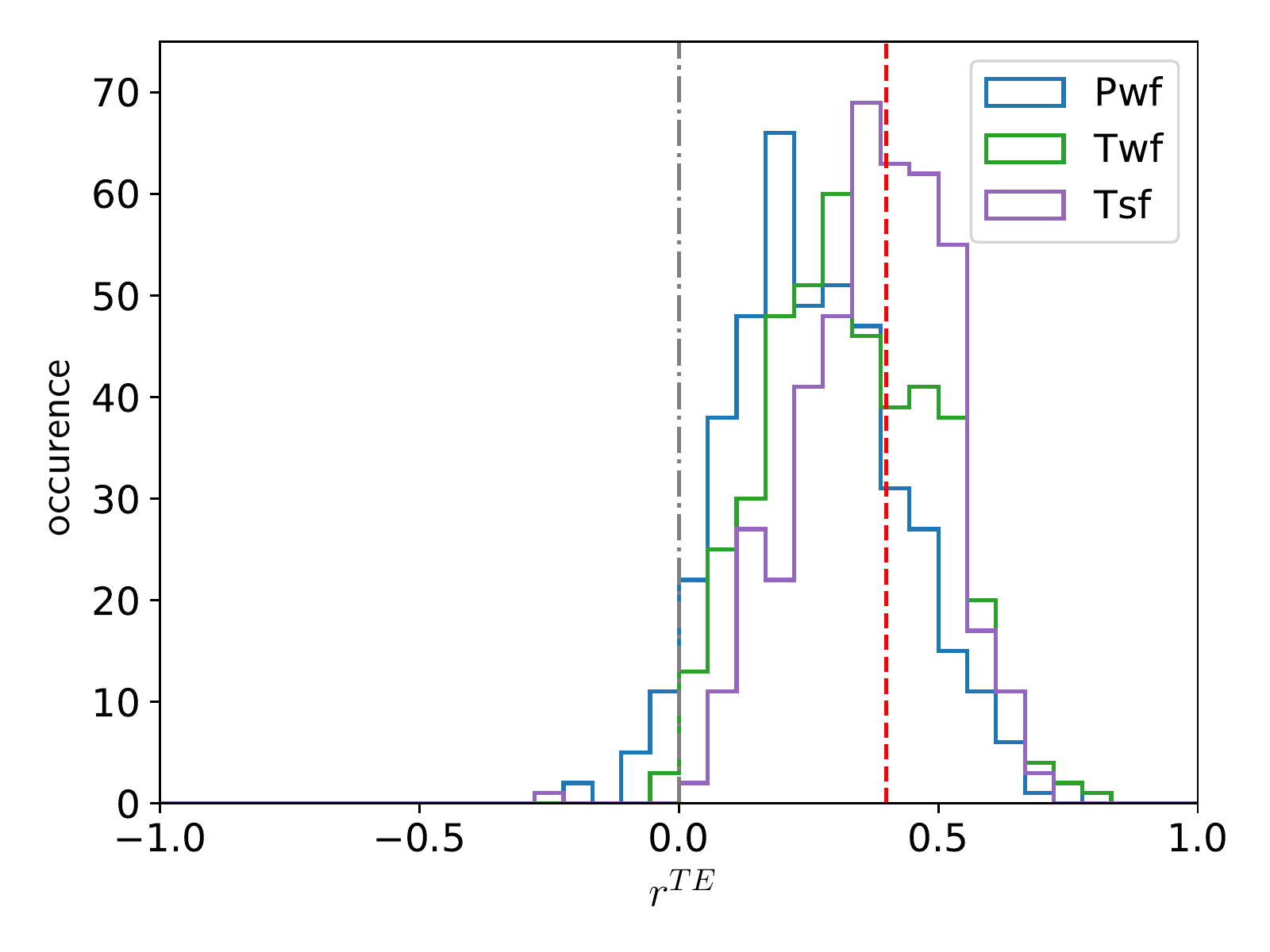}
    \\
    \includegraphics[trim={0.5cm 0.5cm 0.5cm 0.5cm},clip,width=.95\columnwidth]{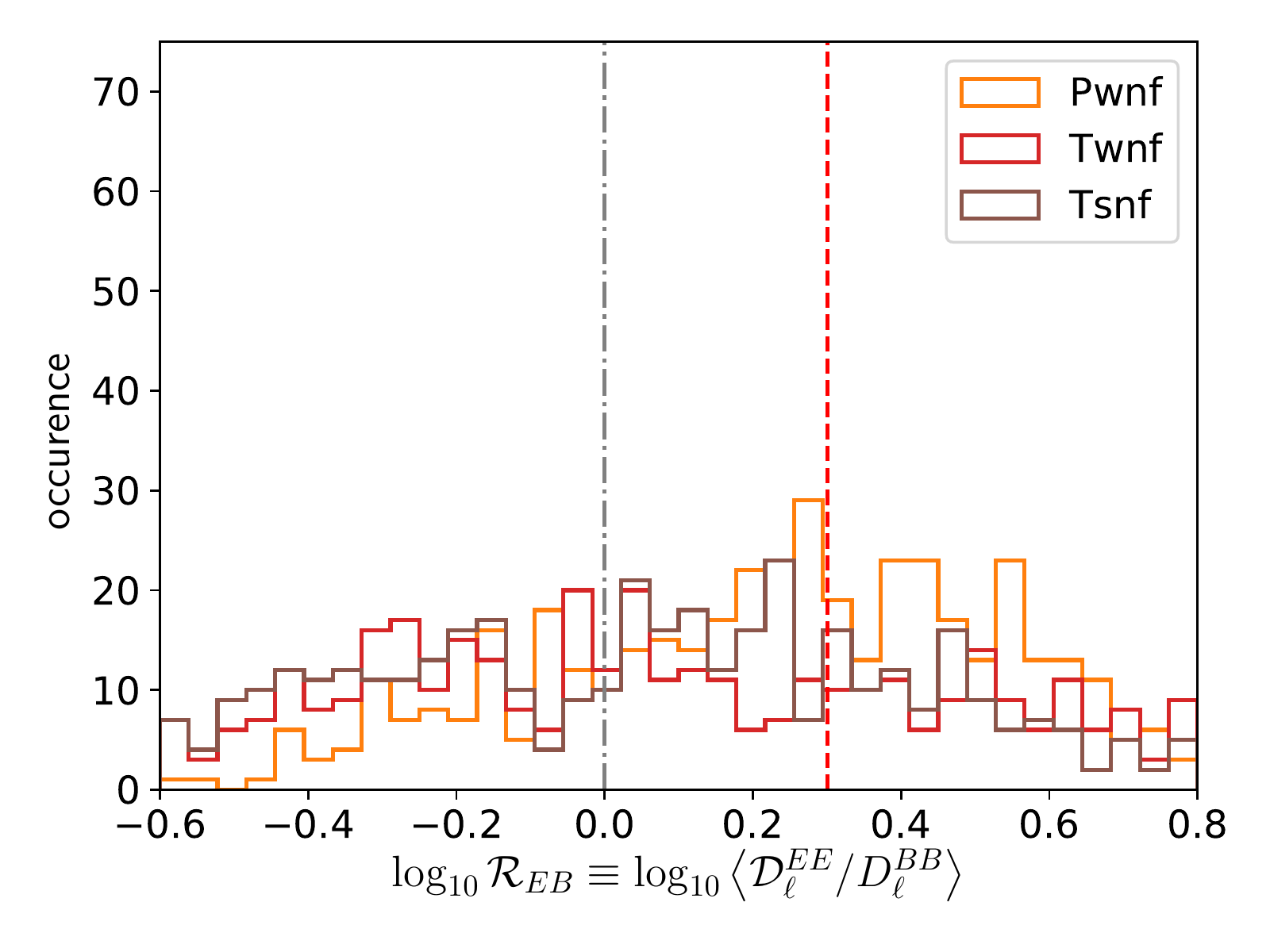}
    & \includegraphics[trim={0.5cm 0.5cm 0.5cm 0.5cm},clip,width=.95\columnwidth]{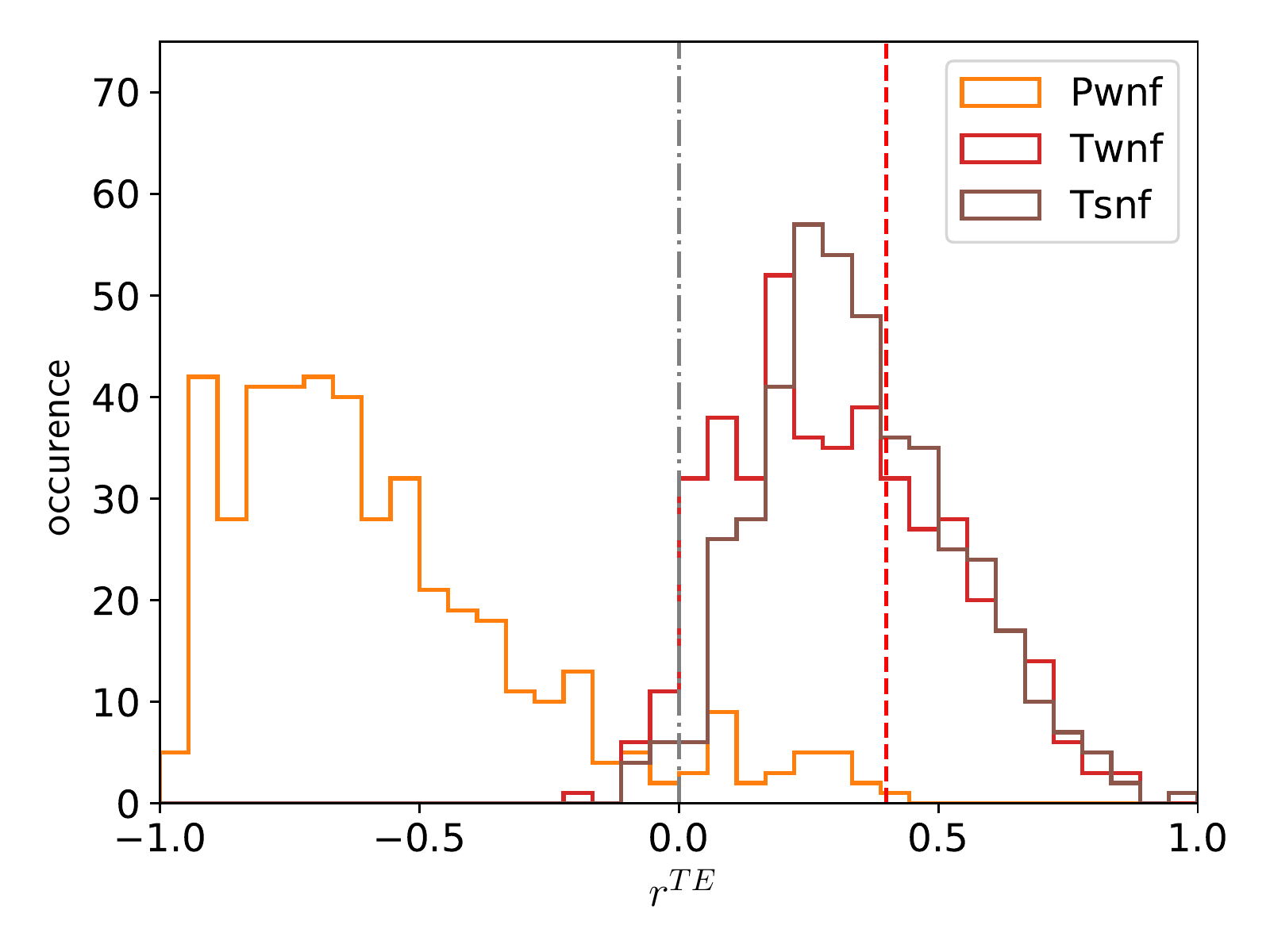}
    \end{tabular}
    \caption{Histograms of $\log_{10}(\mathcal{R}_{EB})$ (left) and $r^{TE}$ (right) as measured on polarization maps synthesized from the three models with feedback (top) and without feedback (bottom). Power spectra corresponding to all values of $f_{\rm{sky}}$ are put together, making an ensemble of 432 measurements per galaxy setup.
    The vertical gray dot-dashed (red dashed) lines mark the values of $\mathcal{R}_{EB} = 1$ ($2$) on the left and of $r^{TE} = 0$  ($0.4$) on the right, for reference.}
    \label{fig:REB-rTE_hist}
\end{figure*}

The increase of $\log_{10}(\mathcal{R}_{EB})$ with $f_{\rm{sky}}$, however, is not significant with respect to the variance induced by the observer's location for each $f_{\rm{sky}}$, shown as error bars in the same figure.
Therefore, for each galaxy model, we can group measurements from different $f_{\rm{sky}}$ values together to generate the histograms shown in Fig.~\ref{fig:REB-rTE_hist}. The large difference of $\log_{10}(\mathcal{R}_{EB})$ distributions between galaxies that include feedback (top panel) from those that do not (bottom panel) is clearly demonstrated.
The presence of feedback reduces the range of $\mathcal{R}_{EB}$ values, inducing a peak between $1\lesssim \mathcal{R}_{EB}\lesssim2$. 
Table~\ref{tab:REB-rTE_dist} contains summarizing statistics of these distributions.

We quantify the similarity of $\mathcal{R}_{EB}$ distributions obtained from the different galaxies using a 2-sample Kolmogorov-Smirnov (KS2S) test, computing the probabilities that distributions in pairs are drawn from the same parent distributions. These probabilities are reported in the lower left triangle of Table~\ref{tab:KS2S_REB-rTE_full}.

The distributions of $\log_{10}(\mathcal{R}_{EB})$ in model Pwf and Twf are very similar to each other. $\mathcal{R}_{EB}$ does not seem to be related to the initial topology of the magnetic field, at least when the field strength is low (models Twf, Twnf, Pwf, Pwnfb).
Comparison of the $\log_{10}(\mathcal{R}_{EB})$ distributions of Twf and Tsf reveals a possible effect of the magnetic field strength. The stronger the field, the narrower the distribution and the higher the central value.

The $\mathcal{R}_{EB}$ distributions are more difficult to compare between the galaxies without feedback.
However, the $\log_{10}(\mathcal{R}_{EB})$ distributions of Twnf and Tsnf appear more similar to one another than to that of Pwnf, which extends to larger values.

\subsubsection{The $r^{TE}$ parameter}
Figure~\ref{fig:REB-rTE_fsky} (right) shows that the synthetic polarization maps from all the simulations except Pwnf have mainly positive $r^{TE}$ values, independently of the $f_{\rm{sky}}$. For these five models, $r^{TE}$ increases with $f_{\rm{sky}}$. This trend stems from the inclusion of more structures as $f_{\rm{sky}}$ increases. 
Nevertheless, the $r^{TE}$ values at $f_{\rm{sky}} = 0.2$ and $0.7$ agree within uncertainties and, as for $\mathcal{R}_{EB}$ we can combine measurements obtained with different $f_{\rm{sky}}$ values to compare characteristics of the power spectra between different galaxies. The right panel of Fig.~\ref{fig:REB-rTE_hist} shows the histograms of $r^{TE}$ for different galaxies. Summarizing statistics are provided in Table~\ref{tab:REB-rTE_dist}.
We quantify the similarity of the $r^{TE}$ distributions obtained from the different galaxies using a KS2S test, computing the probabilities that distributions in pairs are drawn from the same parent distributions. Those probabilities are reported in the upper right triangle of Table~\ref{tab:KS2S_REB-rTE_full}.

In general (as seen from Fig.~\ref{fig:REB-rTE_hist}, right panels), the $r^{TE}$ histograms indicate that a toroidal magnetic field topology (i.e. field lines parallel to the disk) leads to positive $TE$ correlations, and that a stronger magnetic field strengthens the correlation.
As for the $\mathcal{R}_{EB}$ ratio, the presence of feedback in the simulations appears to shrink the distribution of possible $r^{TE}$ values. However, here the effect is milder.

The $r^{TE}$ values from the Pwnf galaxy are consistently negative for all $f_{\rm{sky}}$ values and present no significant trend with $f_{\rm{sky}}$.
This anticorrelation of $T$ and $E$ modes is caused by the peculiar magnetic field topology of this model. In Pwnf, the field lines are by construction perpendicular to the disk, and they remain so because of the absence of feedback and the short integration time of the simulation.  As \cite{Kon2021} demonstrate, a configuration where density structures and magnetic field are perpendicular to each other leads to a negative $r^{TE}$.
In the cases without feedback and a toroidal field geometry, the field lines remain parallel to the density structures of the ISM, a configuration that leads to positive $r^{TE}$.
Interestingly, the rare occurrences with positive $r^{TE}$ in model Pwnf correspond to observers located at the edges of the spiral arms. In these cases magnetic field and density structures locally follow coherent orientations and cover a large area of the sky. In other cases, the polarization sky is very quiescent.

\begin{table*}
    \caption{Results of KS2S tests on global $\mathcal{R}_{EB}$ and $r^{TE}$ distributions}
    \label{tab:KS2S_REB-rTE_full}
    \centering
    \begin{tabular}{r ccc ccc}
    \hline
    \hline \\[-1.5ex]
$\mathcal{R}_{EB}$ / $r^{TE}$ & Pwf & Twf       & Tsf             & Pwnf            & Twnf            & Tsnf            \\ \\[-1.5ex]
    \hline\\[-.5ex]
    Pwf     & $1$             & $1.0 10^{-6}$   & $1.2 10^{-21}$  & $\ll$           & $6.0 10^{-5}$   & $6.7 10^{-8}$   \\ \\[-1.5ex]
    Twf     & $1.8 10^{-2}$   & $1$             & $1.4 10^{-6}$   & $\ll$           & $8.2 10^{-4}$   & $0.28$          \\ \\[-1.5ex]
    Tsf     & $3.7 10^{-10}$  & $8.1 10^{-5}$   & $1$             & $\ll$           & $1.7 10^{-12}$  & $1.6 10^{-5}$   \\ \\[-1.5ex]
    Pwnf    & $1.3 10^{-16}$  & $1.2 10^{-13}$  & $6.2 10^{-18}$  & $1$             & $\ll$           & $\ll$           \\ \\[-1.5ex]
    Twnf    & $1.0 10^{-12}$  & $2.1 10^{-13}$  & $1.5 10^{-22}$  & $6.9 10^{-7}$   & $1$             & $8.1 10^{-5}$   \\ \\[-1.5ex]
    Tsnf    & $3.7 10^{-10}$  & $6.2 10^{-18}$  & $9.0 10^{-25}$  & $1.5 10^{-9}$   & $1.4 10^{-4}$   & $1$             \\ \\[-1.5ex]
    \hline
    \end{tabular}
    \tablefoot{KS2S probabilities that $\mathcal{R}_{EB}$ and $r^{TE}$ distributions from different galaxy are drawn from the same parent distribution. Values for $\mathcal{R}_{EB}$ are given in the lower left triangle and values for $r^{TE}$ in the upper right triangle.}
\end{table*}

\subsection{The viewpoint-induced variance}

Both $\mathcal{R}_{EB}$ and $r^{TE}$ show an increase with $f_{\rm{sky}}$. As $f_{\rm{sky}}$ increases, lines of sight with higher column density are included in the power spectrum computation. The positive correlations of $\mathcal{R}_{EB}$ and $r^{TE}$ with $f_{\rm{sky}}$ may thus reflect a correlation with density and or line-of-sight complexity of the probed ISM. The probed ISM density can be quantified through the weighted average of the column density or the weighted average of the emission intensity ($\bar{I}$), where the weights come from the mask.

\begin{SCfigure*}[.22]
  \caption{Variations of the averaged weighted emission intensity ($\bar{I}$) (top left), $\mathcal{R}_{EB}$ (bottom left) and $r^{TE}$ (bottom right) as a function of the azimuthal position of the observer ($\phi_\odot$ [$^\circ$]) for the different galaxies (indicated in the captions) and for different values of sky fraction ($f_{\rm{sky}}$, indicated by the colors in the legend).
  (top right) shows the evolution of the averaged dust density (normalized to the sample median) in spheres of radii (100 pc, 200 pc and 500 pc) around an observer as a function its azimuthal coordinate ($\phi_\odot$).
  The colors represent different values of the radius of the sphere surrounding the observer as informed by the legend. Spheres are independent if their $\phi_\odot$ differ by at least $1.43^\circ$, $2.86^\circ$ and $7.16^\circ$ for radius $R_{\rm{sph}}$ of $100$ pc, $200$ pc, and $500$ pc, respectively.
  The vertical gray bands indicate the range of $\phi_\odot$ values in which we consider the observer to be located in the arms (see Sect.~\ref{subsec:arm_interarm})
  \vspace{.8cm}
  \label{fig:I-nd-REB-rTE_phirot}}
    \begin{tabular}{cc}
        \hspace{.8cm} {\small $\bar{I}$}  & \hspace{.8cm} {\small $\tilde{n}_d$}\\
        \includegraphics[trim={1.33cm 1.cm .95cm .2cm},clip,width=.37\textwidth]{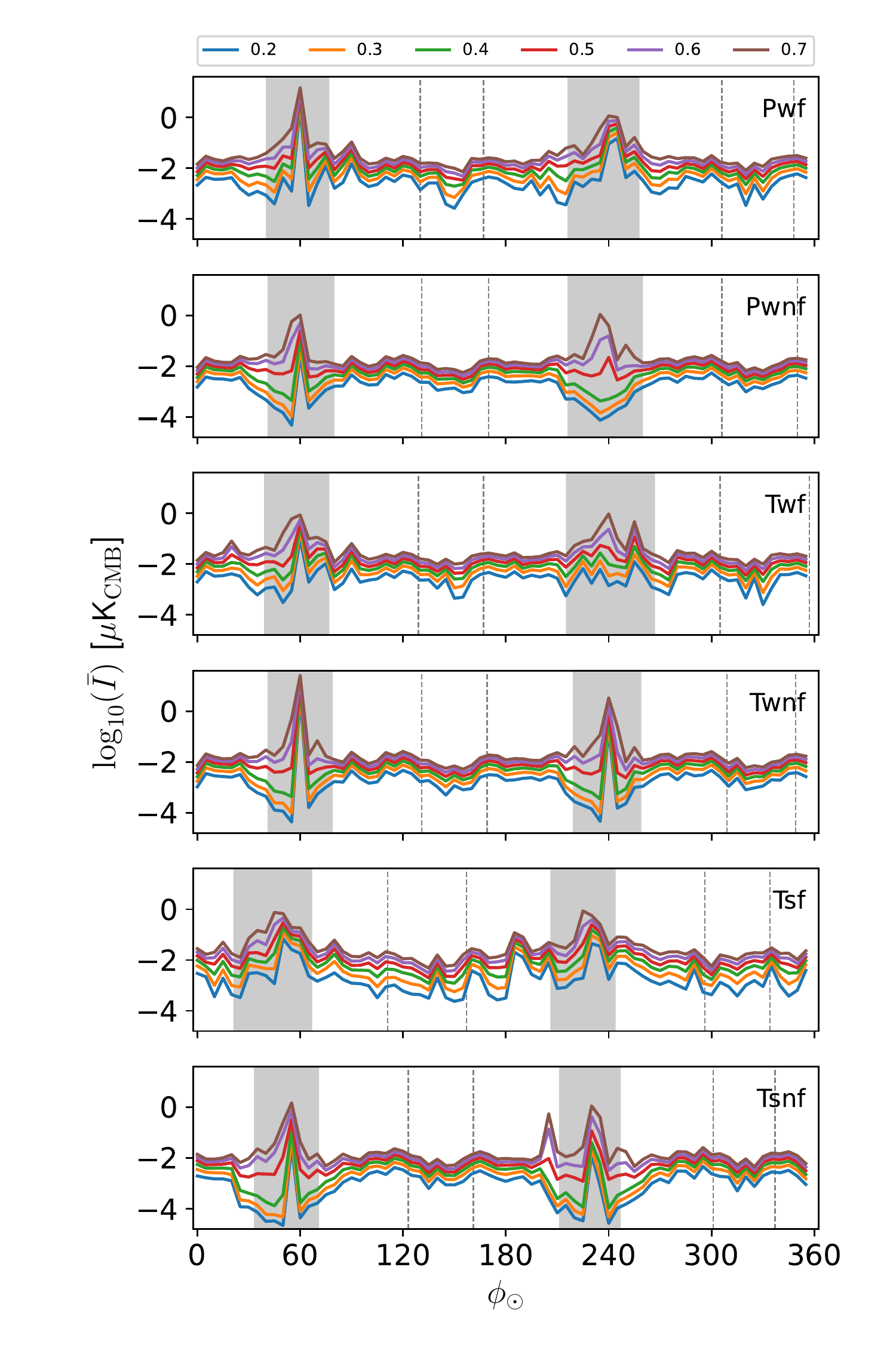}
        &
        \includegraphics[trim={1.33cm 1.cm .9cm 0.2cm},clip,width=.37\textwidth]{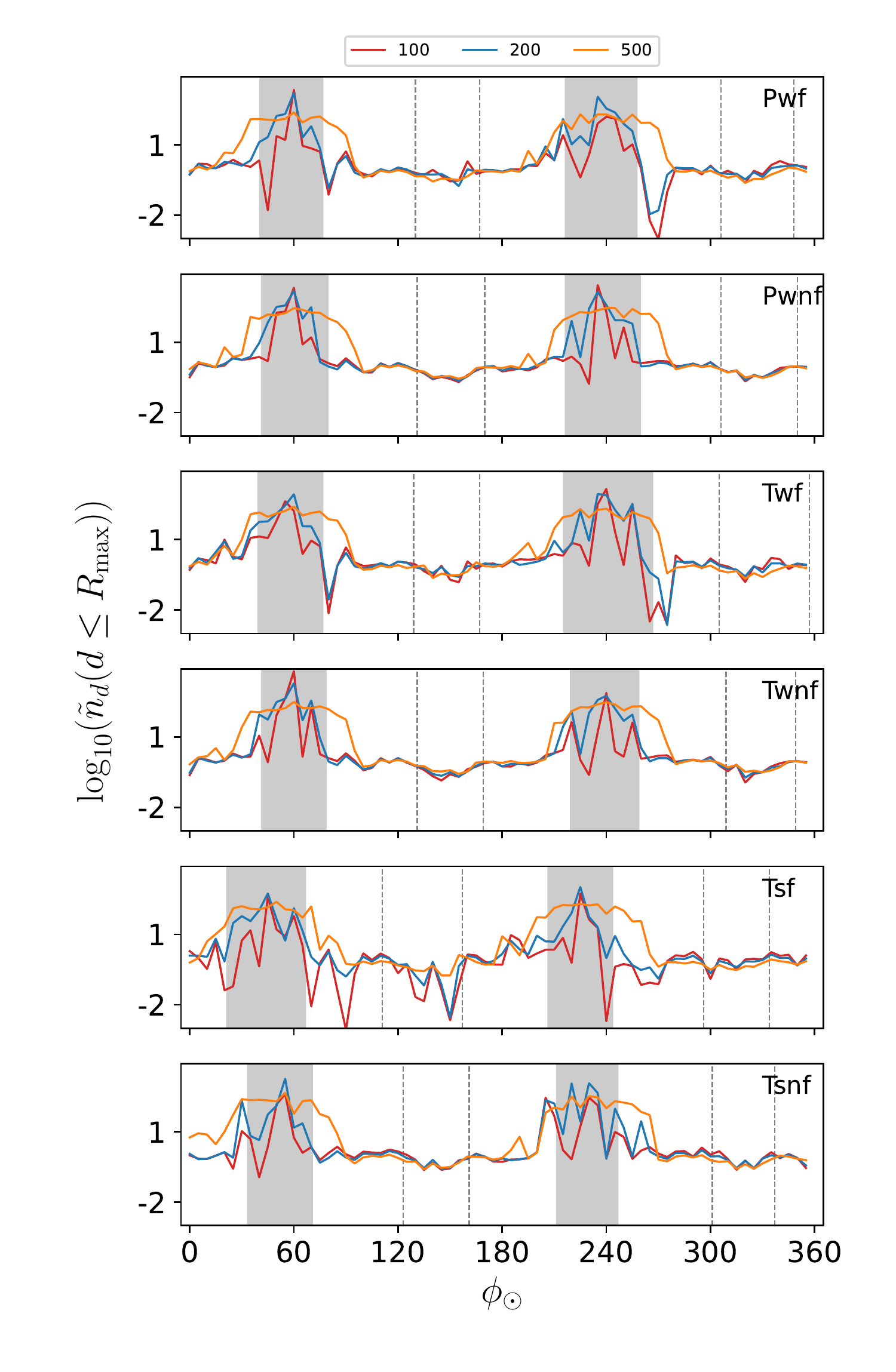}
        \\[.5ex]
        \hspace{.8cm} {\small $\mathcal{R}_{EB}$}  & \hspace{.8cm} {\small $r^{TE}$}\\
        \includegraphics[trim={1.33cm 1.cm .95cm .2cm},clip,width=.37\textwidth]{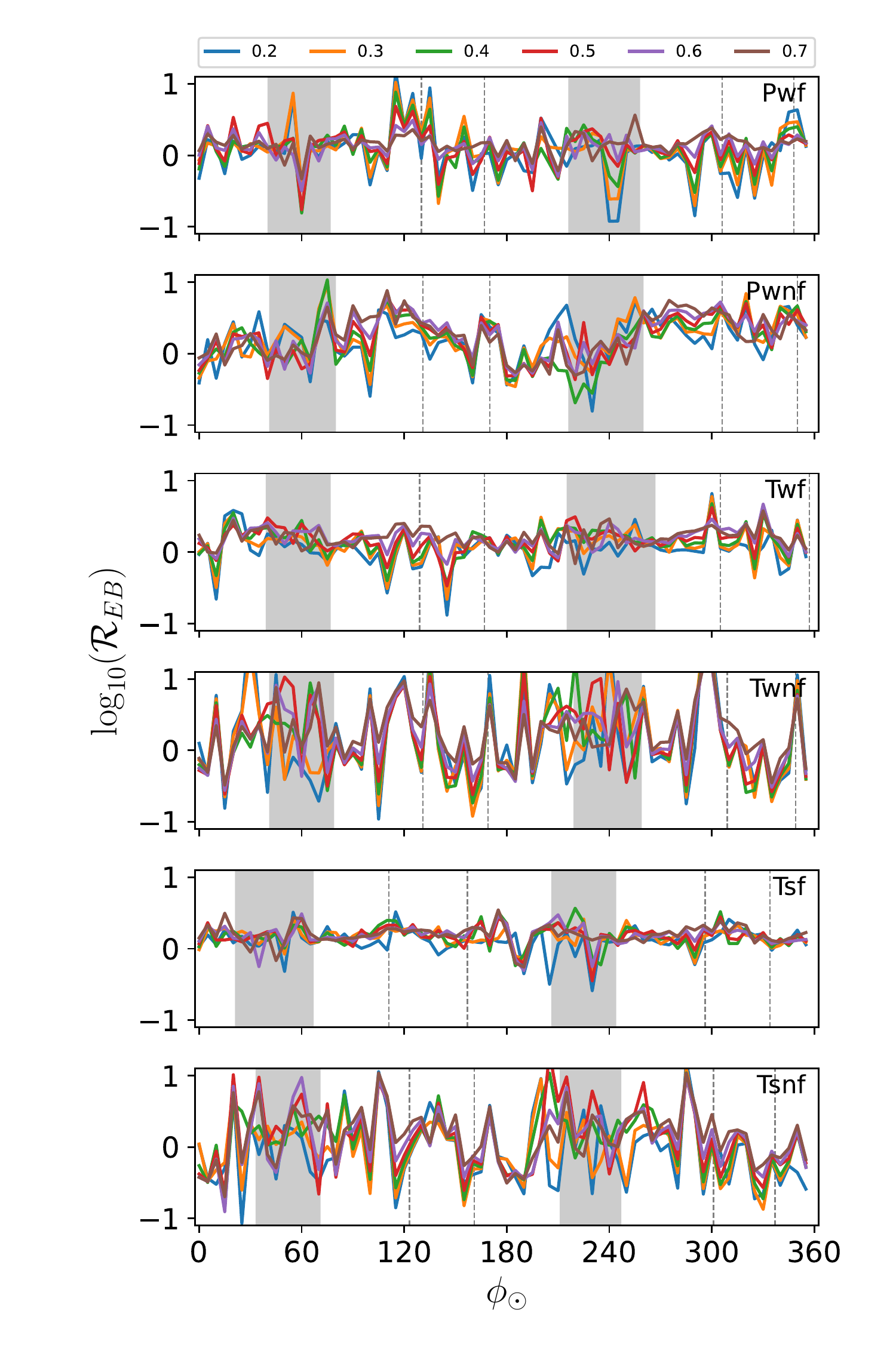}
        &
        \includegraphics[trim={1.33cm 1.cm .9cm .2cm},clip,width=.37\textwidth]{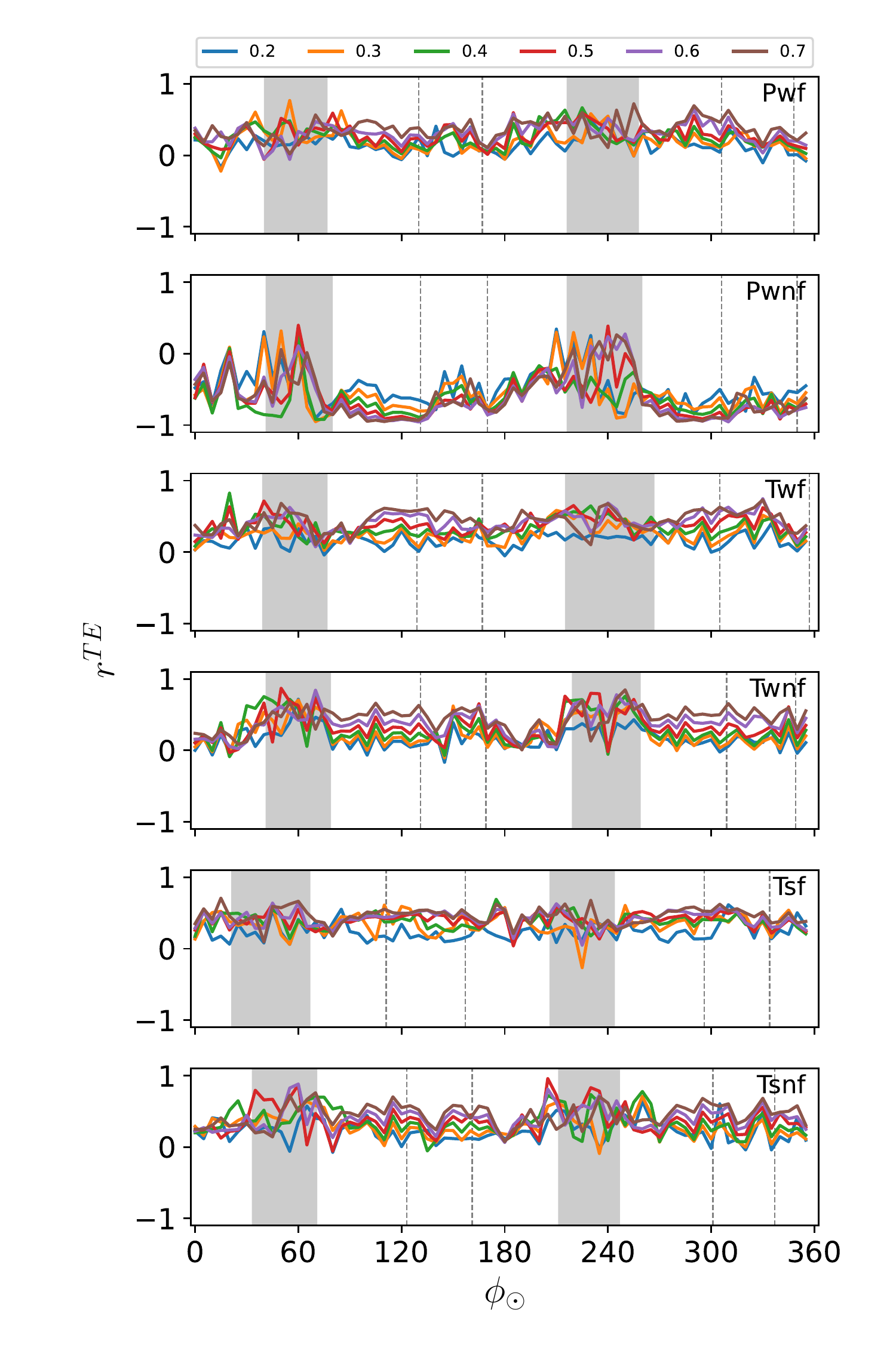}
    \end{tabular}
\end{SCfigure*}

For a given simulated galaxy the weighted average of emission intensity depends on the observer's location and on the retained sky fraction in the analysis. Figure~\ref{fig:I-nd-REB-rTE_phirot} illustrates this complex dependence. The dependence of the mean intensity, $\mathcal{R}_{EB}$ and $r^{TE}$ with azimuth angle $\phi_\odot$ (the angular coordinate of the observer in the galactic disk as introduced in Appendix~\ref{app:conventions}) is shown on the left, middle, and right panels, respectively. The difference between galaxies with and without feedback is again striking: the presence of feedback creates much more constrained distributions of $\mathcal{R}_{EB}$ and $r^{TE}$.
However, based on these figures it is difficult to conclude whether or not the $\mathcal{R}_{EB}$ and $r^{TE}$ azimuthal modulations are correlated with those of $\bar{I}$.

To address this question we carry out Spearman rank order correlation tests to quantify whether or not $\mathcal{R}_{EB}$ and or $r^{TE}$ are correlated with $\bar{I}$ either when all galaxies and $f_{\rm{sky}}$ are considered together or when galaxies are considered separately. The results are reported in Table~\ref{tab:Imean_REB-rTE_Spearman}.

The correlation of $\mathcal{R}_{EB}$ with $\bar{I}$ is strong when measurements from all sky fraction values and all galaxies are considered together.
There is a relatively strong correlation between $\mathcal{R}_{EB}$ and $\bar{I}$ for Twf and Twnf, and a weak correlation for Pwf and Pwnf. There is no correlation between $\mathcal{R}_{EB}$ and $\bar{I}$ for Tsf and Tsnf.
The absence of correlation for Tsf is likely due to the smaller range of $\mathcal{R}_{EB}$ values encountered in this galaxy.

The correlation of $r^{TE}$ with $\bar{I}$ is strong and significant when gathering all sky fraction and all galaxies into a unique sample. It also persists when examining each model individually. The strongest correlations are found for galaxies with feedback and low value of the initial magnetic field strength.

In general, $\mathcal{R}_{EB}$ and $r^{TE}$ are well correlated with $\bar{I}$ for the Twf galaxy whereas Tsf (an identical model with stronger field) shows almost no correlation.
Again, the loss of correlation is likely due to the smaller range of $\mathcal{R}_{EB}$ and $r^{TE}$ observed for Tsf as compared to Twf.

\smallskip

We conclude that $\mathcal{R}_{EB}$ and $r^{TE}$ depend strongly on the observer's position. This dependence is generally reflected by a correlation of both $\mathcal{R}_{EB}$ and $r^{TE}$ with the weighted averaged of the emission intensity.

\begin{table}
    \caption{Results of Spearman correlation tests.}
    \label{tab:Imean_REB-rTE_Spearman}
    \centering
    \begin{tabular}{r cc | cc}
    \hline
    \hline \\[-1.5ex]
\multirow{2}{*}{Setup} & \multicolumn{2}{c}{$\bar{I} - \log_{10}(\mathcal{R}_{EB})$} & \multicolumn{2}{c}{$\bar{I} - r^{TE}$} \\
        &   $\rho$  &  p-value &  $\rho$  &  p-value    \\
    \\[-.5ex]
    \hline
    \\
full    & $0.12$    & $10^{-9}$     & $0.20$    & $10^{-25}$    \\ \\[-1.ex]
Pwf     & $0.10$    & $3\,10^{-2}$	& $0.33$	& $2\, 10^{-12}$\\ \\[-1.5ex]
Pwnf	& $0.15$	& $1\,10^{-3}$	& $-0.16$	& $9\, 10^{-4}$	\\ \\[-1.5ex]
Twf		& $0.25$	& $8\, 10^{-8}$	& $0.41$	& $3\, 10^{-19}$\\ \\[-1.5ex]
Twnf	& $0.19$	& $5\, 10^{-5}$	& $0.20$	& $3\, 10^{-5}$	\\ \\[-1.5ex]
Tsf		& $0.03$	& $0.50$		& $0.18$	& $2\, 10^{-4}$	\\ \\[-1.5ex]
Tsnf	& $0.05$	& $0.26$		& $0.24$	& $4\, 10^{-7}$	\\ \\[-1.5ex]
\hline
    \end{tabular}
    \tablefoot{Spearman rank order correlation coefficients and p-values. Testing correlation between $\bar{I}$ and $\mathcal{R}_{EB}$ and $r^{TE}$, gathering all galaxies and retained sky fractions in one sample (full, top line) or dividing the sample by galaxy model.}
\end{table}

\medskip

\subsection{The observer environment}
Given that the summarizing characteristics of the polarization power spectra appear to depend sensitively on the observer's location, we want to check whether placing the observer in similar environments yields similar $\mathcal{R}_{EB}$ and $r^{TE}$ values independently of the galaxy model.
In the following, we start by investigating possible differences in the power spectra obtained by observers located within or outside spiral arms (Sect.~\ref{subsec:arm_interarm}).
Since we do find substantial differences between these two subgroups, we proceed to search for any correlations of the power spectrum characteristics with the mean dust density and magnetic field strength in the observer's vicinity (Sect.~\ref{subsec:local_nd-B}). We cannot find any, but caution against the fact that, due to masking, those local estimates are not fully representative of the ISM regions that are imprinted in the power spectra.
Finally, we also investigate the effect of placing the observer within supernova-driven bubbles (Sect.~\ref{subsec:bubbles}). We find that power spectra taken from within bubbles are peculiar within each galaxy model, even if the scatter remains large. 
However, the effect of a bubble environment generally supersedes the imprint of the underlying galaxy model.

\subsubsection{Arm vs inter-arm regions}
\label{subsec:arm_interarm}
We first want to investigate the possible effects of having the observer in an arm or in an inter-arm region. Figure~\ref{fig:I-nd-REB-rTE_phirot} (top right) shows the normalized angular profile ($\tilde{n}_d(\phi_\odot)$) of the mean density $\bar{n}_d$ in a sphere of given radius ($R_{\rm{sph}}$) surrounding the observer which we normalize by the median of the profile.
The location of the two arms are well spotted by eye. Within the arms they are several substructures with local maxima and minima, possibly due to the presence of large aggregates of clumps and large under-densities or bubbles.
To automate the determination of the angular coordinates of the edges of the arms at the galacto-centric radius of $8$ kpc, we consider the mean-density curves ($\tilde{n}_d(\phi_\odot)$) obtained with $R_{\rm{sph}} = 200$ pc measured at an angular step of $1^\circ$. We smooth those curves with a wide Gaussian kernel (FWHM of 10$^{\circ}$) to eliminate small-scale features and consider the successive derivatives of $\tilde{n}_d(\phi_\odot)$. The edges of the arms are defined as the locii where the third derivatives ($\partial_{\phi_\odot}^3 (\tilde{n}_d)$) vanish on either sides of the two main maxima of $\tilde{n}_d(\phi_\odot)$. For each galaxy, the range of angular coordinates span by the arms are marked by gray bands on the panels of Fig.~\ref{fig:I-nd-REB-rTE_phirot}.

It is interesting to note that on the right side of most of the arms one can observe sharp dips that reveal the trails of the density waves that propagate through the disk toward decreasing $\phi_{\odot}$. This feature is best seen with $R_{\rm{sph}}=200$ pc (and with angular sampling of $1^\circ$, not shown here).

\smallskip

In order to study whether or not the fact that the observer is in an arm influences the characteristics of the polarization power spectra, we create subsamples of measurements corresponding (i) to observers within the arms (i.e. with $\phi_\odot$ falling in the gray bands of Fig.~\ref{fig:I-nd-REB-rTE_phirot}) and (ii) to observers located away from the arms. The latter are obtained by shifting the limits of the arms by 90$^\circ$. This ensures that, for a given galaxy, the sub-samples of spectra taken within and away from the arms have the same size.
The number of observer locations thus obtained ranges from 15 to 17 per model.
We visually check (on column density maps like the one in Fig.~\ref{fig:circularexcursion}) that our arm/inter-arm determination is effective.
In Table~\ref{tab:REB-rTE_inoutArms} we report descriptive statistics of $\mathcal{R}_{EB}$ and $r^{TE}$ distributions obtained for the six galaxies, merging measurements obtained with all $f_{\rm{sky}}$.
\begin{table}
    \caption{Statistics of $\mathcal{R}_{EB}$ and $r^{TE}$ within/away from the arms.}
    \label{tab:REB-rTE_inoutArms}
    \centering
    \begin{tabular}{r cc | cc}
    \hline
    \hline \\[-1.5ex]
\multirow{2}{*}{Setup} & \multicolumn{2}{c}{$\log_{10}(\mathcal{R}_{EB})$} & \multicolumn{2}{c}{$r^{TE}$} \\
    &   in  &   out &   in  &   out \\
    \\[-.5ex]
    \hline
    \\
full	& $0.17_{-0.26}^{+0.26}$	& $0.15_{-0.38}^{+0.21}$	& $0.34_{-0.32}^{+0.23}$    & $0.26_{-0.42}^{+0.49}$	\\ \\[-1.ex]
Pwf	    & $0.12_{-0.20}^{+0.15}$	& $0.10_{-0.34}^{+0.18}$	& $0.28_{-0.12}^{+0.24}$    & $0.22_{-0.11}^{+0.17}$	\\ \\[-1.ex]
Pwnf	& $0.06_{-0.26}^{+0.35}$	& $0.33_{-0.19}^{+0.23}$	& $-0.38_{-0.37}^{+0.47}$   & $-0.72_{-0.13}^{+0.17}$	\\ \\[-1.ex]
Twf	    & $0.20_{-0.13}^{+0.14}$	& $0.13_{-0.21}^{+0.19}$	& $0.39_{-0.17}^{+0.16}$	& $0.31_{-0.17}^{+0.22}$	\\ \\[-1.ex]
Twnf	& $0.28_{-0.55}^{+0.51}$	& $-0.15_{-0.34}^{+0.31}$	& $0.54_{-0.24}^{+17}$	    & $0.27_{-0.16}^{+0.25}$	\\ \\[-1.ex]
Tsf	    & $0.18_{-0.14}^{+0.12}$	& $0.22_{-0.09}^{+0.09}$	& $0.38_{-0.16}^{+0.16}$	& $0.44_{-0.18}^{+0.09}$	\\ \\[-1.ex]
Tsnf	& $0.23_{-0.34}^{+0.30}$	& $0.06_{-0.48}^{+0.23}$	& $0.41_{-0.19}^{+0.25}$	& $0.32_{-0.21}^{+0.19}$	\\ \\[-1.ex]
\hline
\end{tabular}
    \tablefoot{Medians of $\mathcal{R}_{EB}$ and $r^{TE}$ as measured on spectra corresponding to observer located within an arm (in) or away from the arms (out) for each galaxy setup and merging data from all $f_{\rm{sky}}$ values. The $\pm$ values are such that the interval around the median encodes 68\% of the data points.}
\end{table}
For most of the cases, we see that both $\mathcal{R}_{EB}$ and $r^{TE}$ are larger when observed from within the arms. Only models Pwnf and Tsf do not show this trend.

We test the null hypothesis that distributions of $\mathcal{R}_{EB}$ ($r^{TE}$) as measured from inside or away from the arms are drawn from the same parent distribution using a KS2S test. The results are reported in Table~\ref{tab:REB-rTE_inoutArms_KS2s}.
\begin{table}
    \caption{Results of KS2S tests between inside/outside-arm distributions.}
    \label{tab:REB-rTE_inoutArms_KS2s}
    \centering
    \begin{tabular}{r cc | cc}
    \hline
    \hline \\[-1.5ex]
\multirow{2}{*}{Setup} & \multicolumn{2}{c}{$\log_{10}(\mathcal{R}_{EB})$} & \multicolumn{2}{c}{$r^{TE}$} \\
    &   ks  &   p-value &   ks  &   p-value \\
    \\[-.5ex]
    \hline
    \\
full	& $0.07$	& $0.13$		& $0.14$	& $2\,10^{-5}$	\\ \\[-1.ex]
Pwf	    & $0.13$	& $0.44$		& $0.22$	& $2\,10^{-2}$	\\ \\[-1.ex]
Pwnf	& $0.42$	& $2\,10^{-8}$	& $0.54$	& $6\,10^{-14}$	\\ \\[-1.ex]
Twf	    & $0.23$	& $1\,10^{-2}$	& $0.21$	& $3\,10^{-2}$	\\ \\[-1.ex]
Twnf	& $0.41$	& $2\,10^{-7}$	& $0.45$	& $5\,10^{-9}$	\\ \\[-1.ex]
Tsf	    & $0.16$	& $0.16$		& $0.21$	& $3\,10^{-2}$	\\ \\[-1.ex]
Tsnf	& $0.3$	    & $6\,10^{-4}$	& $0.23$	& $1\,10^{-2}$	\\ \\[-1.ex]
\hline
    \end{tabular}
    \tablefoot{We report the measured statistic (ks) and corresponding p-value that the distributions of $\mathcal{R}_{EB}$ (left) and $r^{TE}$ (right) as measured for an observer inside or away from the galaxy arms are drawn from the same parent distribution. We merge data values from all $f_{\rm{sky}}$ values.}
\end{table}
Generally, the arm and inter-arm distributions differ. This effect is stronger for galaxies without feedback and is not observed for the Tsf galaxy. The latter is consistent with the fact that the variance of both $\mathcal{R}_{EB}$ and $r^{TE}$ distributions are smaller for the Tsf galaxy.
The distributions of $\mathcal{R}_{EB}$ and $r^{TE}$ within and away from the arms still differ according to the underlying galaxy simulations.

\smallskip

According to the latest observational evidence, the Milky Way probably has four spiral arms  (e.g., \citealt{Reid2019}), while our input models only include two. Doubling the number of arms would certainly affect our results.
According to our findings, the general trend is for both $\mathcal{R}_{EB}$ and $r^{TE}$ to increase for maps obtained from within the arms compared to other locations. Therefore, we may speculate that we would observe more frequently large values of $\mathcal{R}_{EB}$ and $r^{TE}$ in histograms like the ones in Fig.~\ref{fig:REB-rTE_hist} for a galaxy with four instead of two spiral arms.
However, the difference seen between arm and inter-arm regions should persist. In future work, this conjecture needs to be tested against new simulations because other non-trivial effects could impact the power spectra.

\subsubsection{Local density and magnetic field strength}
\label{subsec:local_nd-B}
In this section we search for correlation of $\mathcal{R}_{EB}$ and $r^{TE}$ with the local density and the local strength of the magnetic field.
We determine the local values for each observer location in measuring the mean values within a sphere of a given radius ($R_{\rm{sph}}$) around the observer.
We consider two values for the radius: 200 and 500 pc. These values are chosen such that the spheres encompass most of the dusty ISM matter that contribute to the simulated polarization skies at high galactic latitudes. However, because the brightest regions are masked, these local values may not be fully representative of the part of the ISM that is included in the power spectrum estimates.

\medskip

The ranges of local density values encountered in all the six galaxy models are similar and follow the variation about the coordinate $\phi_\odot$ shown in Fig.~\ref{fig:I-nd-REB-rTE_phirot} (top right).
We are not able to find convincing correlation of $\mathcal{R}_{EB}$ or $r^{TE}$ with the local density. Only the no-feedback galaxies do show weak correlations (with p-values at the order of 1\%) between $r^{TE}$ and the local density. This is because both $r^{TE}$ and the density increase within the arms (see also previous section) as it can be inferred by inspecting the bottom right panel of Fig.~\ref{fig:I-nd-REB-rTE_phirot}.

\medskip

\begin{figure}
    \centering
    \includegraphics[trim={0.4cm .4cm 0.4cm 0cm},clip,height=6cm]{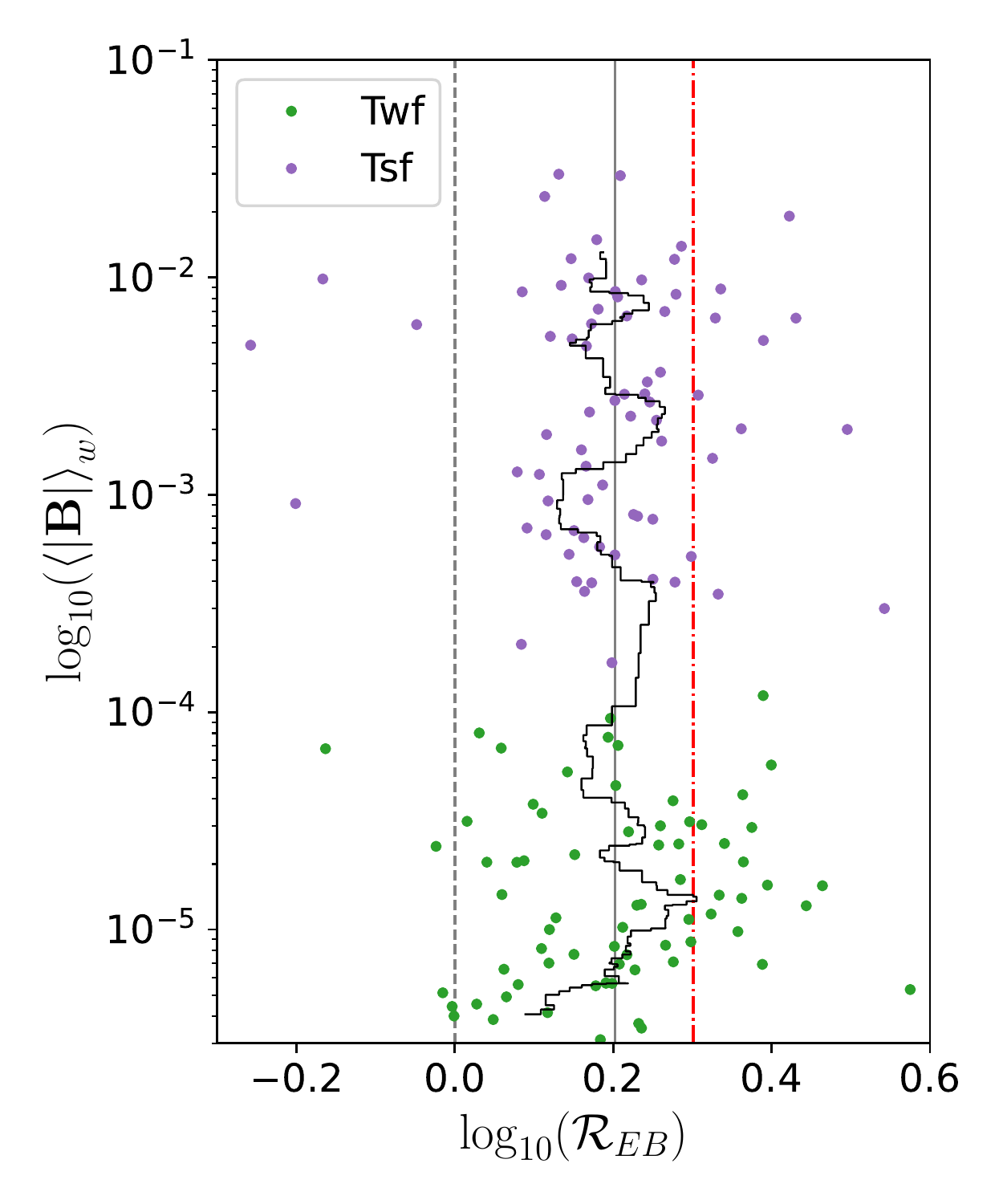}
    \includegraphics[trim={2.6cm .4cm .4cm 0cm},clip,height=6cm]{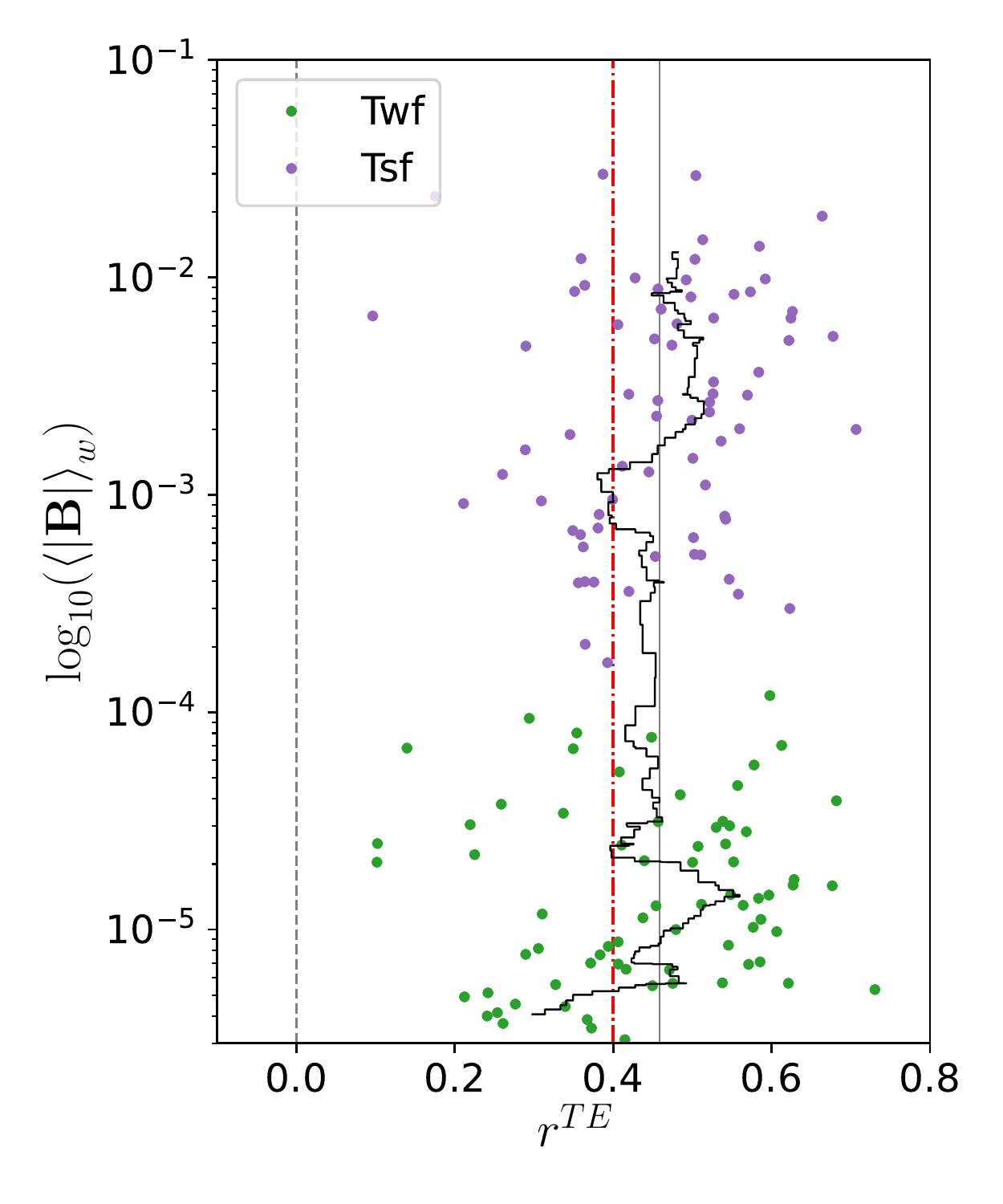}
    \caption{Scatter plots of strength of the local magnetic field (in $\mu$G) versus $\mathcal{R}_{EB}$ (left) and $r^{TE}$, measured from the Twf (green) and Tsf (purple) galaxies for $f_{\rm{sky}} = 0.7$.
    The strength of the field is the volume-weighed average of the field of all cells having their centers within a sphere radius of 500 pc centered on the observer. Gray and red vertical lines are the same as in Fig.~\ref{fig:REB-rTE_hist}. The gray tine solid line show the medians of the data points and the shacked black lines show running means along $|\mathbf{B}|$ considering 10 data points at a time.}
    \label{fig:Bf500_REB-rTE_fsky070}
\end{figure}

The range of values of the strength of the magnetic field that is local to the observer ($\left< |\mathbf{B}| \right>_{R_{\rm{sph}}}$) encountered in all the six galaxy models spans several orders of magnitude.
We concentrate on the Twf and Tsf galaxies in order to infer the effect of the strength of the magnetic field while keeping as much similar as possible the other parameters. In this case, $\left< |\mathbf{B}| \right>_{R_{\rm{sph}}}$ spans four orders of magnitude, as shown in Fig.~\ref{fig:Bf500_REB-rTE_fsky070}. Despite this large coverage we find no correlation of $\mathcal{R}_{EB}$ or $r^{TE}$ with $\left< |\mathbf{B}| \right>_{R_{\rm{sph}}}$ as also illustrated in that figure and as confirmed through Spearman rank order tests that we performed for all values of the sky fraction and for both $R_{\rm{sph}} = 200$ and $500$ pc.
For $R_{\rm{sph}} = 500$, the probability that any observed correlation happens by chance are 55\% and 12\% for the pairs $(\left< |\mathbf{B}| \right>_{R_{\rm{sph}}},\, \mathcal{R}_{EB})$ and $(\left< |\mathbf{B}| \right>_{R_{\rm{sph}}},\, r^{TE})$, respectively.
Therefore, based on this analysis we cannot conclude that the strength of the local field to the observer is an important factor in shaping the polarization power spectra.
We emphasize that this estimate of $\left< |\mathbf{B}| \right>_{R_{\rm{sph}}}$ does not necessarily represent adequately the strength of the field permeating the medium that is imprinted in the polarization maps. The test carried in Sect.~\ref{sec:BinCone} addresses this point in particular.

\subsubsection{Within bubbles}
\label{subsec:bubbles}
There is ample observational evidence that the Sun resides in a special environment called the Local Bubble. The Local Bubble is a cavity of hot plasma created by supernova explosions, surrounded by a magnetized shell of cold, dusty gas (e.g., \citealt{Lal2018}; \citealt{Pel2020}). Guided by this observational fact, we create maps as before, this time placing observers inside supernova-driven bubbles, self-consistently generated in the simulations.
We visually identify 14, 14, and 11 bubbles within the Pwf, Twf, and Tsf galaxies in various locations (center, arm, and inter-arm regions) of the galactic midplane.
We do not impose any other selection criteria on our bubble sample.
Their sizes range from $\sim 100$ pc to $\sim 1.3$ kpc.
Interestingly, we find that the strength of the magnetic field within the bubbles ranges from 0.5 to 1 $\mu$G independently of their environment (center, arm, inter-arm) and their underlying galaxy model. This coincidence may be related to the physical conditions necessary to lead to the formation and explosion of stars. However, a detailed study of this feature is outside the scope of this paper.
\begin{figure*}
    \centering
    \begin{tabular}{cc}
    \hspace{1cm} $\mathcal{R}_{EB}$  & \hspace{1cm} $r^{TE}$\\
    \includegraphics[trim={0.5cm 0cm 0.5cm 0.2cm},clip,width=.95\columnwidth]{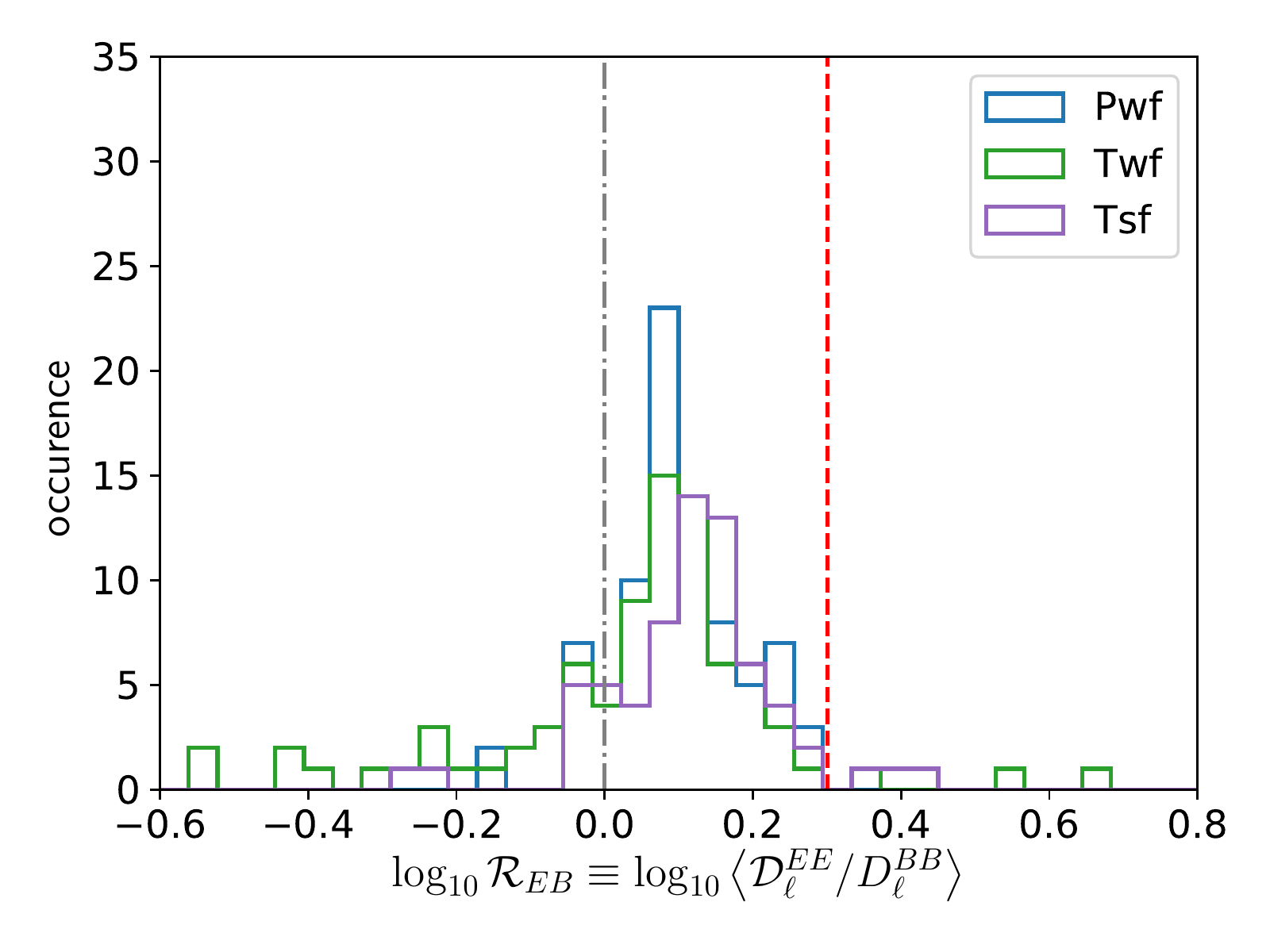}
    & \includegraphics[trim={0.5cm 0cm 0.5cm 0.2cm},clip,width=.95\columnwidth]{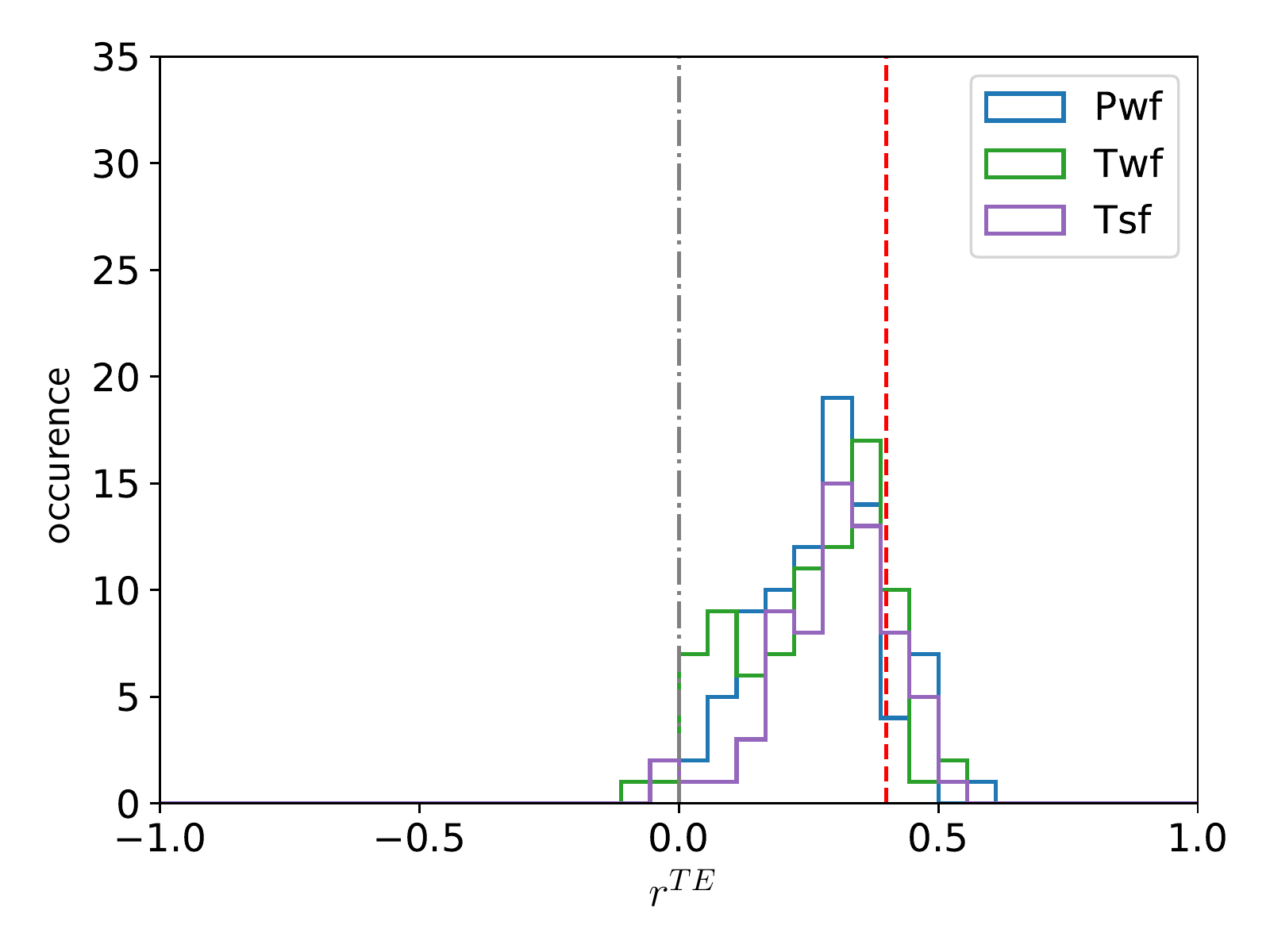}
    \end{tabular}
    \caption{Normalized histograms of $\log_{10}(\mathcal{R}_{EB})$ (left) and $r^{TE}$ (right) as measured on polarization maps synthesized from the inside of bubbles detected in the three models with feedback. Power spectra corresponding to all values of $f_{\rm{sky}}$ are put together. Vertical lines are the same as in Fig.~\ref{fig:REB-rTE_hist}.}
    \label{fig:REB-rTE_hist-Bubbles}
\end{figure*}
In what follows we compare the distributions of $\mathcal{R}_{EB}$ and $r^{TE}$ as obtained from the inside of bubbles (shown in Fig.~\ref{fig:REB-rTE_hist-Bubbles}) to those from the main samples (see top row of Fig.~\ref{fig:REB-rTE_hist}).

First, for each galaxy model, we investigate whether the $\mathcal{R}_{EB}$ and $r^{TE}$ distributions from the bubbles can be considered random occurrences within the main samples. We use a KS2S test to address this question.
The probabilities that the distributions corresponding to the bubble locations and those of the main sample are drawn from the same parent distribution --for the same galaxy model-- are found to be low, as reported in Table~\ref{tab:REB-rTE_bb-full_KS2s}.
Therefore, it is unlikely that the distributions of $\mathcal{R}_{EB}$ and $r^{TE}$ measured from the inside of bubbles are
random realizations of the main sample. The polarization skies and power spectra, from inside bubbles are peculiar realizations in their respective galaxy: Both the $\mathcal{R}_{EB}$ and $r^{TE}$ distributions from within bubbles are shifted toward lower values. The effect is more pronounced for Twf and Tsf.
\begin{table}
    \caption{Results of KS2S tests between global and inside-bubble distributions.}
    \label{tab:REB-rTE_bb-full_KS2s}
    \centering
    \begin{tabular}{r c | c}
    \hline
    \hline \\[-1.5ex]
{Setup} & {$\log_{10}(\mathcal{R}_{EB})$} & {$r^{TE}$} \\
    \\[-.5ex]
    \hline
    \\
Pwf	    & $7.9\,10^{-3}$    & $0.138$	        \\ 
Twf	    & $1.9\,10^{-6}$    & $3.4\,10^{-4}$	\\ 
Tsf	    & $6.2\,10^{-4}$    & $5.3\,10^{-5}$	\\ \\[-1.ex]
\hline
    \end{tabular}
    \tablefoot{We report the p-value that the distributions of $\mathcal{R}_{EB}$ (left) and $r^{TE}$ (right) as measured for an observer within a bubble have the same parent distributions than from the main sample (full circular excursion). We merge data values from all $f_{\rm{sky}}$ values.}
\end{table}

Second, we study whether the distributions corresponding to the bubbles are statistically similar irrespective of their underlying galaxy. We compute the KS statistics ($ks^\star$) and the corresponding probabilities ($p_{KS}^\star$) for each comparison in pairs. The values are reported in the upper blocks of Table~\ref{tab:KS2S_REB-rTE_Bubbles}.
We see that the distributions are generally more similar to each other than the main samples are to one another (see Table~\ref{tab:KS2S_REB-rTE_full}).
The distributions of $\mathcal{R}_{EB}$-bubbles and $r^{TE}$-bubbles from Twf and Tsf are those that keep deviating the most with a minimum probability for $\mathcal{R}_{EB}$ of $\lesssim 1$\% of being drawn from the same parent distribution.

To assess whether this similarity between the distributions is not simply an effect of decreased sample size we proceed as follows.
We generate 10.000 bootstrap sub-samples from the main sample, each with the size of the bubble samples. The bootstrap selection is performed by means of observer location. Then, we compare bootstrapped samples from two different galaxy setups and for each compute the KS2S statistics $ks$. This leads to distributions of $ks$ for the comparison of two sub-samples drawn from, say, Pwf and Twf. We then compare the same statistics but between Pwf-bubble and Twf-bubble  ($ks^\star$) to the distribution obtained above in computing the number of draws with $ks\leq ks^\star$ and divide by the total number of draws. This gives the one-sided p-value that the bubbles samples look similar by chance, i.e. that random sub-samples from the main samples lead to at least this level of similarity due to the loss of statistic. These probabilities are reported in the bottom block of Table~\ref{tab:KS2S_REB-rTE_Bubbles}.

We see that the probability of randomly observing a $\mathcal{R}_{EB}$ from the main sample that is similar to one from the bubble sample is high. However, that is not true of the $r^{TE}$ distributions. There, the probability reaches a minimum $\lesssim 1$\% for the comparison of Pwf-bubble and Tsf-bubble.

This observation is consistent with our earlier finding that, for the toroidal field geometry, taking power spectra from within a bubble reduces the $r^{TE}$ values. This effect is more pronounced the stronger the magnetic field and negligible for the poloidal field topology, where the observer's position induces a larger scatter.

\smallskip

In conclusion, we observe that given an underlying galaxy model, the polarization power spectra taken from within bubbles are peculiar. The distribution of $\mathcal{R}_{EB}$ and $r^{TE}$ taken from within bubbles are generally similar, independently of the galaxy setup, but the scatters remain large.
Furthermore, the influence of the underlying galaxy model seems small.
More specifically, observing from inside a bubble seems to decouple the observed sky from the underlying galaxy. The information about the topology of the magnetic field at large scales is lost (comparing Pwf-bubbles and Twf-bubbles) while a mild effect from the strength of the field may survive (comparing Twf-bubbles and Tsf-bubbles).
However, we find no correlation between $\mathcal{R}_{EB}$ or $r^{TE}$ with the (local) field strength in bubbles, the size of the bubble, or with the environment of the bubbles within the galaxy.
Additionally, for the bubble samples, we do not observe a correlation between the averaged weighted intensity and the $\mathcal{R}_{EB}$ and $r^{TE}$ values. Combining all bubbles and all values of $f_{\rm{sky}}$ in one sample, the Spearman rank order coefficients are $0.04$ and $0.16$, to which p-values of $56\%$ and $1.2\%$ correspond, respectively.

\begin{table}
    \centering
    \begin{tabular}{c  c}
    \hline\hline \\[-.5ex]
    $p_{KS}^{\star}$    &
        \begin{tabular}{r ccc}
    $\mathcal{R}_{EB}$ $\backslash$ $r^{TE}$ & Pwf & Twf & Tsf \\
    \hline\\[-1.5ex]
       Pwf            &    1            &    0.360   &    0.358  \\
       Twf             &    0.095  &    1             &    0.146  \\
       Tsf              &   0.052   &    0.009   &    1           \\
    \end{tabular} \\ \\ \hline \\[-.5ex]
    $ks^{\star}$     &
        \begin{tabular}{r ccc}
$\mathcal{R}_{EB}$ $\backslash$ $r^{TE}$ & Pwf & Twf   & Tsf \\
    \hline\\[-1.5ex]
    Pwf & 0     & 0.143 & 0.147 \\
    Twf & 0.190 & 0     & 0.183 \\
    Tsf & 0.216 & 0.264 & 0     \\
    \end{tabular} \\ \\ \hline \\[-.5ex]
    p-value     &
        \begin{tabular}{r ccc}
$\mathcal{R}_{EB}$ $\backslash$ $r^{TE}$ & Pwf & Twf & Tsf \\
    \hline\\[-1.5ex]
    Pwf & 1     & 0.091 & 0.0068    \\
    Twf & 0.430 & 1     & 0.1460    \\
    Tsf & 0.229 & 0.666 & 1         \\ \\
    \end{tabular}\\ \\ [-1.5ex] \hline
    \end{tabular}
    \caption{KS2S results on bubble sub-samples.
    From top to bottom, three blocks present respectively $p_{KS}^{\star}$, $ks^{\star}$ and the p-value. $p_{KS}^{\star}$ and $ks^{\star}$ are the KS probability and statistic for the comparison of bubble distributions from different galaxy simulations. The p-value is the probability that random sub-samples drawn from the main samples and having the size of the bubble samples look as similar as the bubble sample do. In each block, lower left triangle corresponds to comparison of $\mathcal{R}_{EB}$ distributions, upper right triangle corresponds to comparison of $r^{TE}$ distributions.
    }
    \label{tab:KS2S_REB-rTE_Bubbles}
\end{table}

\subsection{Strength of $\mathbf{B}$ in the cone of observation}
\label{sec:BinCone}
An important result from the previous subsections is that the distributions of $\mathcal{R}_{EB}$ and $r^{TE}$ for Twf and Tsf differ significantly. This suggests that the overall strength of the magnetic field in the galaxy may play a role in shaping the polarization power spectra. This observation holds true for the spectra taken from within bubbles, even though we have shown that the bubble environment has the general effect of regularizing the distributions.

\begin{figure}
    \centering
    \includegraphics[trim={.4cm 0.4cm 0.4cm 0cm},clip,height=6.cm]{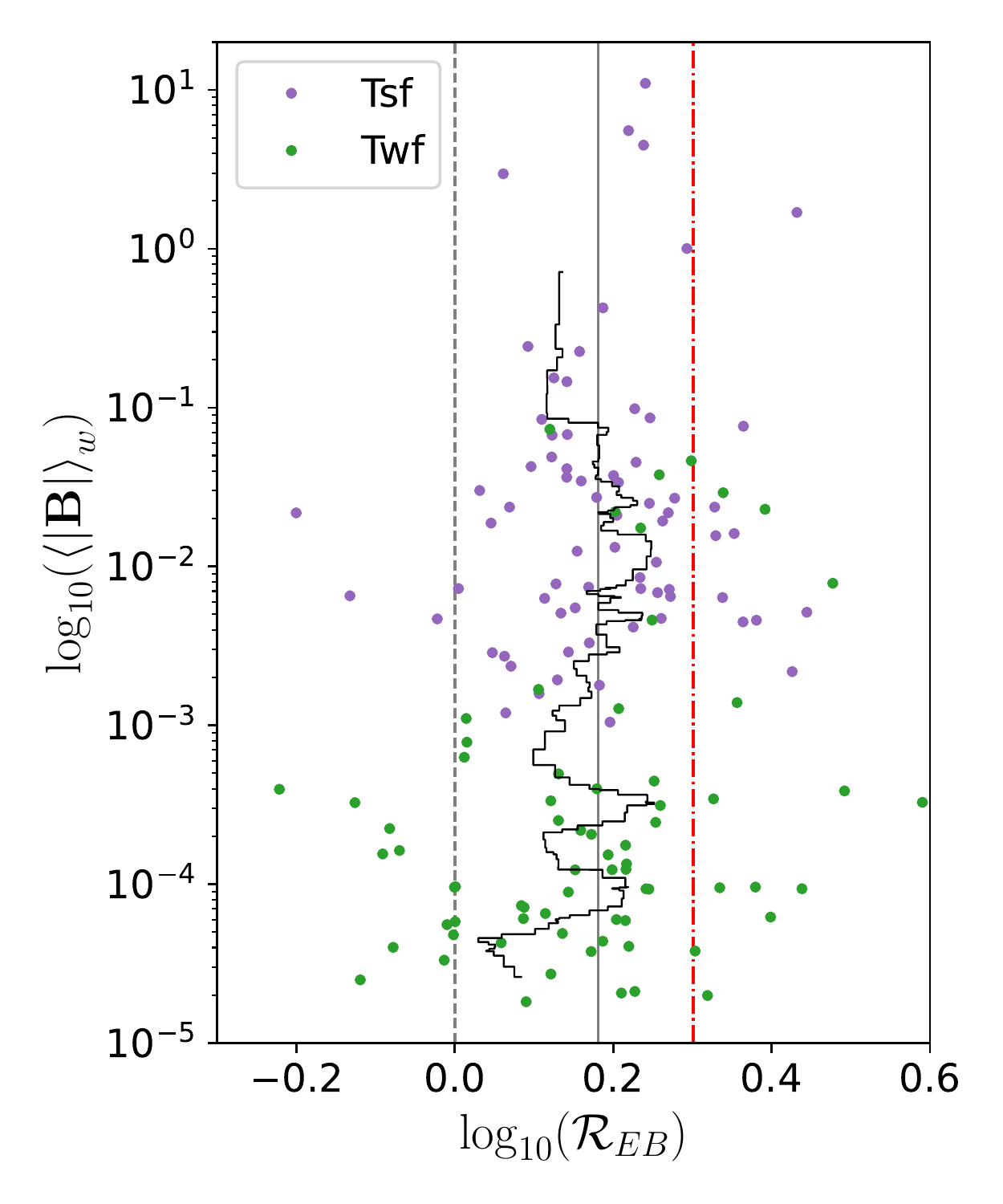}
    \includegraphics[trim={2.6cm 0.4cm 0.4cm 0cm},clip,height=6.cm]{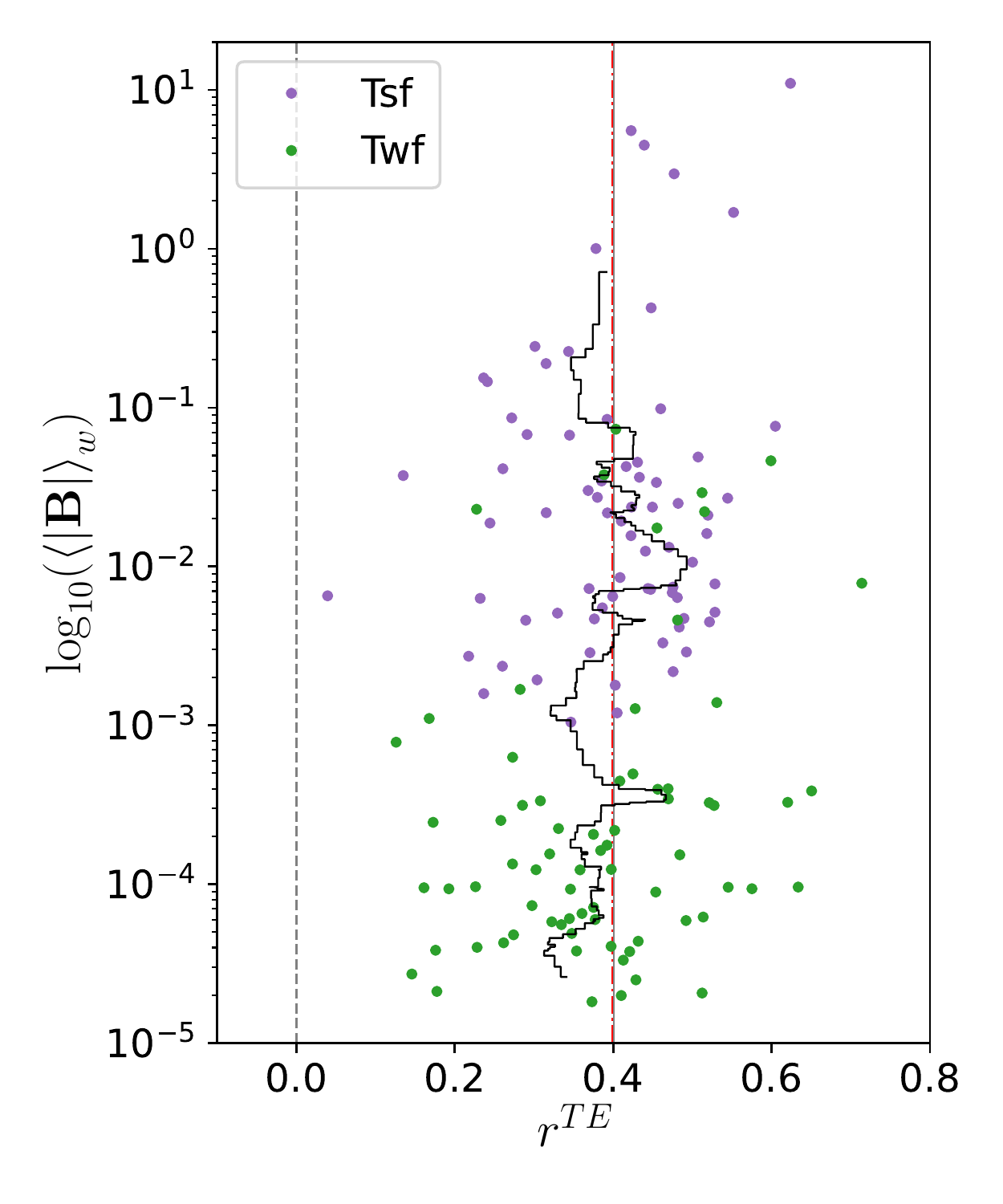}
    \caption{Scatter plots of strength of the magnetic field (in $\mu$G) in the cone of observation versus $\mathcal{R}_{EB}$ (left) and $r^{TE}$ (right), measured from the Twf (green) and Tsf (purple) galaxies for $f_{\rm{sky}} = 0.5$. The strength of the field is the mass-weighed average of the field for all cells having their centers in the cone centered on the observer, with an opening angle of $60^\circ$ around the $z$ axis and with height lower than 1 kpc.
    Vertical and shacked lines are built the same way as in Fig.~\ref{fig:Bf500_REB-rTE_fsky070}.}
    \label{fig:BfCone_REB-rTE_fsky050}
\end{figure}

Here, we quantify the dependence of the $\mathcal{R}_{EB}$ and $r^{TE}$ on the magnetic field strength in the retained regions of the ISM.
For this purpose, for each observer position, we consider a double (two hemispheres) vertical cone with an apex angle of 120$^\circ$, a height of 1 kpc, and with the observer at the tip (the polar caps defined by the intersection of such cone and the celestial sphere cover half of the sky). Then, we calculate the mass-weighed average strength of the field using all the simulation cells located in the cone. This process is repeated for all observer positions in the Twf and Tsf galaxies and for $f_{\rm{sky}} = 0.5$\footnote{We notice that the sky area encompassed by the double cone or defined by the mask may not exactly match.} 
Fig.~\ref{fig:BfCone_REB-rTE_fsky050} shows $\mathcal{R}_{EB}$ and $r^{TE}$ as a function of these magnetic field strength estimates for each observer. Although the field strength measurements span several orders of magnitude, we do not find any correlation with $\mathcal{R}_{EB}$ or $r^{TE}$ (a Spearman correlation test shows that the probability that these scatter plots are random are 5\% and 3\% respectively).
Therefore, we conclude that the magnetic field strength in the retained ISM regions is not directly responsible for the differences of distributions seen in Figs.~\ref{fig:REB-rTE_hist} and~\ref{fig:REB-rTE_hist-Bubbles}.

\section{Summary and conclusion}
\label{sec:ccl}

In this work, we synthesized full-sky observations of polarized dust emission at 353~GHz from a set of global, Milky-Way-sized galaxy simulations. We studied the resulting $T$, $E$, and $B$ power spectra in terms of the $E/B$ power asymmetry ($\mathcal{R}_{EB}$) and the correlation between $T$ and $E$ modes ($r^{TE}$), aiming to identify the physical parameters that affect these correlations. Specifically, we investigated the dependence of
$\mathcal{R}_{EB}$ and $r^{TE}$ on (i) the physical conditions in the galaxy simulations (presence of feedback, topology, and strength of the large-scale magnetic field) and (ii) the particular location of the observer in the galaxy.
Our intention is not to reproduce or model actual observations of the Galactic magnetized ISM but rather to identify the dominant factors shaping the polarization power spectra.

We found that the distributions of $\mathcal{R}_{EB}$ and $r^{TE}$ from different galaxy models differ significantly: The presence of feedback, the topology, and the strength of the magnetic field at large scales all impact the shape of the distributions. 
However, the cosmic variance dominates: the $\mathcal{R}_{EB}$ and $r^{TE}$ parameters measured from polarization power spectra depend sensitively on the observer's location.

In particular, we showed that measuring the  $\mathcal{R}_{EB}$ and $r^{TE}$ distributions from within a spiral arm as opposed to an inter-arm region leads to a measurable difference.
However, we did not find any correlation between the local matter density or the strength of the magnetic field and the measured $\mathcal{R}_{EB}$ and $r^{TE}$ values.
We further demonstrated that taking power spectra from the inside of supernovae-driven superbubbles minimizes the dependence on the underlying galaxy model. However, it does not eliminate it: the distributions of $\mathcal{R}_{EB}$ and $r^{TE}$ from within superbubbles still differ between models.

We found evidence that the strength of the global magnetic field in the galaxy model (which is below equipartition values in all of the models) may play a role in shaping the distribution of power spectrum characteristics; it decreases the viewpoint-induced variances and increases both the $E/B$ asymmetry and the $TE$ correlation.
However, we found no correlation between the power spectra characteristics and the strength of the magnetic field in the regions of the ISM that enter the maps we analyzed.

In summary, we have shown that the statistics of the sky in polarization depend predominantly on the observer's point of view. However, no correlation with the local mean values of physical quantities such as density or magnetic field strength was found. Therefore,
any statistical properties of the polarized sky from intermediate to high-Galactic latitude regions may reflect the complexity of the line-of-sight structure of the magnetized ISM rather than mean local or global properties of the ISM.
This result reinforces previous works that demonstrate the importance of local structures in the study and characterization of the observed sky in polarization (e.g., see \citealt{Alv2018,Bra2019a,Ska2019,Pel2020}).
This evidence suggests that modeling the dynamical history of the Solar neighborhood and its 3D structure may be necessary to obtain a better description of the polarization sky at intermediate and high-Galactic latitudes.
However, restricting our study to the local Solar neighborhood may not be sufficient: the statistical properties of the polarization sky also depend on more distant regions of the ISM probed by the lines of sight. This motivates modeling efforts to connect models of the local ISM to large-scale models of the magnetized Galaxy.

\begin{acknowledgements}
The authors thank Andrea Bracco, Fabio Del Sordo, and Vincent Guillet for helpful comments and discussions and their anonymous referee for providing a constructive and thorough report.
This project has received funding from the European Research Council (ERC) under the European Unions Horizon 2020 research and innovation programme under grant agreements No. 771282 and 740120. E.N. acknowledges funding from a Marie Curie Action of the European Union (grant agreement No. 749073).
\end{acknowledgements}

\bibliographystyle{aa}
\bibliography{myBiblio}

\begin{thebibliography}{48}
\expandafter\ifx\csname natexlab\endcsname\relax\def\natexlab#1{#1}\fi

\bibitem[{{Alves} {et~al.}(2018){Alves}, {Boulanger}, {Ferri{\`e}re}, \&
  {Montier}}]{Alv2018}
{Alves}, M.~I.~R., {Boulanger}, F., {Ferri{\`e}re}, K., \& {Montier}, L. 2018,
  \aap, 611, L5

\bibitem[{{Balsara} \& {Kim}(2004)}]{Balsara_2004}
{Balsara}, D.~S. \& {Kim}, J. 2004, \apj, 602, 1079

\bibitem[{{Bracco} {et~al.}(2019{\natexlab{a}}){Bracco}, {Candelaresi}, {Del
  Sordo}, \& {Brandenburg}}]{Bra2019a}
{Bracco}, A., {Candelaresi}, S., {Del Sordo}, F., \& {Brandenburg}, A.
  2019{\natexlab{a}}, \aap, 621, A97

\bibitem[{{Bracco} {et~al.}(2019{\natexlab{b}}){Bracco}, {Ghosh}, {Boulanger},
  \& {Aumont}}]{Bra2019}
{Bracco}, A., {Ghosh}, T., {Boulanger}, F., \& {Aumont}, J. 2019{\natexlab{b}},
  \aap, 632, A17

\bibitem[{{Brandenburg} {et~al.}(2019){Brandenburg}, {Bracco}, {Kahniashvili},
  {Mandal}, {Roper Pol}, {Petrie}, \& {Singh}}]{Bran2019}
{Brandenburg}, A., {Bracco}, A., {Kahniashvili}, T., {et~al.} 2019, \apj, 870,
  87

\bibitem[{{Caldwell} {et~al.}(2017){Caldwell}, {Hirata}, \&
  {Kamionkowski}}]{Cal2017}
{Caldwell}, R.~R., {Hirata}, C., \& {Kamionkowski}, M. 2017, \apj, 839, 91

\bibitem[{{Clark} {et~al.}(2021){Clark}, {Kim}, {Hill}, \& {Hensley}}]{Cla2021}
{Clark}, S.~E., {Kim}, C.-G., {Hill}, J.~C., \& {Hensley}, B.~S. 2021, \apj,
  919, 53

\bibitem[{{Clark} {et~al.}(2014){Clark}, {Peek}, \& {Putman}}]{Cla2014}
{Clark}, S.~E., {Peek}, J.~E.~G., \& {Putman}, M.~E. 2014, \apj, 789, 82

\bibitem[{{Ferri{\`e}re} \& {Terral}(2014)}]{Ferriere_Terral_2014}
{Ferri{\`e}re}, K. \& {Terral}, P. 2014, \aap, 561, A100

\bibitem[{{Finkbeiner} {et~al.}(1999){Finkbeiner}, {Davis}, \&
  {Schlegel}}]{Fin1999}
{Finkbeiner}, D.~P., {Davis}, M., \& {Schlegel}, D.~J. 1999, \apj, 524, 867

\bibitem[{{Fromang} {et~al.}(2006){Fromang}, {Hennebelle}, \&
  {Teyssier}}]{Fromang_2006}
{Fromang}, S., {Hennebelle}, P., \& {Teyssier}, R. 2006, \aap, 457, 371

\bibitem[{{G{\'o}rski} {et~al.}(2005){G{\'o}rski}, {Hivon}, {Banday}, {Wand
  elt}, {Hansen}, {Reinecke}, \& {Bartelmann}}]{Gor2005}
{G{\'o}rski}, K.~M., {Hivon}, E., {Banday}, A.~J., {et~al.} 2005, \apj, 622,
  759

\bibitem[{{Herv{\'\i}as-Caimapo} \& {Huffenberger}(2021)}]{Her2021}
{Herv{\'\i}as-Caimapo}, C. \& {Huffenberger}, K. 2021, arXiv e-prints,
  arXiv:2107.08317

\bibitem[{{Hu} \& {White}(1997)}]{Hu1997}
{Hu}, W. \& {White}, M. 1997, \prd, 56, 596

\bibitem[{{Kamionkowski} {et~al.}(1997{\natexlab{a}}){Kamionkowski},
  {Kosowsky}, \& {Stebbins}}]{Kam1997b}
{Kamionkowski}, M., {Kosowsky}, A., \& {Stebbins}, A. 1997{\natexlab{a}}, \prl,
  78, 2058

\bibitem[{{Kamionkowski} {et~al.}(1997{\natexlab{b}}){Kamionkowski},
  {Kosowsky}, \& {Stebbins}}]{Kam1997a}
{Kamionkowski}, M., {Kosowsky}, A., \& {Stebbins}, A. 1997{\natexlab{b}}, \prd,
  55, 7368

\bibitem[{{Kandel} {et~al.}(2017){Kandel}, {Lazarian}, \& {Pogosyan}}]{Kan2017}
{Kandel}, D., {Lazarian}, A., \& {Pogosyan}, D. 2017, \mnras, 472, L10

\bibitem[{{Kim} {et~al.}(2019){Kim}, {Choi}, \& {Flauger}}]{Kim2019}
{Kim}, C.-G., {Choi}, S.~K., \& {Flauger}, R. 2019, \apj, 880, 106

\bibitem[{{Kim} \& {Ostriker}(2017)}]{Kim2017}
{Kim}, C.-G. \& {Ostriker}, E.~C. 2017, \apj, 846, 133

\bibitem[{{Konstantinou} {et~al.}(2021){Konstantinou}, {Pelgrims}, {Fuchs}, \&
  {Tassis}}]{Kon2021}
{Konstantinou}, A., {Pelgrims}, V., {Fuchs}, F., \& {Tassis}, K. 2021, \aap,
  submitted

\bibitem[{{Kritsuk} {et~al.}(2018){Kritsuk}, {Flauger}, \&
  {Ustyugov}}]{Kri2018}
{Kritsuk}, A.~G., {Flauger}, R., \& {Ustyugov}, S.~D. 2018, \prl, 121, 021104

\bibitem[{{Kritsuk} {et~al.}(2017){Kritsuk}, {Ustyugov}, \& {Norman}}]{Kri2017}
{Kritsuk}, A.~G., {Ustyugov}, S.~D., \& {Norman}, M.~L. 2017, New Journal of
  Physics, 19, 065003

\bibitem[{{Lallement} {et~al.}(2018){Lallement}, {Capitanio}, {Ruiz-Dern},
  {Danielski}, {Babusiaux}, {Vergely}, {Elyajouri}, {Arenou}, \&
  {Leclerc}}]{Lal2018}
{Lallement}, R., {Capitanio}, L., {Ruiz-Dern}, L., {et~al.} 2018, \aap, 616,
  A132

\bibitem[{{Lee} \& {Draine}(1985)}]{Lee85}
{Lee}, H.~M. \& {Draine}, B.~T. 1985, \apj, 290, 211

\bibitem[{{Navarro} {et~al.}(1996){Navarro}, {Frenk}, \& {White}}]{NFW96}
{Navarro}, J.~F., {Frenk}, C.~S., \& {White}, S.~D.~M. 1996, \apj, 462, 563

\bibitem[{{Ntormousi}(2018)}]{EN2018}
{Ntormousi}, E. 2018, \aap, 619, L5

\bibitem[{{Pelgrims} {et~al.}(2021){Pelgrims}, {Clark}, {Hensley},
  {Panopoulou}, {Pavlidou}, {Tassis}, {Eriksen}, \& {Wehus}}]{Pel2021a}
{Pelgrims}, V., {Clark}, S.~E., {Hensley}, B.~S., {et~al.} 2021, \aap, 647, A16

\bibitem[{{Pelgrims} {et~al.}(2020){Pelgrims}, {Ferri{\`e}re}, {Boulanger},
  {Lallement}, \& {Montier}}]{Pel2020}
{Pelgrims}, V., {Ferri{\`e}re}, K., {Boulanger}, F., {Lallement}, R., \&
  {Montier}, L. 2020, \aap, 636, A17

\bibitem[{{Perret}(2016)}]{Perret_2016}
{Perret}, V. 2016, {DICE: Disk Initial Conditions Environment}, Astrophysics
  Source Code Library

\bibitem[{{Perret} {et~al.}(2014){Perret}, {Renaud}, {Epinat}, {Amram},
  {Bournaud}, {Contini}, {Teyssier}, \& {Lambert}}]{Perret_2014}
{Perret}, V., {Renaud}, F., {Epinat}, B., {et~al.} 2014, \aap, 562, A1

\bibitem[{{Planck Collaboration I}(2020)}]{PlaI2020}
{Planck Collaboration I}. 2020, \aap, 641, A1

\bibitem[{{Planck Collaboration Int. XLIV}(2016)}]{PlaXLIV2016}
{Planck Collaboration Int. XLIV}. 2016, \aap, 596, A105

\bibitem[{{Planck Collaboration Int. XX}(2015)}]{PlaXX2015}
{Planck Collaboration Int. XX}. 2015, \aap, 576, A105

\bibitem[{{Planck Collaboration XI}(2014)}]{PlaXI2014}
{Planck Collaboration XI}. 2014, \aap, 571, A11

\bibitem[{{Planck Collaboration XI}(2020)}]{PlaXI2020}
{Planck Collaboration XI}. 2020, \aap, 641, A11

\bibitem[{{Planck Collaboration XXX}(2016)}]{PlaXXX2016}
{Planck Collaboration XXX}. 2016, \aap, 586, A133

\bibitem[{{Planck Collaboration XXXVIII}(2016)}]{PlaXXXVIII2016}
{Planck Collaboration XXXVIII}. 2016, \aap, 586, A141

\bibitem[{{Reid} {et~al.}(2019){Reid}, {Menten}, {Brunthaler}, {Zheng}, {Dame},
  {Xu}, {Li}, {Sakai}, {Wu}, {Immer}, {Zhang}, {Sanna}, {Moscadelli}, {Rygl},
  {Bartkiewicz}, {Hu}, {Quiroga-Nu{\~n}ez}, \& {van Langevelde}}]{Reid2019}
{Reid}, M.~J., {Menten}, K.~M., {Brunthaler}, A., {et~al.} 2019, \apj, 885, 131

\bibitem[{{Reissl} {et~al.}(2019){Reissl}, {Brauer}, {Klessen}, \&
  {Pellegrini}}]{Rei2019}
{Reissl}, S., {Brauer}, R., {Klessen}, R.~S., \& {Pellegrini}, E.~W. 2019,
  \apj, 885, 15

\bibitem[{{Reissl} {et~al.}(2020){Reissl}, {Guillet}, {Brauer}, {Levrier},
  {Boulanger}, \& {Klessen}}]{Rei2020}
{Reissl}, S., {Guillet}, V., {Brauer}, R., {et~al.} 2020, \aap, 640, A118

\bibitem[{{Seifried} {et~al.}(2019){Seifried}, {Walch}, {Reissl}, \&
  {Ib{\'a}{\~n}ez-Mej{\'\i}a}}]{Sei2019}
{Seifried}, D., {Walch}, S., {Reissl}, S., \& {Ib{\'a}{\~n}ez-Mej{\'\i}a},
  J.~C. 2019, \mnras, 482, 2697

\bibitem[{{Skalidis} \& {Pelgrims}(2019)}]{Ska2019}
{Skalidis}, R. \& {Pelgrims}, V. 2019, \aap, 631, L11

\bibitem[{{Teyssier}(2002)}]{Teyssier_02}
{Teyssier}, R. 2002, \aap, 385, 337

\bibitem[{{Tristram} {et~al.}(2005){Tristram}, {Mac{\'\i}as-P{\'e}rez},
  {Renault}, \& {Santos}}]{Tri2005}
{Tristram}, M., {Mac{\'\i}as-P{\'e}rez}, J.~F., {Renault}, C., \& {Santos}, D.
  2005, \mnras, 358, 833

\bibitem[{{Vandenbroucke} {et~al.}(2021){Vandenbroucke}, {Baes}, {Camps},
  {Kapoor}, {Barrientos}, \& {Bernard}}]{Van2021}
{Vandenbroucke}, B., {Baes}, M., {Camps}, P., {et~al.} 2021, \aap, 653, A34

\bibitem[{{Wang} {et~al.}(2020){Wang}, {Jaffe}, {En{\ss}lin}, {Ullio}, {Ghosh},
  \& {Santos}}]{Wan2020}
{Wang}, J., {Jaffe}, T.~R., {En{\ss}lin}, T.~A., {et~al.} 2020, \apjs, 247, 18

\bibitem[{{Zaldarriaga}(2001)}]{Zal2001}
{Zaldarriaga}, M. 2001, \prd, 64, 103001

\bibitem[{{Zaldarriaga} \& {Seljak}(1997)}]{Zal1997}
{Zaldarriaga}, M. \& {Seljak}, U. 1997, \prd, 55, 1830

\end{thebibliography}
\label{lastpage}

\clearpage
\begin{appendix}
\section{Observer excursion: Angular convention}
\label{app:conventions}
\begin{figure}
    \centering
    \includegraphics[trim={2.8cm 1.6cm 2.8cm 1.cm},clip,width=.98\linewidth]{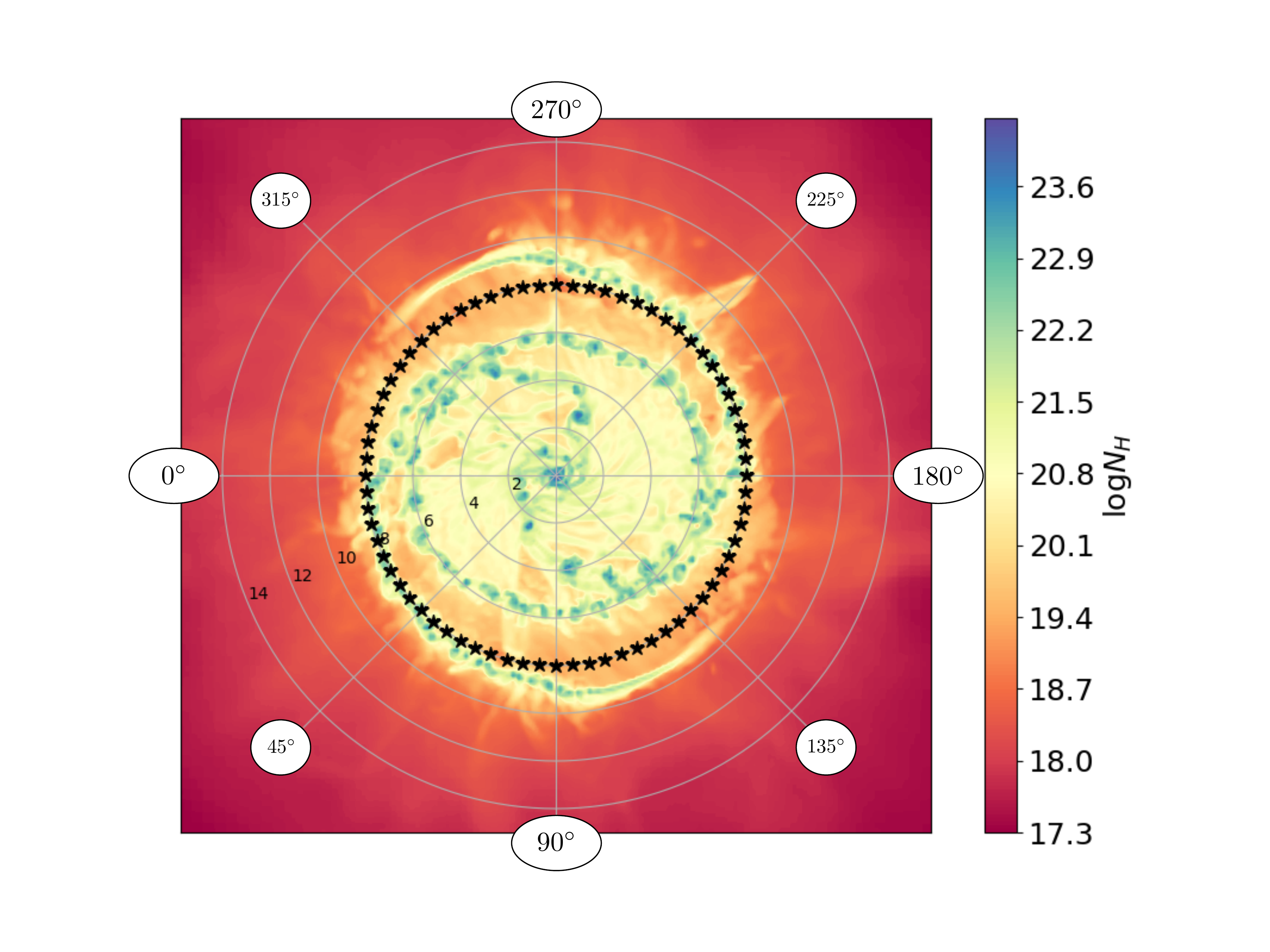}
    \caption{Illustration of the observer's circular excursion around the galaxy center on a top-down view of the Tsf galaxy model considered in the paper. A black star is placed every $5^\circ$ at 8 kpc from the galaxy center. The background color trace the gas density.}
    \label{fig:circularexcursion}
\end{figure}

\begin{figure}
\begin{center}
\includegraphics[trim={1cm 0.cm 1cm -0.65cm},clip,width=.98\linewidth]{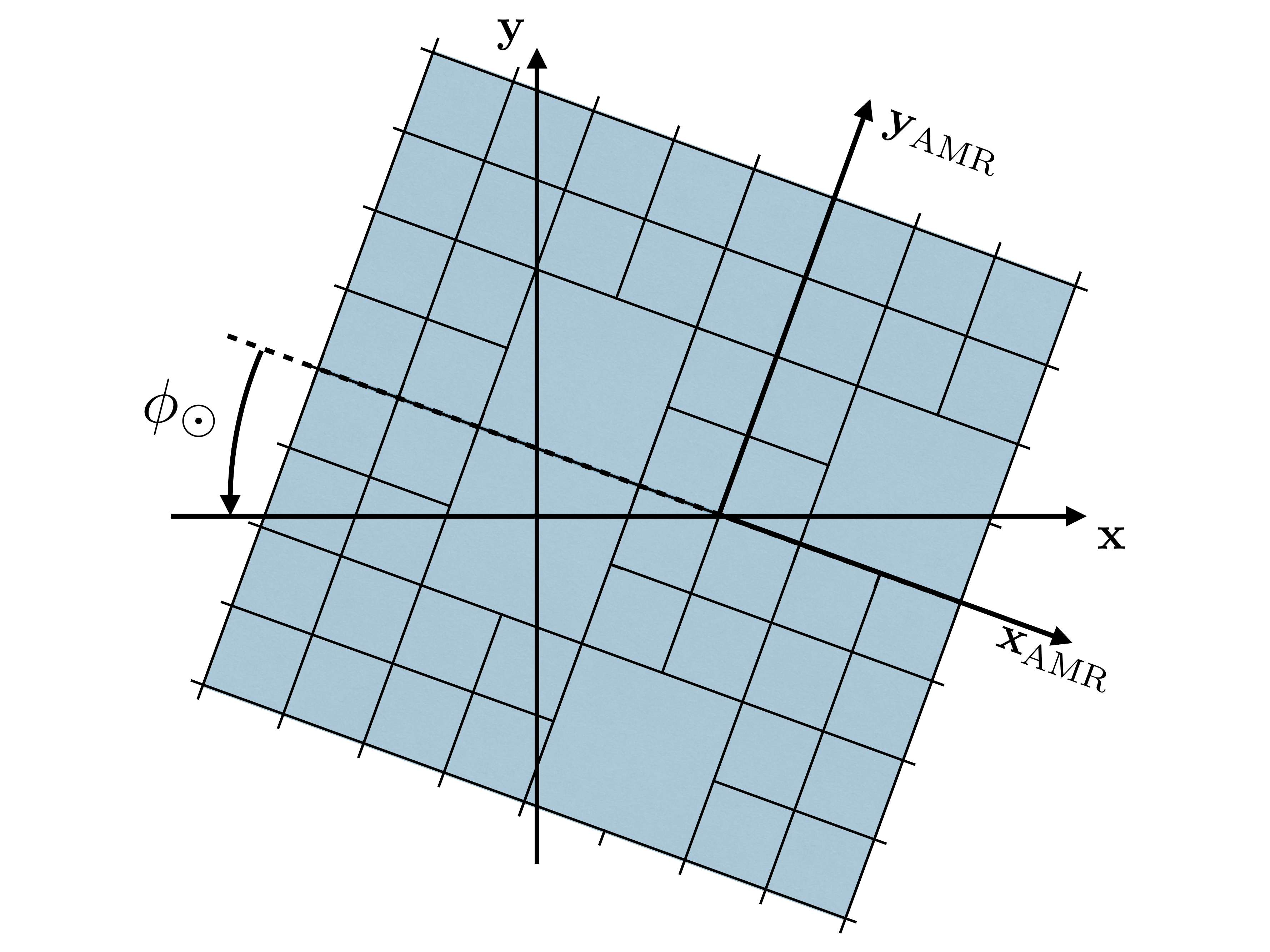}
\caption{Convention for observer angular position and AMR grid rotation. The third dimension ($z$ axis) is dropped as it is left unchanged during our coordinate transformation. The observer is at $(x,\,y) = (0,\,0)$ and the galaxy center at $(x_{\rm{AMR}},\,y_{\rm{AMR}}) = (0,\,0)$. The AMR grid is rotated clockwise by $\phi_\odot$ about the galaxy center and then shifted at $(x,\,y) = (8,\,0)$ to make the circular excursion of the observer.}
\label{appfig:conventions}
\end{center}
\end{figure}

In this work we consider an hypothetical observer on the $xy$ plane of the Cartesian grid of the simulation ($z=0$) at a distance of $R_\odot = 8$ kpc from the galaxy center. To quantify the effect of the observer's location on our inferences of the dust polarization maps, we move the observer in a circle on the $xy$ plane; as illustrated in Fig.~\ref{fig:circularexcursion} on a top-down view of the Tsf galaxy model.
We choose the galactic center to be placed at longitude zero ($l=0^\circ$) on the synthetic maps for any observer location. To achieve this, we rotate the AMR grid according to $\mathbf{x}' = \mathcal{R}_{\phi_\odot} \, \mathbf{x}_{\rm{AMR}}$ where $ \mathcal{R}_{\phi_\odot}$ is the clockwise rotation matrix by an angle $\phi_\odot$ about the $z$ axis. Then we translate the full grid to obtain the coordinates of each cell in the observer Cartesian coordinate system ($\mathbf{x} = \mathbf{x}' + R_\odot \, (1,\,0,\,0)$) where the conversion to spherical coordinates is straightforward. The spherical coordinate system centered on the observer is the natural reference frame in which we want to synthesize the observables.
In the observer reference frame the galactic center is always at coordinates $(x,\,y,\,z) = (8.0,\,0.0,\,0.0)$ kpc. The used conventions and coordinate transformation are sketched in Fig.~\ref{appfig:conventions}.
We apply the same vector transformation to the magnetic vector field ($\mathbf{B}(\mathbf{x}') = \mathcal{R}_{\phi_\odot} \, \mathbf{B}_{\rm{AMR}}(\mathbf{x}_{\rm{AMR}})$) before we convert it to spherical coordinates, thus, centered on the observer:
\begin{equation}
B_r = \mathbf{B} \cdot \mathbf{e}_r \; ; \; B_\theta = \mathbf{B} \cdot \mathbf{e}_\theta \; ; \; B_\phi = \mathbf{B} \cdot \mathbf{e}_\phi
\end{equation}
where $(\mathbf{e}_r,\,\mathbf{e}_\theta ,\,\mathbf{e}_\phi)$ is the orthonormal vector basis of an observer-centered spherical coordinate system with $\mathbf{e}_r$ pointing away from the observer and $\mathbf{e}_\theta$ pointing southward. We discuss the synthetic maps in terms of galactic latitudes and galactic longitudes which are related to the spherical angular coordinates as $(b,\,l) = (90 - \theta,\, \phi)$ when $\theta$ and $\phi$ are expressed in degrees.

\section{Ray-tracing algorithm}
\label{app:raytracing}
We rely on vector formalism to compute the path length through a cubic cell of a line of sight.
This choice is motivated first by the fact that we consider rotated AMR grids by arbitrary angles and second by the very large number of cells in our simulations, from about $16 \times 10^6$ to $40 \times 10^6$. Both call for an automated and homogeneous procedure that can be implemented in an efficient algorithm.

\begin{figure}
    \centering
    \includegraphics[trim={5cm 4.cm 5cm 3.5cm},clip,width=.95\linewidth]{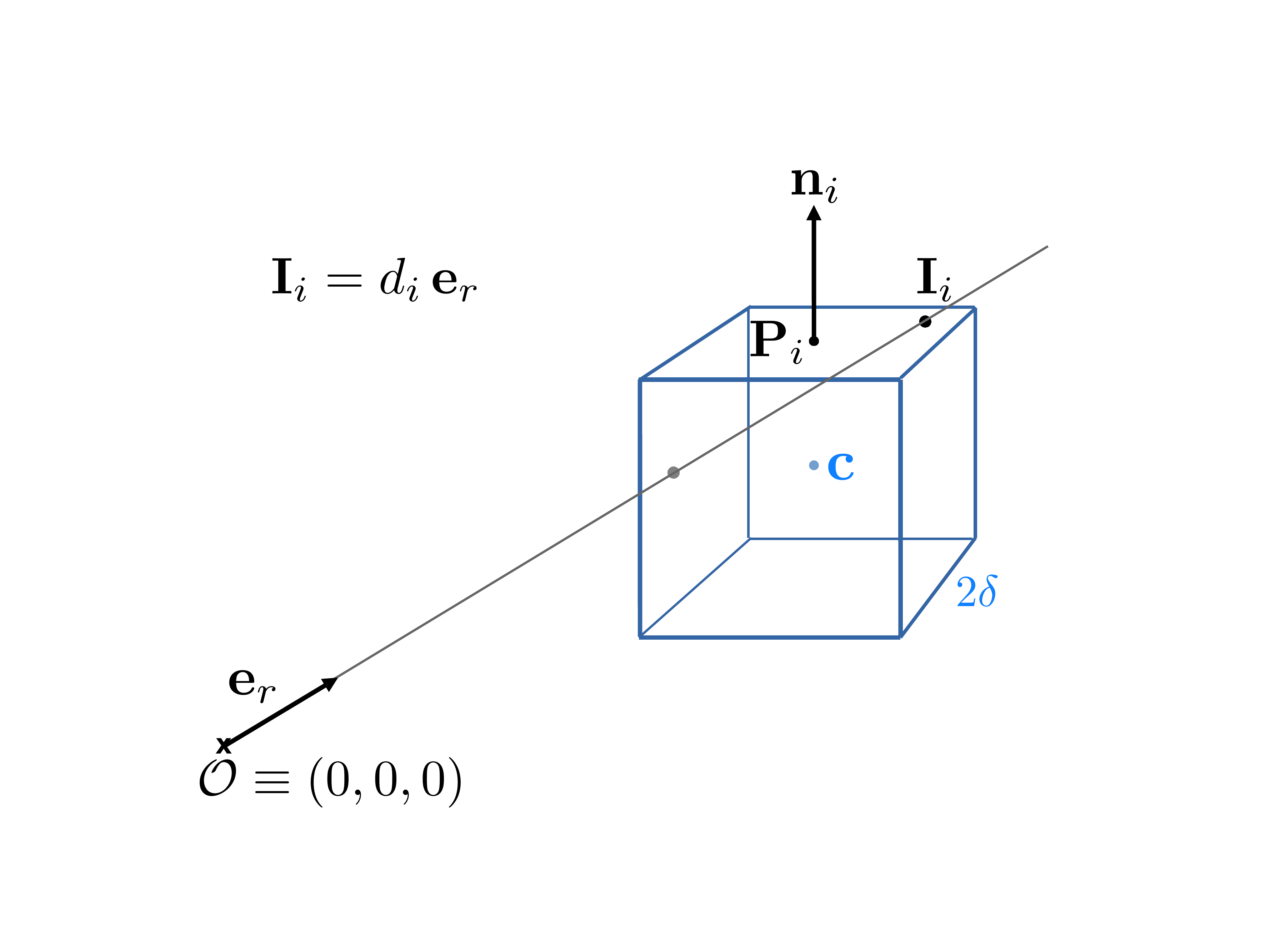}
    \caption{Computing the distance $d_i$ from the observer to the intersecting point ($\mathbf{I}_i$) of a sightline with the bounding plane $i$ of a cubic cell.}
    \label{fig:skect_RTd}
\end{figure}
We consider the equations of the six planes bounding the cell and the equation of the line of sight with its unit direction vector, $\mathbf{e}_r$, oriented outward from the observer. As sketched in Fig.~\ref{fig:skect_RTd}, the distance from the observer at the origin ($\mathcal{O}$) to the intersection (in $\mathbf{I}_i$) between the line of sight with a plane can be written as
\begin{equation}
d_i = \frac{\mathbf{P}_i \cdot \mathbf{n}_i}{\mathbf{e}_r \cdot \mathbf{n}_i}
\label{eq:d_i}
\end{equation}
where $i$ denotes one of the six planes and takes values from one to six, $\mathbf{n}_i$ is the unit vector normal to the plane and points outward from the cell center and $\mathbf{P}_i$ is one arbitrary point on the plane. We specify the latter from the cell center ($\mathbf{c}$) and by a translation along the corresponding median of the cell: $\mathbf{P}_i = \mathbf{c} + \delta\,\mathbf{n}_i$, where $\delta$ is half the cell size.

To compute the path length through a cell, we encounter two cases. The observer is either fully outside the cell, including the faces, (distant cells) or it is inside (or on  face) of the cell (observer cells).
In both cases we compute the six distances $d_i$. We sort them by increasing order, defining the sorted set of distance $d^{[j]}$ ($ j \in [1, 6] $). For a distant cell of the simulation the path length through the cell is always given by $\lambda = d^{[4]} - d^{[3]}$.
For the case of an observer cell, the intersection point at distance $d^{[3]}$ is antipodal to the observer (toward $-\mathbf{e}_r$) or is zero if the observer stands on a face. In those cases, the path length of the line of sight through the cell is simply given by $d^{[3]}$.

It has to be noted that due to the AMR nature of the Cartesian grid that is ray-traced, the number of observer cells can be anything from one to eight depending on whether the observer is inside a cell or is placed on a face, an edge or even at a corner of the cells, and that adjacent cells may correspond to different resolution levels, so have different sizes.
In such cases particular care is needed not to count more than once the path along cell faces since given lines of sight can be attributed to several cells. If not accounted for; this may produce artifacts at constant longitude(s) and or latitude(s) depending on the precise observer location within the grid and the $\phi_{\odot}$ value which specify the grid orientation with respect to the celestial coordinate system of the observer.

\smallskip

In practice, the algorithm to ray-trace the simulation is defined as follows\footnote{At the time of submitting this work, \cite{Her2021} independently proposed a ray-tracing algorithm that is very close to ours.}. First, we fix the resolution of the map. Ideally it has to be sufficiently good to guarantee that all simulation cells are crossed at least once. The resolution fixes a set of pixels, i.e. of lines of sight. From this choice the algorithm then goes as follows:
\begin{itemize}[topsep=0pt]
    \item[] \textbf{For each simulation cell:}
\begin{enumerate}
    \item from cell position and size; select sightlines that may intersect the cell through the selection criteria: $\arccos \left ( \mathbf{e}_r \cdot \mathbf{c}/|\mathbf{c}| \right ) \leq 2 \delta \sqrt{3} / |\mathbf{c}|$. \\[-1.5ex]
    \item \textbf{for each selected lines of sight:}
    \begin{enumerate}
        \item compute the six distances $d_i$,
        \item sort the $d_i$'s and generate the list of $d^{[i]}$,
        \item compute the path length through the cell $\lambda = d^{[4]} - d^{[3]}$,
        \item make sure the intersection are on cell faces,
        \item use $\lambda$ to update observable maps (e.g., Eqs.~\ref{eq:I_model}--\ref{eq:U_model}) or store $\lambda$ and pixel index attached to the cell.
    \end{enumerate}
\end{enumerate}
\end{itemize}

\begin{figure}
\begin{center}
\begin{tabular}{cc}
	$\phi_\odot = 180^\circ$	&	$\phi_\odot = 60^\circ$ \\
\includegraphics[trim={0.4cm .5cm 0.4cm .5cm},clip,width=.45\columnwidth]{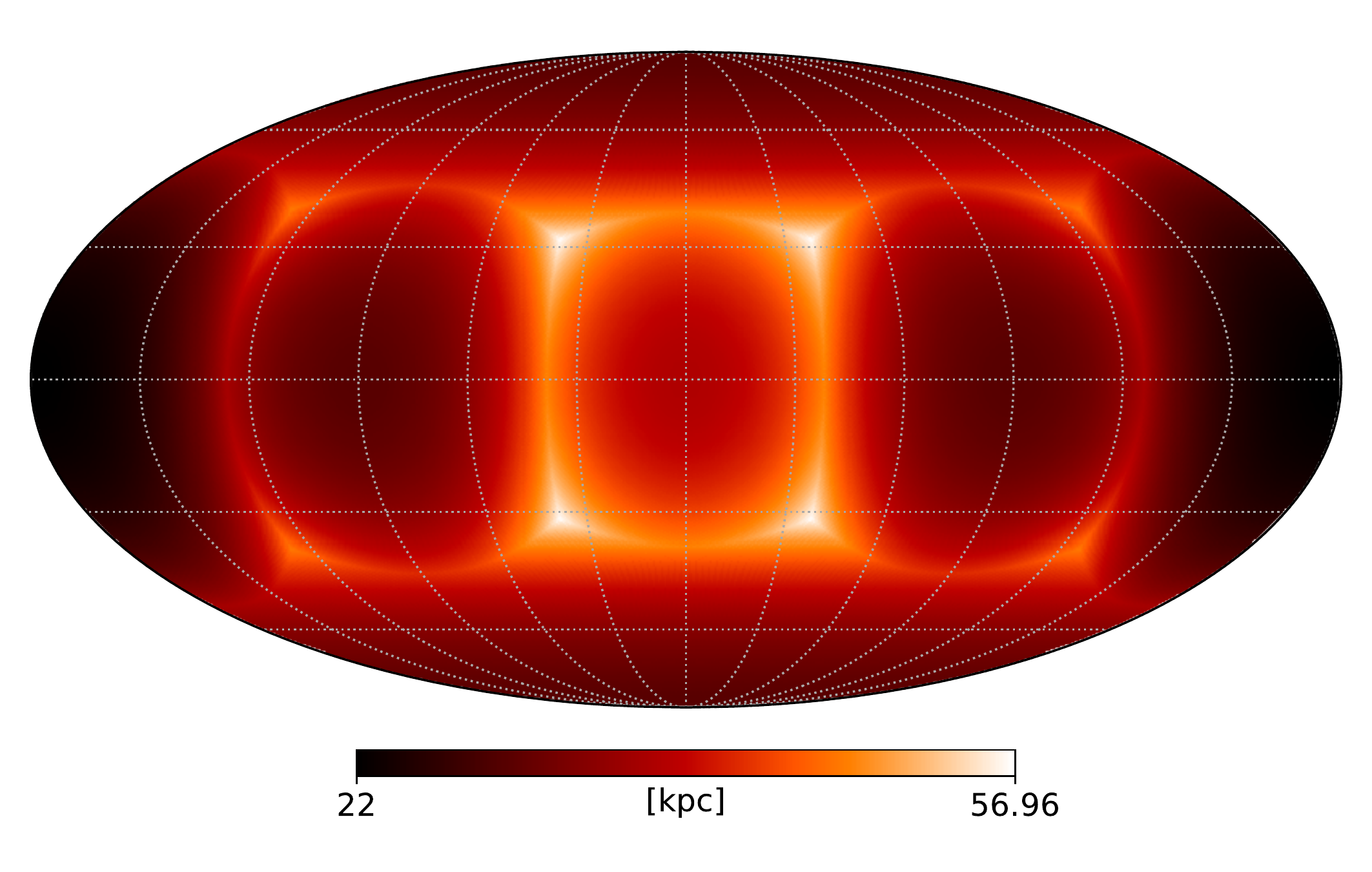}
	&	\includegraphics[trim={0.4cm .5cm 0.4cm .5cm},clip,width=.45\columnwidth]{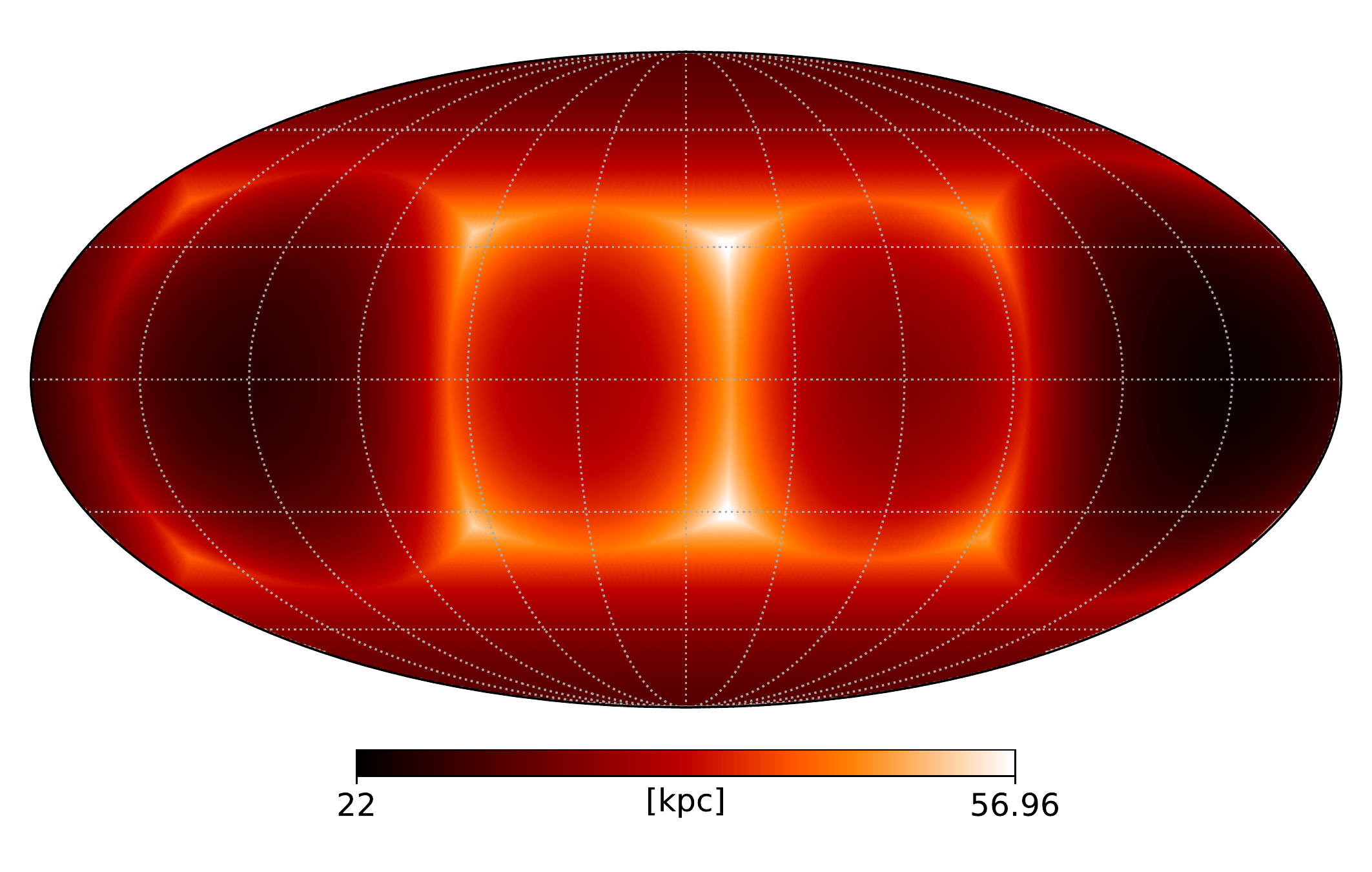}
\\
\includegraphics[trim={0.4cm .5cm 0.4cm .5cm},clip,width=.45\columnwidth]{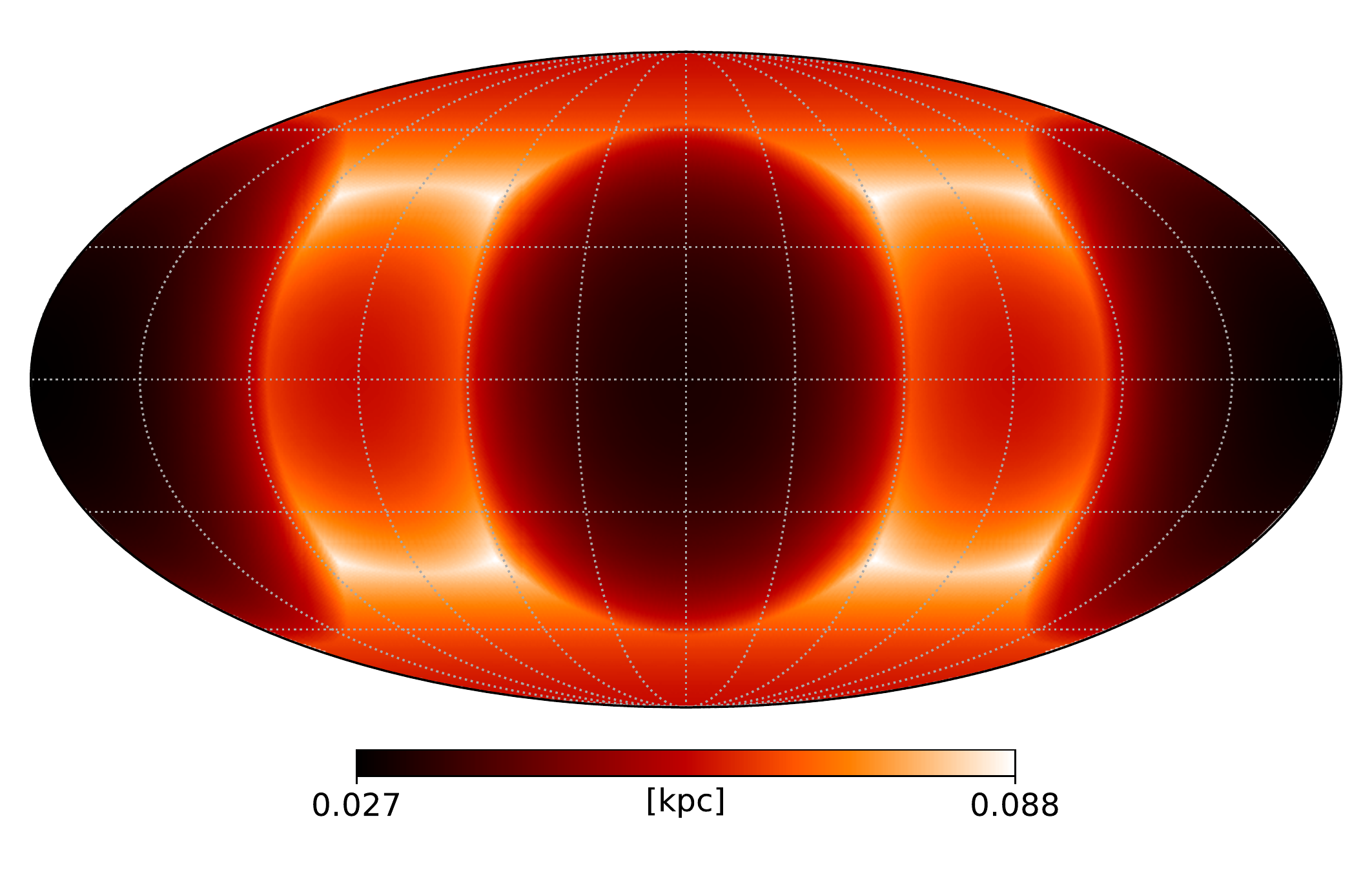}
	&	\includegraphics[trim={0.4cm .5cm 0.4cm .5cm},clip,width=.45\columnwidth]{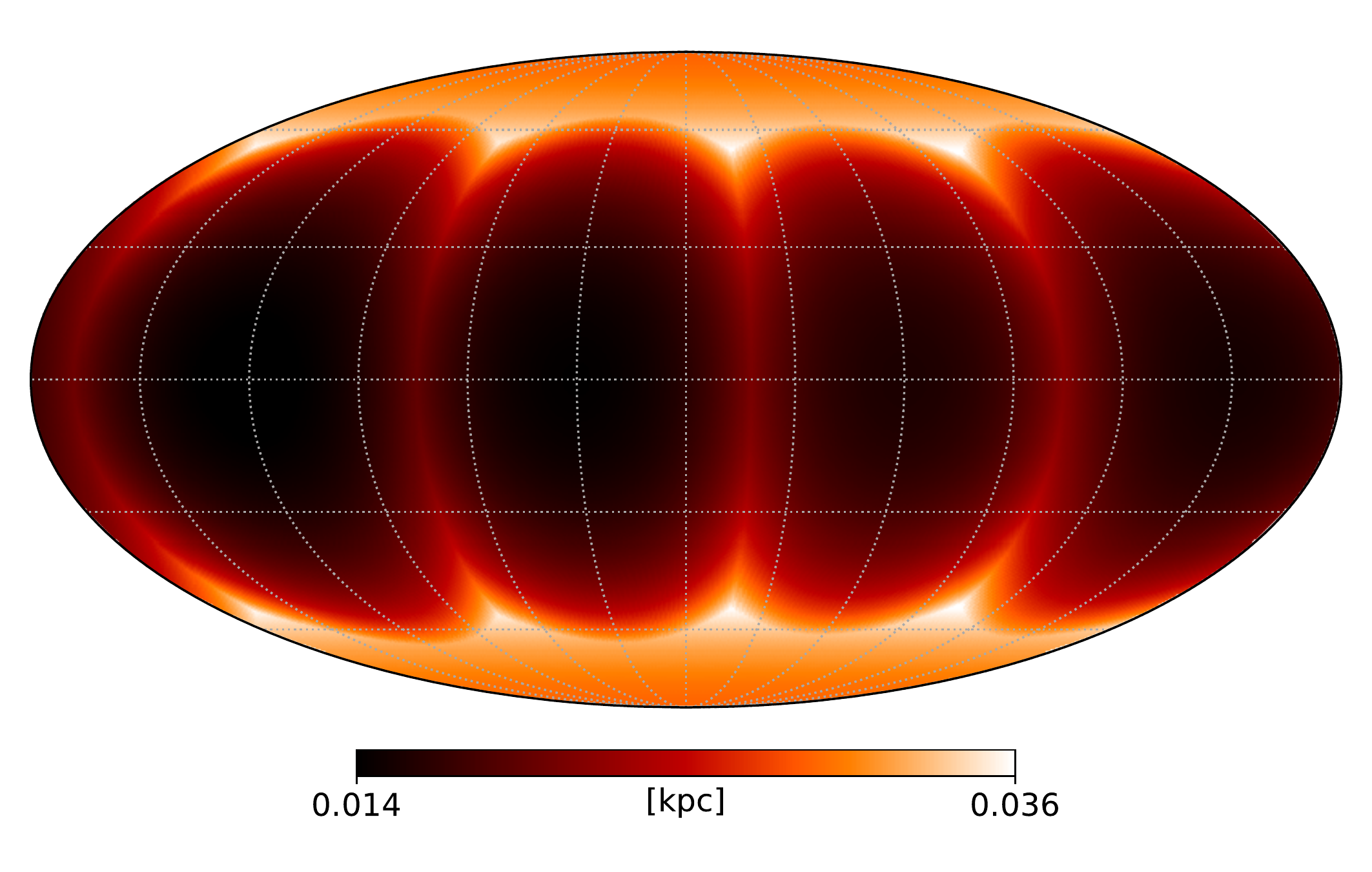}
\\
\end{tabular}
\caption{Integrated path length through the full simulation box (top panels) and through only the observer cells (lower panels).
The full box is of $60$ kpc on a side and is integrated from an observer placed at $R_\odot = 8$ kpc from the center. The observer is immersed in the Tsf galaxy at 8 kpc from the galactic center and his angular coordinate is is $\phi_\odot = 180^\circ$ for the left panels and  $\phi_\odot = 60^\circ$ for the right panels.
}
\label{appfig:lenght_through_box}
\end{center}
\end{figure}

To check the validity of our ray-tracing algorithm, and thus of our synthetic maps, we generate the maps corresponding to the full length integrated up to the simulation box boundaries and corresponding only to the observer cells, for all observer view points and all simulation snapshots considered in this work. Examples of such maps are presented in Fig.~\ref{appfig:lenght_through_box}.

The path length from the observer cells is very small compared to the path length through the full box. However, if we do not include it in the sightline integration, the highly nonuniform features induced by the observer cells may produce strong patterns at the synthetic-map level when, for example, the disk-scale height right above and or right below the observer is small. Accounting for the path spent in the observer cell mitigates such artifacts.

\section{Map making and interpolation details}
\label{app:interpolation}

In this work, we wanted to avoid any effect that could mix angular scales because we are interested in characterizing the angular power spectra. Therefore, we developed the ray-tracing algorithm presented in Appendix~\ref{app:raytracing} to circumvent the use of interpolation techniques that could, in principle, lead to such unwanted artificial effects.
In this appendix, we show that the synthetic polarization maps may indeed depend on the choice of interpolation parameters.
It is not our intention to conduct a detailed analysis of the differences that may appear on a large sample of maps or to study the use of different interpolation schemes. Such a detailed analysis, while important, is beyond the scope of this paper.
Therefore, we carry out our analysis for the case of one interpolation scheme and one observer position in one simulated galaxy (corresponding to the maps shown in Fig.~\ref{fig:IQUmap_example}).

To synthesize maps from a Cartesian grid using interpolation, we proceed as follows. We consider a spherical grid surrounding the observer. This grid has an angular sampling determined by a HEALPix map with resolution $N_{\rm{side}} = 32$, and a radial sampling of 30 pc (approximately the size of the smallest cells in N18 simulations). At each node of the grid, we evaluate the dust density and the magnetic field vector through an interpolation scheme (a multiquadric radial-basis interpolation, working in three dimensions) and seek the information from the $n$ closest neighbors. The observables are then obtained through line-of-sight integration according to the mid-point rule.

\begin{figure}
    \centering
    \includegraphics[trim={0.4cm .4cm 0.4cm 0cm},clip,width=.95\linewidth]{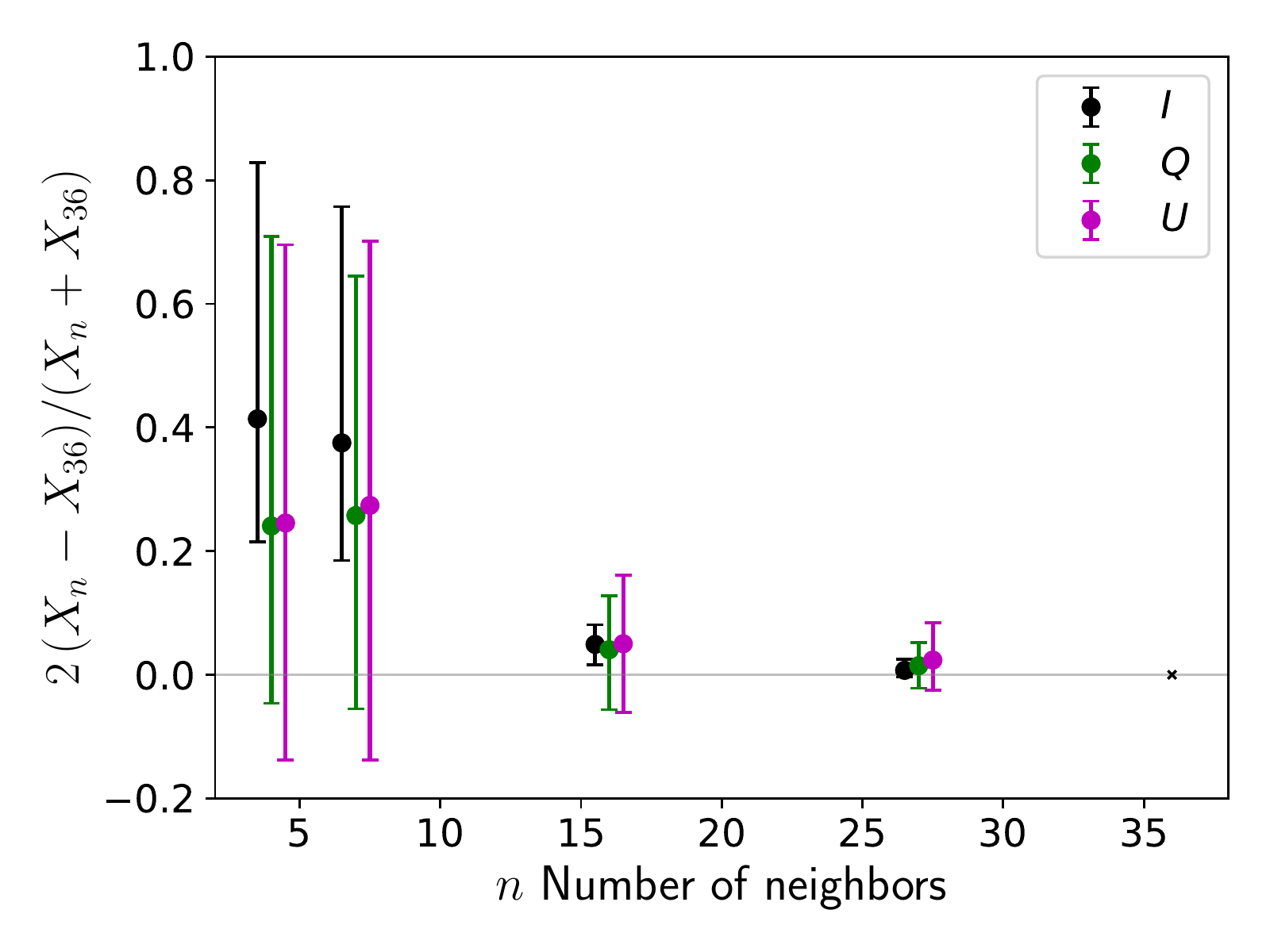}
    \caption{Summarizing statistics of the relative differences between maps obtained using $n$ neighbors in the interpolation scheme and $n=$36. For each value of $n$ we show the median of the relative differences and the error bars are such that they contain 64 per cent of the data. The observables ($X$) are shown as indicated by the legend. We slightly shift horizontally the values for $I$ and $U$ for a better visualization.
    }
    \label{fig:reldiff_n}
\end{figure}

We repeat the synthesis of the $I$, $Q$, and $U$ maps for different values of $n$: 4, 7, 16, 27, and 36.
First, we notice that the individual pixel values of the Stokes parameters change with the choice of $n$. Indications for convergence appear for large values of the closest neighbors involved in the interpolation. This is best illustrated in Fig.~\ref{fig:reldiff_n} where we present the relative difference of the $I$, $Q$, and $U$ maps as estimated for different $n$ and arbitrarily compared to $n = 36$.

\medskip

\begin{figure}
    \centering
    \includegraphics[trim={0.55cm 1.8cm 0.4cm 0cm},clip,width=.95\linewidth]{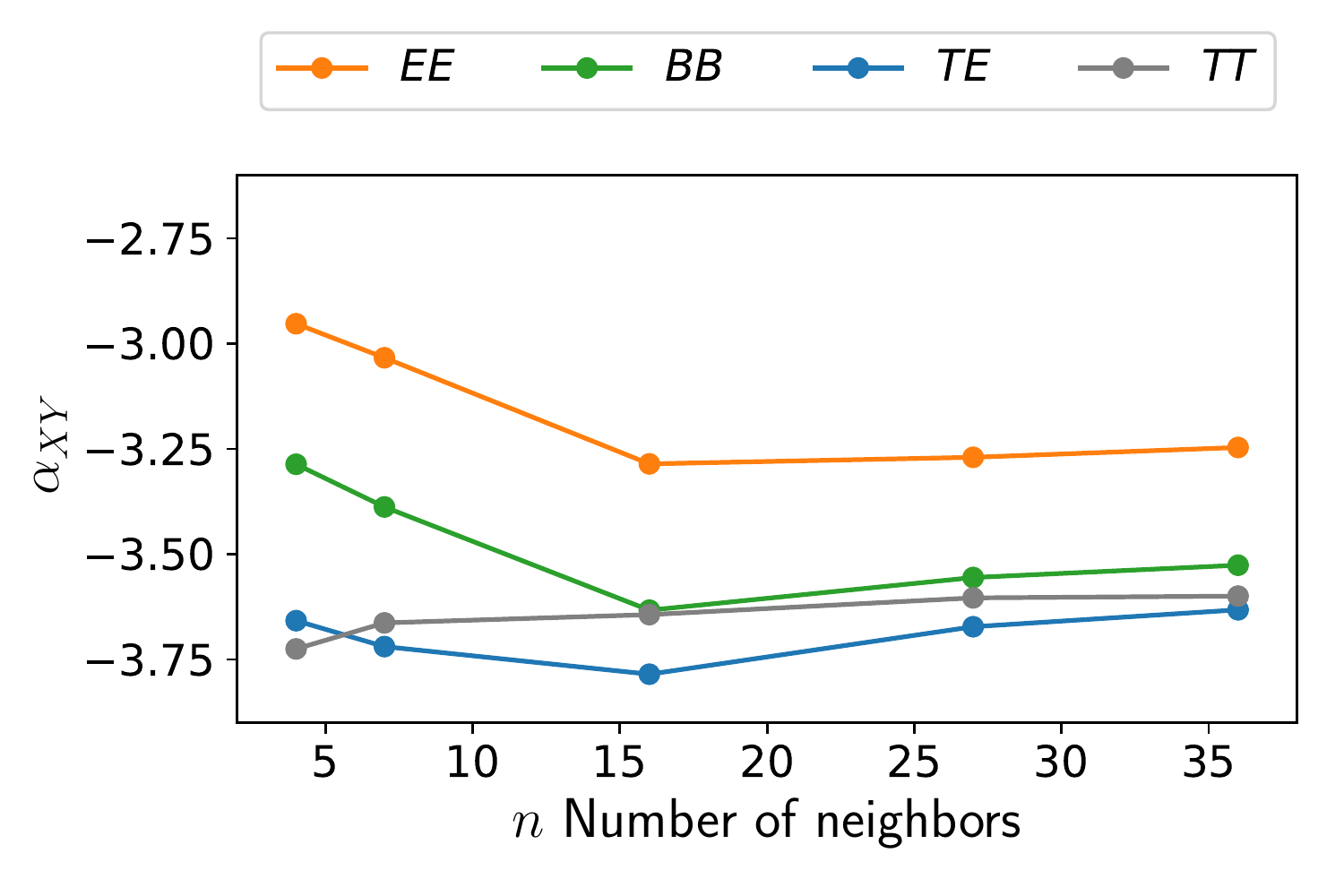} \\
    \includegraphics[trim={0.4cm 1.8cm 0.4cm 0.4cm},clip,width=.95\linewidth]{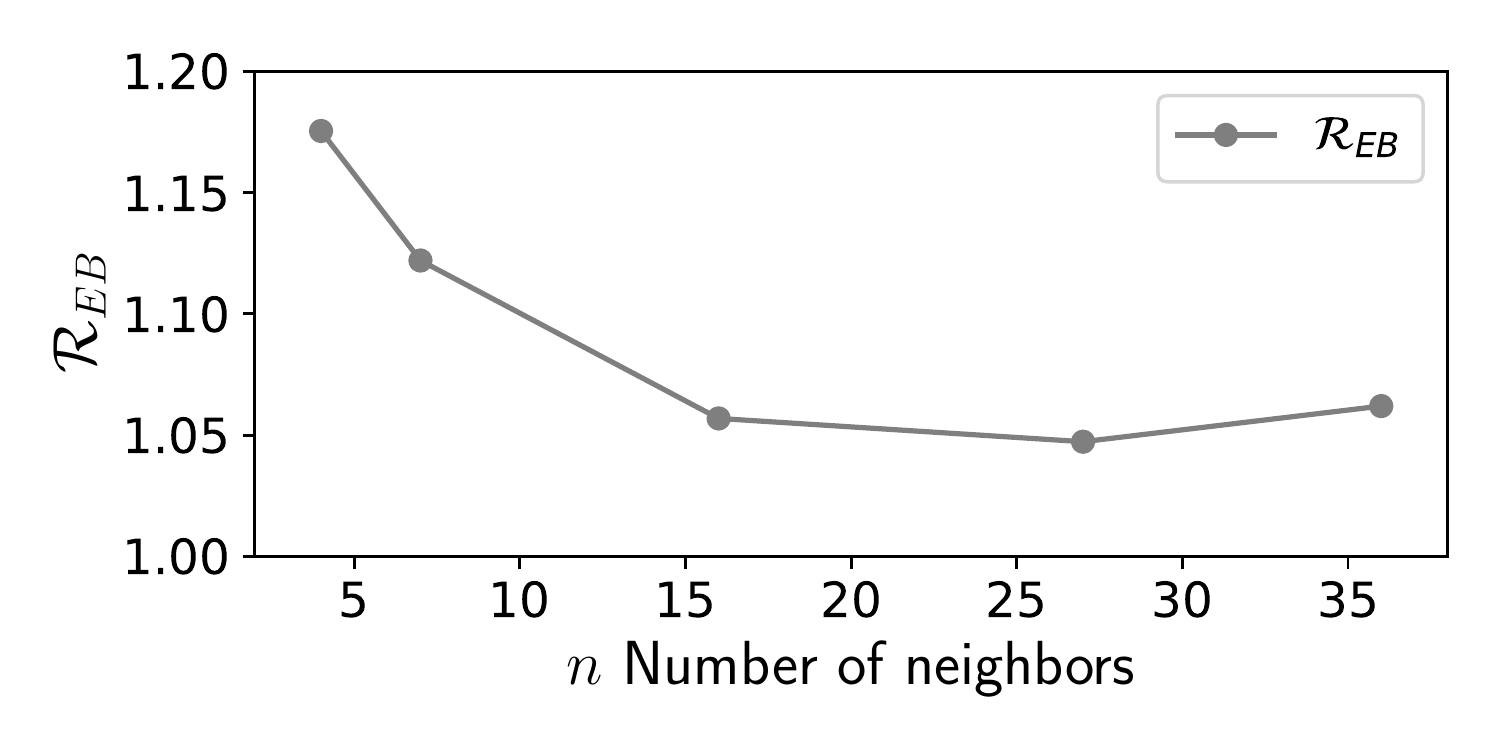} \\ \includegraphics[trim={0.4cm 0cm 0.4cm .4cm},clip,width=.95\linewidth]{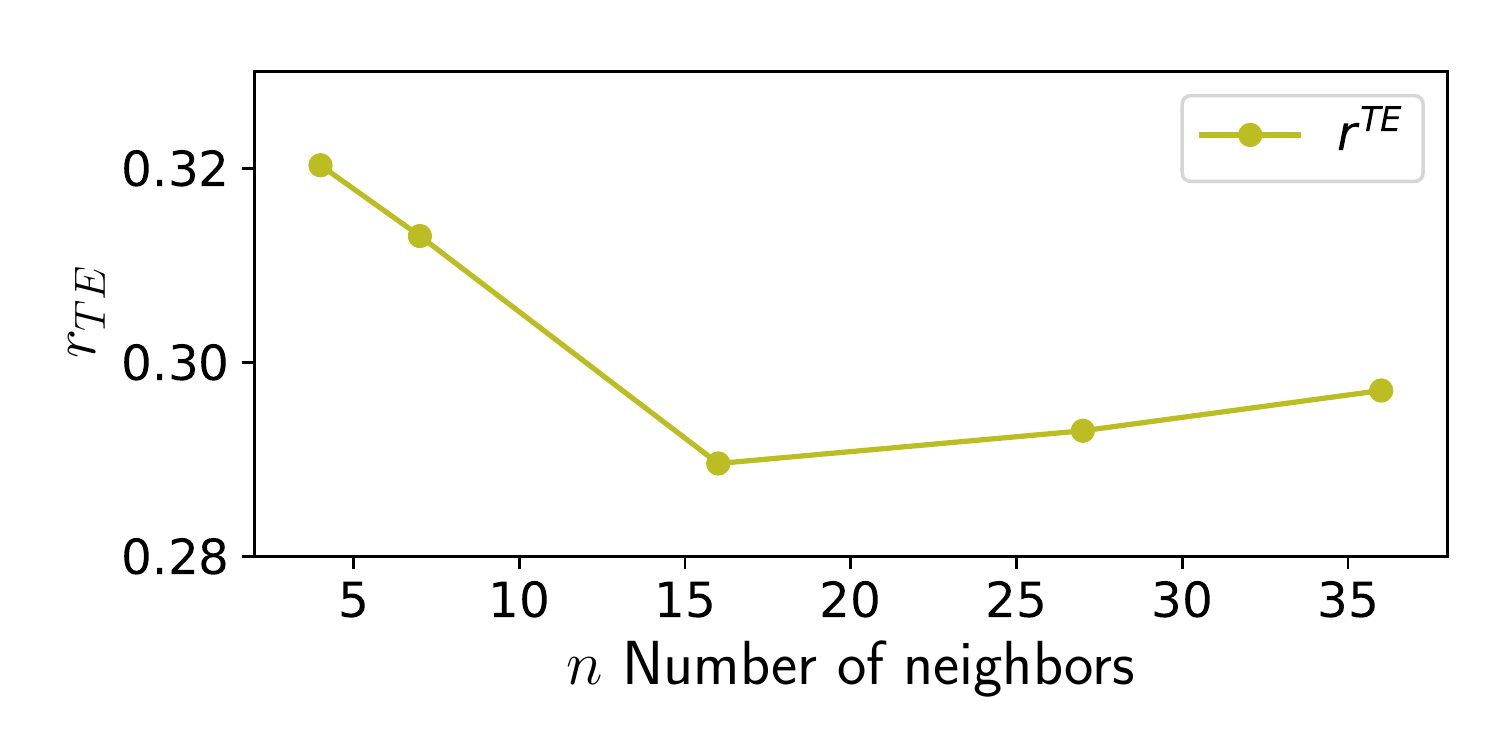} \\
    \caption{Power spectrum characteristics and number of neighbors in the interpolation scheme. We show the slopes of the $TT$, $EE$, $BB$, and $TE$ spectra (top), $\mathcal{R}_{EB}$ (middle) and $r^{TE}$ (bottom) as a function of $n$.
    }\label{fig:PS-vs-Nneigb}
\end{figure}

In the following, we explore the effect of these differences on the polarization power spectra.
We choose a sky fraction $f_{\rm{sky}} = 0.5$ and generate the mask corresponding to each intensity map following the same procedure as in the core of the paper.
We then define a common mask for all $n$ by multiplying the individual masks. We apply this mask to the five sets of polarization maps, and compute the polarization power spectra using \texttt{Xpol}. Because the maps are at a lower resolution than the sample presented in the main text, we consider the multipole bins between 40 and 80 with a bin width of 5.
Each set of maps leads to a set of auto- and cross-power spectra, for which we compute the $\mathcal{R}_{EB}$ ratio and the $r^{TE}$ correlation coefficients, and we fit the spectra with power-law functions. We show these values as a function of $n$ in Fig.~\ref{fig:PS-vs-Nneigb}. The power spectrum characteristics vary depending on the specific choice of the used interpolation scheme. This effect justifies our choice not to use a map-making algorithm that involves interpolation.

\end{appendix}
\end{document}